\documentclass[smallextended,final]{svjour3}

\usepackage{graphicx}
\usepackage{amssymb}
\usepackage{amsmath}
\usepackage{commath}
\usepackage{color}
\usepackage{subfig}
\usepackage{arydshln}
\usepackage{hyperref}
\usepackage{enumitem}
\usepackage{textcomp}
\usepackage{grffile}
\usepackage{multirow}

\usepackage[final,margin]{fixme}
\fxsetup{theme=color,mode=multiuser}
\FXRegisterAuthor{db}{dabi}{\color{blue}DABI}
\FXRegisterAuthor{ae}{apek}{\color{red}APEK}
\FXRegisterAuthor{ce}{clae}{\color{magenta}CLAE}
\FXRegisterAuthor{rv}{reviewer}{\color{green}REV}
\begin{document}

% \listoffixmes

%\tableofcontents

% \begin{frontmatter}

%% Title, authors and addresses

%% use the tnoteref command within \title for footnotes;
%% use the tnotetext command for the associated footnote;
%% use the fnref command within \author or \address for footnotes;
%% use the fntext command for the associated footnote;
%% use the corref command within \author for corresponding author footnotes;
%% use the cortext command for the associated footnote;
%% use the ead command for the email address,
%% and the form \ead[url] for the home page:
%%
%% \title{Title\tnoteref{label1}}
%% \tnotetext[label1]{}
%% \author{Name\corref{cor1}\fnref{label2}}
%% \ead{email address}
%% \ead[url]{home page}
%% \fntext[label2]{}
%% \cortext[cor1]{}
%% \address{Address\fnref{label3}}
%% \fntext[label3]{}

%\title{Generalized Polynomial Chaos for efficient Uncertainty Quantification in Nonlinear Water Wave Simulations}
% \title{Spectral Approach to Efficient Uncertainty Quantification in Nonlinear Water Wave Simulations}
\title{Efficient uncertainty quantification of\\a fully nonlinear and dispersive\\ water wave model with random inputs}
\titlerunning{Uncertainty quantification of a water wave model with random inputs}
% \title{A Stochastic Nonlinear Water Wave Model \\for Efficient Uncertainty Quantification}
% \titlerunning{A Stochastic Water Wave Model for Uncertainty Quantification}

%% use optional labels to link authors explicitly to addresses:
%% \author[label1,label2]{<author name>}
%% \address[label1]{<address>}
%% \address[label2]{<address>}

\author{Daniele Bigoni \and Allan P. Engsig-Karup \and Claes Eskilsson}

% \address[engsig-karup]{apek@dtu.dk}
%\address[bigoni]{dabi@dtu.dk}
\institute{D. Bigoni \and A.P. Engsig-Karup \at Department of Applied Mathematics and Computer Science, Technical University of Denmark, 2800 Kgs. Lyngby, Denmark \\\email{dabi@dtu.dk}
  \and
  C. Eskilsson \at Department of Shipping and Marine Technology, Chalmers University of Technology, SE--412 96 Gothenburg, Sweden}

\date{Received: date / Accepted: date}
% The correct dates will be entered by the editor

\maketitle

\begin{abstract}
A major challenge in next-generation industrial applications is to improve numerical analysis by quantifying uncertainties in predictions. In this work we present a formulation of a fully \rvnote*{\#1-1}{nonlinear} and dispersive potential flow water wave model with random inputs for the probabilistic description of the evolution of waves. The model is analyzed using random sampling techniques and non-intrusive methods based on generalized Polynomial Chaos (PC). These methods allow to accurately and efficiently estimate the probability distribution of the solution and require only the computation of the solution in different points in the parameter space, allowing for the \rvnote*{\#1-1}{reuse} of \rvnote*{\#1-2}{existing simulation software}. The choice of the applied methods is driven by the number of uncertain input parameters and by the fact that finding the solution of the considered model is computationally intensive.
We revisit experimental benchmarks often used for validation of deterministic water wave models. Based on numerical experiments and assumed uncertainties in boundary data, our analysis reveals that some of the known discrepancies from deterministic simulation in comparison with experimental measurements could be partially explained by the variability in the model input. We finally present a synthetic experiment studying the \rvnote*{\#1-3}{variance based} sensitivity of the wave load on an off-shore structure to a number of input uncertainties. In the numerical examples presented the PC methods have exhibited fast convergence, suggesting that the problem is amenable to being analyzed with such methods.
\end{abstract}



%% main text
\section{Introduction}
\label{sec:introdution}

%On uncertainties in wave kinematics and force estimations.
%\url{http://www.onepetro.org/mslib/servlet/onepetropreview?id=SUT-AUTOE-v29-075}

% Stress water waves and today's use...
In coastal and off-shore engineering it is important to design maritime structures that can withstand critical failures due to wave-induced loadings. The most extreme wave induced-loadings can be estimated from direct measurements, laboratory experiments and simulation-based tools\rvnote*{\#2-1}{,} which can account for the wave kinematics sufficiently accurately. It is still common to predict wave kinematics using numerical tools \rvnote*{\#2-2}{that} have been validated by a single or few deterministic simulations and compared to idealized physical experiments, e.g., in wave tanks. These validation procedures, and at a greater scale real field simulations, are subject to a number of uncertainties that could lead to unexpected results. The study of the influence of these uncertainties on the resulting engineering analysis and decisions requires ultimately a shift from a deterministic approach to a probabilistic approach \cite{Wojtkiewicz2001}. Such engineering analysis are useful for risk management aimed at reducing risk in design and operations.

The research field of Uncertainty Quantification (UQ) includes all the techniques used to rigorously study uncertainties entering a systems and their influence on its dynamics. These techniques deliver tolerance intervals and probability distributions of system outputs, denoted Quantities of Interest (QoIs). Upon the knowledge of a deterministic model describing the dynamical system, denoted the forward model, uncertainty quantification can be split in four steps:
\begin{enumerate}[label=(\alph*)]
\item identification of sources of uncertainty and QoIs,
\item quantification of uncertainty sources by means of probability distributions,
\item uncertainty propagation through the system,
\item sensitivity analysis.
\end{enumerate}

Uncertainties in coastal and off-shore engineering are either related to weather conditions or to structural/bathymetry characteristics. In the first case, the weather conditions are commonly grouped into a number of sea states characterized by a number of parameters, which determine particular probability distributions associated to waves, wind, currents, sea level and ice characteristics. Transitions between different sea states are modeled by stationary processes governing these parameters. The characterization of these uncertainties -- step (a-b) -- requires extensive measurements and the current \rvnote*{\#2-4}{state of the art} is presented in \cite{Bitner-Gregersen2014,Bitner-Gregersen2014a}.

A significant part of existing works on the propagation of uncertainty and sensitivity analysis of water waves use the Shallow Water Equations (SWE) as forward model, thus addressing mostly tidal waves, where vertical velocities are negligible. The focus is usually on the characterization of extreme responses or, in other words, the probability of failure of certain safety conditions.
In \cite{Naess2012} such analysis are performed using the MC method. 
In \cite{Ge2008} the Monte Carlo (MC) and the Polynomial Chaos (PC) method in the Galerkin and collocation form are compared when applied on the SWE modeling the propagation of a wave over a submerged hump. 
In \cite{Ge2011} the PC method in its collocation form is used for the study of uncertainties in flood-hazard mapping.
In \cite{Liu2009} random sampling methods, sparse grid, PC in Galerkin and collocation form, and a novel quadrature technique called Compound Uncorrelated Dimension (CUD) quadrature were compared when applied on the SWE modeling flood prediction under an uncertain river bed topography and characteristics. 
In \cite{Ricchiuto2014} a combination of non-intrusive (collocation) PC and ANOVA decomposition was used for the propagation and sensitivity analysis of the uncertain parameters entering the SWE modeling the runup of waves.
Unlike the preceding works, \cite{Yildirim2015} studies the influence of uncertainties on the phase-averaged equation, which is suitable for slowly varying wave fields, e.g. ocean waves in deep water. \cite{Yildirim2015} considers uncertainties entering the source term, the boundary conditions and the current field, adopting PC and ANOVA decomposition approaches.

\subsection{Paper contributions}

We consider a fully \rvnote*{\#1-1}{nonlinear} and dispersive potential flow model, which allows the study of the phase resolved propagation of water waves over varying bathymetry. The model has traditionally been consider computationally very costly, however, recent progresses in the design of scalable linear solvers has made the model much more tractable for improved analysis of predictions via uncertainty quantification methods. UQ analysis require an overall computational cost which often scales badly, due to the \textit{curse of dimensionality}, with the computational cost of the forward model. Thus, efficient numerical solvers for the forward model and efficient UQ techniques must be used. We employ state of the art software \cite{EngsigKarupEtAl2008,EngsigKarupEtAl2011,GlimbergEtAl2011,EGNL13,LEDN13} to obtain efficient and accurate simulations of the forward model, while we turn to recently developed techniques based on polynomial chaos \cite{Xiu09fastnumerical,Maitre2010} to perform uncertainty quantification.

These techniques will be applied to classical benchmarks, such as \cite{BB94,DutykhClamond2013}, where different sources of uncertainty and QoIs will be investigated. Due to the lack of data, some assumptions will be made about the probability distributions of the sources of uncertainty -- step (b), that in hydrodynamics simulations are commonly inlet/outlet conditions (boundary conditions), bathymetry data and structural positions (geometry). In experimental settings, all these uncertainties can be classified as \textit{epistemic} \cite{Kiureghian2009}, because they can in principle be reduced either by better measurements and/or by more precise experimental designs.

We will propagate the uncertainties through the dynamical system -- step (c) -- using traditional sampling techniques, such as Monte Carlo methods, and modern techniques based on generalized Polynomial Chaos (gPC) \cite{Xiu09fastnumerical,Maitre2010}. Non-intrusive approaches such as stochastic collocation and sparse grid will be preferred to intrusive approaches, due to the ability of the former of re-using existing code, avoiding the need for re-engineering existing software.

For one of the analyzed test cases we will identify the sources of uncertainty to which the QoI is most sensitive -- step (d). Techniques addressing this problem exist and we will focus on the variance based \rvnote*{\#2-5}{method of Sobol'} \cite{Sobol1993,Rabitz2000,Chan2000,Sudret2008,Crestaux2009,Alexanderian2012}, where the sensitivity of a QoI to a certain input uncertainty is expressed by the amount of variance of the QoI which is due to such input uncertainty. The identification of the least influential inputs allows the refinement of the model, where the uncertainties on these inputs are disregarded. The method will be used to quantify the sensitivity to a number of input uncertainties of the wave load on an off-shore structure. The sensitivities can be used to refine the forward model, disregarding uncertainties on parameters that do not influence significantly the QoIs.

\subsection{Paper organization}

The paper \rvnote*{\#2-6}{is} organized as follows. In section \ref{sec:forwardModel} we introduce the governing equation for the deterministic description of dispersive and  \rvnote*{\#1-1}{nonlinear} water waves based on potential theory. In section \ref{sec:modelWithRandomInputs} we describe how a model with random inputs can be formulated and parametrized.  Section \ref{sec:UQ} presents the theory of random sampling methods and of polynomial chaos methods for the forward propagation of uncertainty, and the description of the method of Sobol' for sensitivity analysis. In section \ref{sec:UQonWaterWaves}, the effect of parametric uncertainty in bathymetry and wave input are studied and numerical experiments are compared for different approaches. Section \ref{sec:Conclusions} contains concluding remarks.

\section{Mathematical formulation of the forward model}\label{sec:forwardModel}

We consider unsteady water waves described by a potential model for three-dimensional fully \rvnote*{\#1-1}{nonlinear} and dispersive free surface flows under the influence of gravity. The flow is assumed inviscid and irrotational. It can, without simplifications, be used for short and long wave propagation in both shallow and deep water where viscous and rotational effects are negligible. The sea bed is assumed variable and impermeable. 
% A review of the model is given in [Yu?Wu?].

We introduce a Cartesian coordinate system $(x,y,z)$ with $(x,y)$ the horizontal and $z$ the vertical dimensions, where the $z$ coordinate points upwards. The functions $h(x,y)$ and $\zeta(t,x,y)$ describe respectively the depth of the sea bed and the free surface. The still water level is given by $z=0$.

\subsection{The deterministic potential flow model}
The evolution of water waves over an arbitrary sea bed are described by the kinematic and dynamic free surface boundary conditions\footnote{The gravitational acceleration constant, $g$, is set to be 9.81 $m/s^2$. }
\begin{subequations}
\label{FreeSurfaceEqs}
\begin{align}
\partial_t\zeta({\bf x},t) &= -\boldsymbol{\nabla}\zeta\cdot\boldsymbol{\nabla}\tilde{\phi}+\tilde{w}(1+\boldsymbol{\nabla}\zeta\cdot\boldsymbol{\nabla}\zeta), \label{FSeta} \\
\partial_t \tilde{\phi}({\bf x},t) &= -g\zeta - \frac{1}{2}\left(\boldsymbol{\nabla}\tilde{\phi}\cdot\boldsymbol{\nabla}\tilde{\phi}-\tilde{w}^2(1+\boldsymbol{\nabla}\zeta\cdot\boldsymbol{\nabla}\zeta)\right),
\label{FSphi}
\end{align}
\end{subequations}
where $\boldsymbol{\nabla}=(\partial_x,\partial_y)$. We will consider waves in a spatial domain $D\in\mathbb{R}^l$ (fluid volume), $l=2,3$ and a time domain $t\in[0,T]$ with final time $T>0$.
For the fluid volume, a Laplace problem defines the scalar velocity potential
\begin{subequations}
\label{eq:laplaceproblem}
\begin{align}
\phi & =  \tilde{\phi}, \quad z = \zeta(\boldsymbol{x},t), \\
\boldsymbol{\nabla}^2\phi + \partial_{zz}\phi & = 0, \quad -h(\boldsymbol{x})\leq z<\zeta(\boldsymbol{x},t), \label{Laplace} \\
\partial_z \phi + \boldsymbol{\nabla}h\cdot\boldsymbol{\nabla}\phi &= 0, \quad z=-h(\boldsymbol{x}). \label{KB}
\end{align}
\end{subequations}
Using a classical $\sigma$-transformation
\begin{align}
\label{sigtrans}
\sigma(\boldsymbol{x},t) \equiv \frac{z+h(\boldsymbol{x})}{d(\boldsymbol{x},t)}, \quad 0\leq \sigma \leq 1,
\end{align}
the Laplace problem can be written as
\begin{subequations}
\label{TransformedLaplace}
\begin{align}
  \Phi & =  \tilde{\phi}, \quad \sigma = 1, \label{FSDirichlet} \\
  \boldsymbol{\nabla}^2\Phi+\boldsymbol{\nabla}^2\sigma(\partial_\sigma\Phi) +2\boldsymbol{\nabla}\sigma\cdot\boldsymbol{\nabla}(\partial_\sigma\Phi) &+ \nonumber \\ (\boldsymbol{\nabla}\sigma\cdot\boldsymbol{\nabla}\sigma+(\partial_z\sigma)^2)\partial_{\sigma\sigma}\Phi&=0, \quad 0\leq\sigma<1, \label{convlaplace}\\
{\bf n} \cdot (\boldsymbol{\nabla},\partial_z\sigma\partial_\sigma)\Phi & = 0, \quad ({\bf x},\sigma)\in\partial\Omega, \label{strucbcs}
\end{align}
\end{subequations}
where
\begin{subequations}
\label{timedepcoef}
\begin{align}
\boldsymbol{\nabla}\sigma & = \tfrac{1-\sigma}{d}\boldsymbol{\nabla}h
- \tfrac{\sigma}{d}\boldsymbol{\nabla}\zeta, \\
\boldsymbol{\nabla}^2\sigma & = \tfrac{1-\sigma}{d}\left(\boldsymbol{\nabla}^2
h-\tfrac{\boldsymbol{\nabla} h\cdot\boldsymbol{\nabla} h}{d}\right)
- \tfrac{\sigma}{d}\left(\boldsymbol{\nabla}^2 \zeta-\tfrac{\boldsymbol{\nabla}
\zeta\cdot\boldsymbol{\nabla}\zeta}{d}\right) 
-\nonumber \\ 
&\tfrac{1-2\sigma}{d^2}\boldsymbol{\nabla} h\cdot\boldsymbol{\nabla}\zeta
-\tfrac{\boldsymbol{\nabla}\sigma}{d}\cdot\left(\boldsymbol{\nabla} h 
+ \boldsymbol{\nabla}\zeta\right),\\
\partial_z \sigma & = 
\tfrac{1}{d}.
\end{align}
\end{subequations}
The relation between the scalar velocity potential function and velocity field is 
\begin{align}
({\bf u},w) = (\boldsymbol{\nabla}+\boldsymbol{\nabla}\sigma\partial_\sigma, \partial_z \sigma \partial_\sigma) \Phi.
\end{align}

The governing equations can be solved in the setting of a numerical wave tank and is then subject to initial and boundary conditions
\begin{align}
\zeta({\bf x},t=0)=\phi(\boldsymbol{x},t=0)=0, \quad \partial_n \zeta = \partial_n \phi = 0, \quad \boldsymbol{x}\in \partial D\backslash \bar{D}^{FS}, 
\end{align}
where wave generation and absorption is done using a line relaxation method \cite{LD83}. A complete derivation of the equations are given in \cite{EGNL13}. These model equations can be solved numerically using flexible-order finite differences \cite{EngsigKarupEtAl2008,EngsigKarupEtAl2011} and the massively parallel implementation \cite{LEDN13} enables fast hydrodynamics computations \cite{EGNL13}. A fast solver is a prerequisite for enabling UQ within acceptable time frames for realistic engineering applications.

\subsection{Calculation of loads on structures}\label{sec:calculationLoads}
The estimation of the pressure in the water column is needed for load calculations. We start from the momentum equation
\begin{align}
\partial_z p = \rho ( g + \partial_t w + u \partial_x w + v \partial_y w + w \partial_z w ),
\end{align}
where $p$ is the local pressure and $\rho$ is the fluid density. By integration  of the pressure in the vertical direction and assuming that the flow is inviscid and irrotational it is possible to derive the pressure equation 
\begin{align}
\frac{p(z)}{\rho} = g(\eta-z) + \int_z^\eta \partial_t w \dif z + \frac{1}{2} ( \tilde{u}^2 - u(z)^2 + \tilde{v}^2 - v(z)^2 + \tilde{w}^2 - w(z)^2).
\label{preseq}
\end{align}
The remaining integral term can be estimated numerically, e.g. by using a sufficiently accurate numerical quadrature rule.

For the estimation of structural forces, we can integrate the expression for the pressure given in Eq. (\ref{preseq}) in the vertical direction, to obtain the force $F$ as
\begin{align}
{\bf F} = -\int_S p {\bf n} \cdot \dif {\bf S},
\end{align}
which for a vertical structure takes the form
\begin{subequations}
\begin{align}
& \int_{-d}^\eta p(z) \dif z = F_{static} + F_{dynamic}\;, \\
& F_{static} = \rho(\eta + d) g \eta - \frac{1}{2} \rho g (\eta^2 - d^2)\;,  \\
&
  \begin{aligned}
    F_{dynamic} & = \rho \int_{-d}^\eta \left( \int_z^\eta \partial_t w(s) \dif s \right) \dif z + \frac{1}{2} \rho \left( \tilde{u}^2 + \tilde{v}^2 + \tilde{w}^2\right) (\eta+d) \\
    & - \frac{1}{2} \rho \int_{-d}^\eta (u(z)^2 + v(z)^2 + w(z)^2) \dif z\;.
  \end{aligned}
\end{align}
\end{subequations}

The integral terms can be estimated numerically.

\section{The model with random inputs}\label{sec:modelWithRandomInputs}
\rvnote*{\#3-1}{Despite the effort towards the emulation of idealistic phenomena, experimental settings suffer from unavoidable epistemic uncertainties.} \rvnote*{\#3-2}{In the context of water waves, these uncertainties characterize the experimental geometry as well as the boundary conditions. The parameters describing these uncertainties are collected into a $d$ dimensional vector ${\bf Z}:\Omega \rightarrow \mathbb{R}^d$ of random variables defined on a probability space $(\Omega,\mathcal{F},\mathcal{P})$ describing such uncertainties, where $\Omega$ is the space of elementary events, $\mathcal{F}$ is a $\sigma$-field and $\mathcal{P}:\mathcal{F}\rightarrow [0,1]$ is a probability measure\footnote{For a treatment of measure theoretic probability theory see e.g. \cite{Billingsley2008}}.} \rvnote*{\#3-3}{In the following we will avoid the direct use of the probability space $(\Omega,\mathcal{F},\mathcal{P})$ by the introduction of the image space $(\mathcal{R}^d,\mathcal{B}(\mathbb{R}^d),F_{\bf z})$, where $\mathcal{B}(\mathbb{R}^d)$ is the Borel $\sigma$-algebra over $\mathbb{R}^d$ and $F_{\bf z}({\bf z})=\mathcal{P}({\bf Z} \leq {\bf z})$ is the cumulative distribution function (CDF) of ${\bf Z}$. This will allow the evaluation of integrals of integrable functions of ${\bf Z}$ by noting that $\int_{\Omega}g({\bf Z}(\omega))\mathcal{P}(d\omega)=\int_{\mathbb{R}^d}g({\bf z})dF_{\bf z}({\bf z})$.}

The reformulation with random inputs of \eqref{FreeSurfaceEqs} and \eqref{eq:laplaceproblem} leads to a system of PDEs for the unknowns $\zeta({\bf x},t,{\bf Z}):\bar{D}^{FS}\times[0,T]\times\mathbb{R}^d\to\mathbb{R}$ and $\phi({\bf x},t,{\bf Z}):\bar{D}\times[0,T]\times\mathbb{R}^d\to\mathbb{R}$, which are now random fields:
\begin{subequations}
\label{FreeSurfaceEqsStochastic}
\begin{align}
\partial_t\zeta({\bf x},t,{\bf Z}) &= -\boldsymbol{\nabla}\zeta\cdot\boldsymbol{\nabla}\tilde{\phi}+\tilde{w}(1+\boldsymbol{\nabla}\zeta\cdot\boldsymbol{\nabla}\zeta), \label{FSetaStochastic} \\
\partial_t \tilde{\phi}({\bf x},t,{\bf Z}) &= -g\zeta - \frac{1}{2}\left(\boldsymbol{\nabla}\tilde{\phi}\cdot\boldsymbol{\nabla}\tilde{\phi}-\tilde{w}^2(1+\boldsymbol{\nabla}\zeta\cdot\boldsymbol{\nabla}\zeta)\right),
\label{FSphiStochastic}
\end{align}
\end{subequations}

% $\bar{D}$ is the closed spatial domain volume with FS indicating the restriction to the free surface, $\bar{D}=\{ {\bf x} | {\bf x} \in \xi \}$. 

% A parametrization of this model is required in order to solve it numerically. A set of random variables ${\bf Z}:\Omega \rightarrow \mathbb{R}^d$, is introduced to characterize the random inputs, where $d\geq 1$ is the dimension of the parameter space.
% Figure \ref{fig:BarTest:BottomProfiles} show an example of input uncertainty with $d=1$: the deterministic bottom topography for the submerged bar experiment in figure \ref{fig:BarTest:DetBottomProfile} is subject to uncertainty on the bar's height. One realisation of this uncertainty is shown in fig. \ref{fig:BarTest:UQHeightBottomProfile}.

% The stochastic reformulation of the deterministic system  \eqref{FreeSurfaceEqs} is
% where for any realization of an uncertain sea state or geometry of the bathymetry, the Laplace problem \eqref{eq:laplaceproblem} is fulfilled to obtain closure.

\subsection{Uncertain bathymetries, random fields and Karhunen-Lo\`{e}ve Expansion}\label{subsec:KL}
The bathymetry function describing still-water depth in the considered domain can be uncertain. In laboratory experiments, this can be due to manufacturing errors, whereas in the real settings this is often due to the lack of precise knowledge of the bottom topography or the unknown action of sedimentation. Since this uncertainty is spatially varying, it must be treated as a random field.
We will treat this random field using the Karhunen-Lo\`{e}ve (KL) expansion \cite{1977probability,Schwab2006} which is a useful parametrization technique when the field is characterized by a certain amount of spatial correlation.

Let $h({\bf x},\omega)$ be a spatially varying random field over a spatial domain $D$ with mean $\mu_h({\bf x})$ and covariance function $C({\bf x}_1,{\bf x}_2)=\textbf{Cov}(h({\bf x}_1,\omega),h({\bf x}_2,\omega))$. Then the bathymetry function $h({\bf x},\omega)$ can be parametrized as an infinite series
\begin{align}\label{eq:KL-expansion}
	h({\bf x},\omega) = \mu_h({\bf x}) + \sum_{i=1}^\infty \sqrt{\lambda_i}\psi_i({\bf x})Y_i(\omega),\
	%h({\bf x},\omega) = \mu_h({\bf x}) + \sigma_h \sum_{i=1}^\infty \sqrt{\lambda_i}\psi_i({\bf x})Y_i(\omega)
\end{align}
where the convergence is in $L^2$, $Y_i(\omega)$ are random variables, $ \mathbf{E}[Y_i(\omega)]=0 $, $ \mathbf{Cov}[Y_i,Y_j]=\delta_{ij} $ and $\{\lambda_i, \psi_i\}_{i=1}^\infty$ are the solutions of the generalized eigenvalue problem
\begin{align}\label{eq:KL-expansion-eig-problem}
	\int_D C({\bf x},{\bf s}) \psi_i({\bf s}) \dif {\bf s} = \lambda_i \psi_i({\bf x}).
\end{align}
If $h({\bf x},\omega)$ is a Gaussian random field, then $ Y_i \sim \mathcal{N}(0,1) $.

\begin{figure}
	\centering
	\subfloat[]{\label{fig:BarTest:FieldBottomProfile}\includegraphics[width=0.48\textwidth]{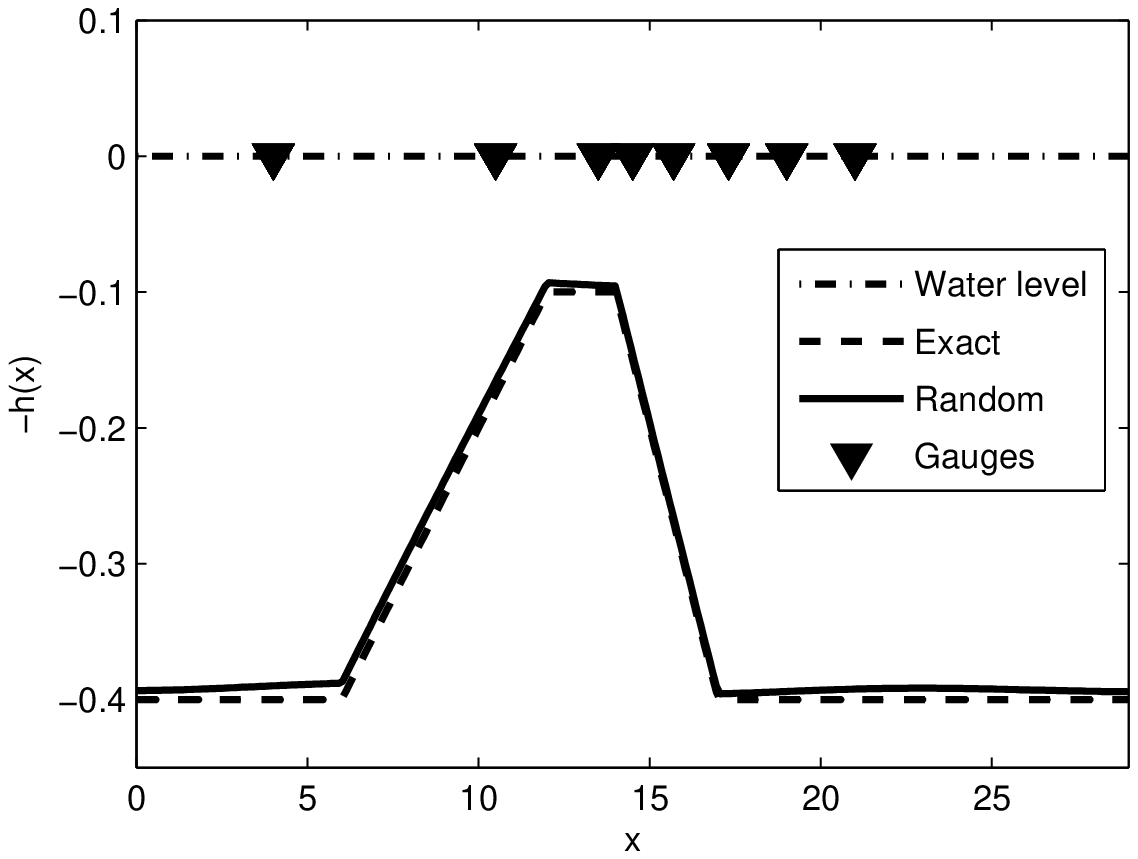}}
	\hspace*{5pt}
	\subfloat[]{\label{fig:BarTest:FieldBottomProfile01}\includegraphics[width=0.48\textwidth]{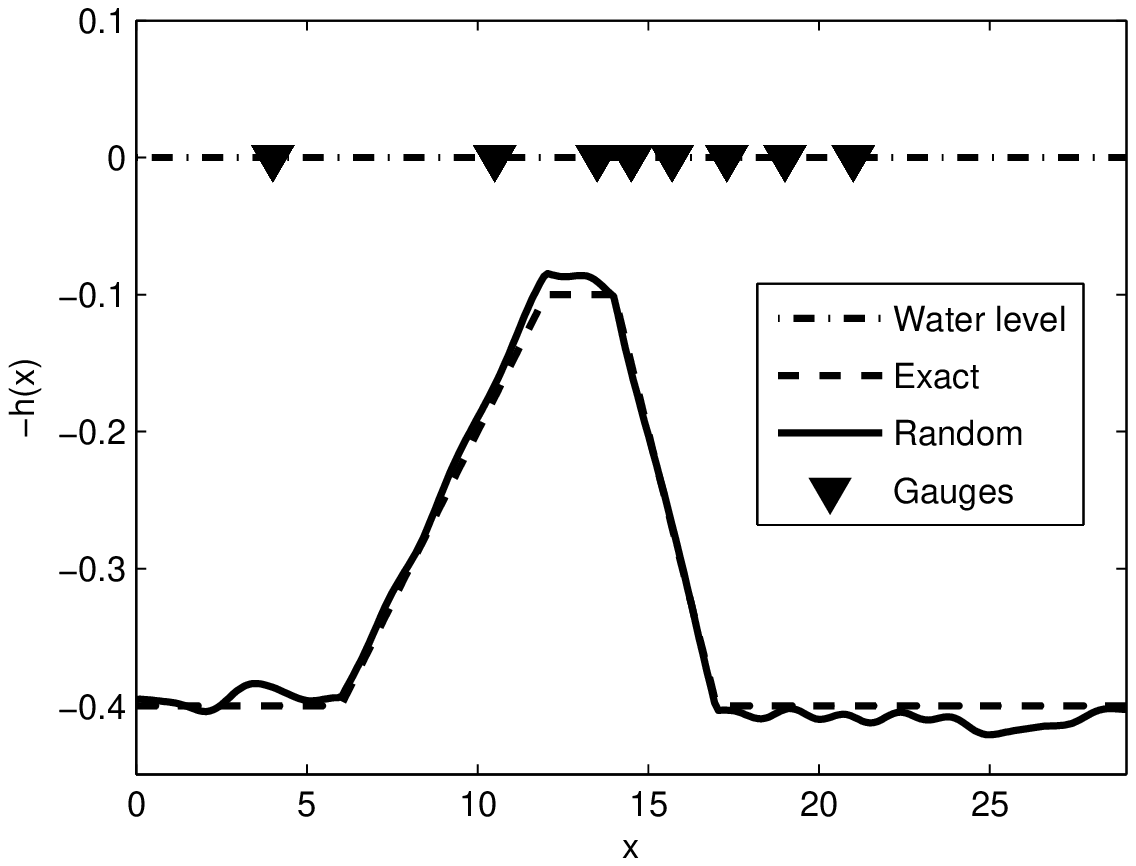}}	
	\caption{Possible topographies of the bottom floor in the submerged bar experiment. Figures \protect\subref{fig:BarTest:FieldBottomProfile} and \protect\subref{fig:BarTest:FieldBottomProfile01} show two realizations of the KL-expanded random fields \eqref{eq:KL-expansion} with different correlation lengths $a=30.0$ and $a=3.0$ respectively, where the total variance represented is $\geq 0.95$ and the mean field is set to be the nominal topography. The resulting numbers of retained terms in the KL-expansion \eqref{eq:KL-expansion} are 5 and 40 respectively.}
	\label{fig:BarTest:BottomProfiles}
        \rvnote{\#3-5}
\end{figure}

For practical computations \eqref{eq:KL-expansion} is truncated at a desired order $N$. It is easy to check how much of the variance of the original random field is retained by such approximation, using that
\begin{align*}
	{\bf Var}\left[h_N({\bf x},\omega)\right] &= {\bf E}\left[ h^2_N({\bf x},\omega) \right] - {\bf E}\left[h_N({\bf x},\omega)\right]^2 \nonumber \\
		&= {\bf E}\left[ \sum_{i,j=1}^N \sqrt{\lambda_i \lambda_j} \psi_i({\bf x}) \psi_j({\bf x}) Y_i(\omega) Y_j(\omega) \right] = \sum_{i=1}^N \lambda_i \psi^2_i({\bf x}),
\end{align*}
where the orthogonality of $ \left\lbrace Y_i \right\rbrace_{i=1}^N $ is exploited.
There are several options regarding the correlation kernel $C({\bf x}_1,{\bf x}_2)$. All these are problem dependent and an appropriate characterization of the random field has to be performed prior to the construction of the KL-expansion. In this work, we will use the exponential covariance kernel
\begin{align}\label{eq:KL-exp:exp-cov-kernel}
	C({\bf x}_1,{\bf x}_2) = \exp\left( -\frac{\left\Vert {\bf x}_1-{\bf x}_2 \right\Vert}{a} \right),
\end{align}
where $a$ is the correlation length. \rvnote*{\#3-6}{By this choice $h$ is then an Ornstein-Uhlenbeck \cite{Uhlenbeck1930} random field.} \rvnote*{\#3-5}{The eigenvalue problem \eqref{eq:KL-expansion-eig-problem} is solved numerically using a spectral discretization of the integral operator.}
Figure \ref{fig:BarTest:BottomProfiles} shows realizations of the KL-expansions of a 1D random field $h(x,\omega)$ for the submerged bar experiment considered in section \ref{sec:submergerbar} with exponential covariance kernel and zero mean for different correlation lengths $a$. The total variance represented by the KL-expansions $h_N(x,\omega)$ is fixed to $0.95$ (the total variance of $h(x,\omega)$ with exponential covariance kernel is $1$). In figure \ref{fig:BarTest:FieldBottomProfile} and \ref{fig:BarTest:FieldBottomProfile01}, fields with different correlation lengths are illustrated. Shorter correlation lengths determine a slower decay of the expansion coefficients in  \eqref{eq:KL-expansion} and thus an expansion retaining an higher number of terms is required to express higher local variability.

\section{Uncertainty Quantification}\label{sec:UQ}
We are interested in studying the propagation of uncertainties through the dynamical system \eqref{FreeSurfaceEqsStochastic}. To reduce the notation used, let $ \mathbf{u}(\mathbf{x},t,\mathbf{Z}) = [\zeta(\mathbf{x},t,\mathbf{Z}),\tilde{\phi}(\mathbf{x},t,\mathbf{Z})]^T $.
%and let \eqref{FreeSurfaceEqsStochastic} be rewritten as
%\begin{align}\label{eq:UQ:StochasticModel}
%	\begin{cases}
%		\mathbf{u}_t(\mathbf{x},t,\mathbf{Z}) = \mathcal{L}(\mathbf{u}), & \bar{D} \times (0,T] \times \mathbb{R}^d,\\
%		\mathcal{B}(\mathbf{u}) = 0, & \partial\bar{D} \times (0,T] \times \mathbb{R}^d, \\
%		\mathbf{u}(\mathbf{x},0,\mathbf{Z}) = \mathbf{u}_0(\mathbf{x},\mathbf{Z}), & \bar{D} \times \mathbb{R}^d,
%	\end{cases}
%\end{align}
%where $\mathcal{L}$ is a differential operator and $\mathcal{B}$ is the boundary condition operator.\\
We describe the results in terms of the probability distribution and/or the first moments, e.g., mean and variance, of ${\bf u}$:
\begin{align}
	\mathbf{E}[\mathbf{u}(\mathbf{x},t,\mathbf{Z})] &= \int_{\mathbb{R}^d} \mathbf{u}(\mathbf{x},t,\mathbf{z}) \dif F_\mathbf{z}(\mathbf{z}) = \int_{\mathbb{R}^d} \mathbf{u}(\mathbf{x},t,\mathbf{z}) \rho_\mathbf{z}(\mathbf{z}) \dif \mathbf{z} = \mu_\mathbf{u}(\mathbf{x},t), \\
	\mathbf{Var}[\mathbf{u}(\mathbf{x},t,\mathbf{Z})] &= \int_{\mathbb{R}^d} \left( \mathbf{u}(\mathbf{x},t,\mathbf{z}) -  \mu_\mathbf{u}(\mathbf{x},t) \right)^2 \rho_\mathbf{z}(\mathbf{z}) \dif \mathbf{z},
\end{align}
where \rvnote*{\#3-3}{$\rho_\mathbf{z}(\mathbf{z})$ is the Probability Density Function (PDF) of the random vector $\mathbf{Z}$}.

In this work we will focus exclusively on non-intrusive methods, which require a minimal development effort. In particular the existing solvers are considered as black boxes and the non-intrusive methods need only to be wrapped around them. On the contrary, intrusive methods require the development of new solvers based on mixed discretization of the stochastic and the spatial models. These methods are usually better in dynamically adapting to time-dependent problems \cite{Cheng2013,Cheng2013a,Boyaval2010,Venturi2006,Sapsis2009} but their implementation is often very demanding -- sometimes prohibitive -- for existing customized solvers.

\subsection{Pseudo-random sampling methods}
Among the existent UQ techniques, pseudo-random sampling methods are the most widely used. The most notable of these techniques, is the Monte Carlo (MC) method. It is based on the law of large numbers which states that given the random vector $\mathbf{u}: \bar{D} \times (0,T] \times \mathbb{R}^d \rightarrow \mathbb{R}^m $ and the functional $g:\mathbb{R}^m \rightarrow \mathbb{R}^m$,
\begin{align}\label{eq:MC}
	\frac{1}{n} \sum_{i=1}^n g\left(\mathbf{u}(\mathbf{x},t,\mathbf{z}^{(i)})\right) \xrightarrow{a.s.} \mathbf{E}\left[g\left(\mathbf{u}(\mathbf{x},t,\mathbf{Z})\right)\right] = \int_{\mathbb{R}^d} g\left(\mathbf{u}(\mathbf{x},t,\mathbf{z})\right) \dif F_{\mathbf{z}}(\mathbf{z}),
\end{align}
for $n \rightarrow \infty$. In the definition above $\left\lbrace \mathbf{z}^{(i)} \right\rbrace_{i=1}^n$ is a ensemble of samples drawn from the probability distribution of $\mathbf{Z}$ and a.s. stands for \textit{almost surely} implying convergence in probability. % For $ g_1:\mathbf{u} \mapsto \mathbf{u} $ and $ g_2:\mathbf{u} \mapsto (\mathbf{u} - \mu_\mathbf{u})^2 $,
% \begin{align}
% 	\mathbf{E}\left[\mathbf{u}(\mathbf{x},t,\mathbf{Z})\right] &= \mathbf{E}\left[g_1\left(\mathbf{u}(\mathbf{x},t,\mathbf{Z})\right)\right] \approx \frac{1}{n} \sum_{i=1}^n g_1\left(\mathbf{u}(\mathbf{x},t,\mathbf{z}^{(i)})\right) =: \bar{\mu}_\mathbf{u}(\mathbf{x},t),\\
% 	\mathbf{Var}\left[\mathbf{u}(\mathbf{x},t,\mathbf{Z})\right] &= \mathbf{E}\left[g_2\left(\mathbf{u}(\mathbf{x},t,\mathbf{Z})\right)\right] \approx \frac{1}{n} \sum_{i=1}^n g_2\left(\mathbf{u}(\mathbf{x},t,\mathbf{z}^{(i)})\right) =: \bar{\sigma}^2_\mathbf{u}(\mathbf{x},t).
% \end{align}
The probabilistic error of these approximations is reduced asymptotically as $ \mathcal{O}\left(1/\sqrt{n}\right) $. In spite of this slow convergence rate, MC method is widely used due to its robustness, ease of use and to the fact that it does not suffer the \textit{curse of dimensionality}, i.e., its convergence rate is independent from $d$. Improvements of the MC method have been proposed in order to cover more uniformly the parameter space, obtaining improved convergence rates.

The Latin \rvnote*{\#4-2}{Hyper-cube Sampling (LHS)} method \cite{Mckay2000} divides the parameter space in $n$ equiprobable bins along each dimension and samples are taken such that each bin contains only one sample. This provides a better convergence rate in the average cases, even if the worst case convergence rate remains $ \mathcal{O}\left(1/\sqrt{n}\right) $.

The Quasi Monte Carlo (QMC) method \cite{Morokoff1995} generates low discrepancy sequences of points such that the domain is uniformly covered. The convergence rate of QMC is improved to $ \mathcal{O}(\ln(n)^d/n) $, but the dimension of the parameter space becomes relevant.

\subsection{Deterministic sampling methods}
In the following we will handle functions with finite variance, i.e. belonging to the weighted $L^2_{\rho_{\mathbf{z}}}$ space defined as
\begin{align}\label{eq:UQ:L2space}
	L^2_{\rho_{\mathbf{z}}} = \left\lbrace f:\mathbb{R}^d \rightarrow \mathbb{R} \left\vert \int_{\mathbb{R}^d} f^2(\mathbf{z}) \rho_{\mathbf{z}} \dif \mathbf{z} < \infty \right. \right\rbrace,
\end{align}
with inner product and norm defined as
\begin{align}
	\left(f,g\right)_{\rho_{\mathbf{z}}} = \int_{\mathbb{R}^d} f(\mathbf{z}) g(\mathbf{z}) \rho_{\mathbf{z}} \dif \mathbf{z} \;, \qquad
	\left\Vert f \right\Vert_{\rho_{\mathbf{z}}} = \sqrt{\left( f,f \right)_{\rho_{\mathbf{z}}}} \;.
\end{align}
For many standard one-dimensional distributions with density $\rho_{z}$, we can find the set of uni-variate polynomials $ \left\lbrace \Phi_i(z) \right\rbrace_{i=0}^\infty$ that form an orthonormal basis for $L^2_{\rho_{z}}$ \cite{Xiu:2002:WPC:587159.587325,Xiu2010}. For instance, orthogonal basis for the space of random variables with Beta, Gaussian or Gamma distributions are the Jacobi, the \rvnote*{\#4-3}{probabilists'} Hermite and the Laguerre polynomials respectively. \rvnote*{\#5-4}{If the distribution is not standard, but admits a probability density function, then one can still use Gram-Schmidt orthogonalization to create suitable polynomials, or use the Stieltjes procedure to compute the recursion coefficients of the polynomials orthogonal with respect to such measure (see \cite{Gautschi2004,Gautschi1994})}. 

For random vectors $\mathbf{Z}:\Omega \rightarrow \mathbb{R}^d$ of mutually independent random variables $Z_1,\dots,Z_d$ with densities $ \rho_{z_1},\dots,\rho_{z_d} $ and basis $\{\phi_{i_d}(z_d) \}_{i_d=1}^N$, it holds that $\rho_{\bf z}(\mathbf{z}) = \prod_{i=1}^d \rho_{z_i}(z_i)$. Thus, a possible set of basis functions for $L^2_{\rho_\mathbf{z}}$ is given by the set of multivariate polynomials $\left\lbrace \Phi_\mathbf{i} \right\rbrace_{\max{\mathbf{i}}\leq N} $ where $\mathbf{i} = (i_1,\dots,i_d)$ is a multi-index and
\begin{align}
	\Phi_\mathbf{i}(\mathbf{z}) = \phi_{i_1}(z_1) \cdot \ldots \cdot \phi_{i_d}(z_d).
\end{align}
This construction includes $N^d$ basis functions and is more accurate than the required order $N$. An alternative set of basis is given by $\left\lbrace \Phi_\mathbf{i} : 0 \leq \vert \mathbf{i} \vert \leq N \right\rbrace $, where $\vert \mathbf{i} \vert = \sum_{j=1}^d i_j$. For this set of basis
\begin{align}
	\mathrm{dim} \left( \mathrm{span} \left\lbrace \Phi_\mathbf{i} \right\rbrace_{ \vert \mathbf{i} \vert = 0}^N \right) = \left(\begin{array}{c} N+d \\ N \end{array} \right) \ll N^d \;, \quad \text{for } d\gg1\;.
\end{align}

Once we have identified these basis function, we can define the projection operator $P_N:L^2_{\rho_{\mathbf{z}}}\rightarrow \mathrm{span}\left(\left\lbrace \Phi_{\bf i}(\mathbf{z}) \right\rbrace_{|{\bf i}|=0}^N\right) $ as
\begin{equation}
\label{eq:UQ:gPCexpansion}
f(\mathbf{z}) \approx \tilde{f}(\mathbf{z}) = P_N f(\mathbf{z}) = \sum_{|{\bf i}|=0}^N \hat{f}_{\bf i}\Phi_{\bf i}(\mathbf{z}), \qquad \hat{f}_{\bf i} = \left(f,\Phi_{\bf i}\right)_{\rho_{\mathbf{z}}}.
\end{equation}
This provides an approximation $\tilde{f}$ of the target function $f$ that is known as the \textit{generalized Polynomial Chaos (gPC) expansion} of $f$. Statistics of $f$ can be computed easily, e.g.
\begin{align}
	\mathbf{E}[f(\mathbf{z})] &\approx \mathbf{E}[\tilde{f}(\mathbf{z})] = \hat{f}_{\bf 0}, \\
	\mathbf{Var}[f(\mathbf{z})] &\approx \mathbf{Var}[\tilde{f}(\mathbf{z})] = \sum_{|{\bf i}| =1}^N \hat{f}_{\bf i}^2 ,
\end{align}
where the orthonormality of the basis $\left\lbrace \Phi_{\bf i}(\mathbf{z}) \right\rbrace_{i=0}^N$ is exploited. 

The convergence of the polynomial approximation \eqref{eq:UQ:gPCexpansion} is spectral (super linear) if $f$ is a smooth function or algebraic otherwise, cf. \cite{Gautschi2004,Canuto2006}.
The coefficients $\hat{f}_i$ in  \eqref{eq:UQ:gPCexpansion} can be obtained by means of two methods: the \textit{Galerkin method}, where a reformulation of  \eqref{FreeSurfaceEqsStochastic} in terms of stochastic modes is required, or the \textit{collocation method}, where approximations of $\hat{f}_i$'s are obtained by solving  \eqref{FreeSurfaceEqsStochastic} on carefully selected points in the parameter space. The Galerkin method is \textit{intrusive}, i.e. the problem needs to be reformulated (see \cite{Xiu2010,Maitre2010} for an introduction to Galerkin methods in this context), whereas the collocation method is \textit{non-intrusive} and thus any existing deterministic solver for  \eqref{FreeSurfaceEqs} can be used without modification.

\subsubsection{Stochastic Collocation method}\label{subsubsec:SCM}

The idea of the Stochastic Collocation (SC) method is to produce an ensemble of solutions $\mathbf{u}^{(j)}$, $i=1,...,M$ obtained by solving the governing equations  \eqref{FreeSurfaceEqsStochastic} at carefully selected points $S_N=\{ \mathbf{z}^{(j)} \}_{j=1}^M$ in the parameter space, in order to enable high accuracy in the evaluation of the coefficients of the gPC-expansion  \eqref{eq:UQ:gPCexpansion}. An alternative approach is the interpolation method, but this is out of the scope of this work (see e.g. \cite{Xiu2010,Maitre2010}). 

Let $\left\lbrace \left(z_i^{(j)},w_i^{(j)} \right) \right\rbrace_{j=0}^P $ be Gauss-type quadrature points and weights associated to $\rho_{z_i}$, which can be obtained using the Golub-Welsch method \cite{Golub1969}. These rules are exact when the integrand have a polynomial order up to $2P+1$. Multidimensional Gauss-type quadrature rules defined by $\left\lbrace \left({\bf z}^{({\bf j})},{\bf w}^{({\bf j})} \right) \right\rbrace_{\max {\bf j}<P}$ can be derived as \rvnote*{\#4-6}{${\bf z}^{({\bf j})}:=\left(z_1^{(j_1)}, \ldots, z_d^{(j_d)}\right)$} and ${\bf w}^{({\bf j})}:= w_1^{(j_1)} \cdot \ldots \cdot w_d^{(j_d)}$\rvnote*{\#4-5}{, where ${\bf j}$ is a multi-index defined as ${\bf j}=(j_1,\ldots,j_d)$}. Then the coefficients $\hat{f}_i$ in \eqref{eq:UQ:gPCexpansion} can be approximated by
\begin{equation}
  \label{eq:UQ:SCM}
  \hat{f}_{\bf i} = \left( f, \Phi_{\bf i} \right)_{\rho_z} = \int_{\mathbb{R}^d} f({\bf z}) \Phi_{\bf i}({\bf z}) \rho_z({\bf z}) \dif {\bf z} \approx \sum_{\max {\bf j} < P} f({\bf z}^{({\bf j})}) \Phi_{\bf i}({\bf z}^{({\bf j})}) {\bf w}^{({\bf j})} \;.
\end{equation}
This procedure differs from the classical Monte Carlo method only by the sampling technique used to select collocation points in the associated parameter space. The construction of the quadrature rule is based on the tensor product of 1-dimensional rules leading to the exponential growth of the total number of collocation points, i.e. it suffers the \textit{curse of dimensionality}.

\begin{figure}
	\centering
	\subfloat[]{\label{fig:UQ:FEJTensorGrid}\includegraphics[width=0.32\textwidth]{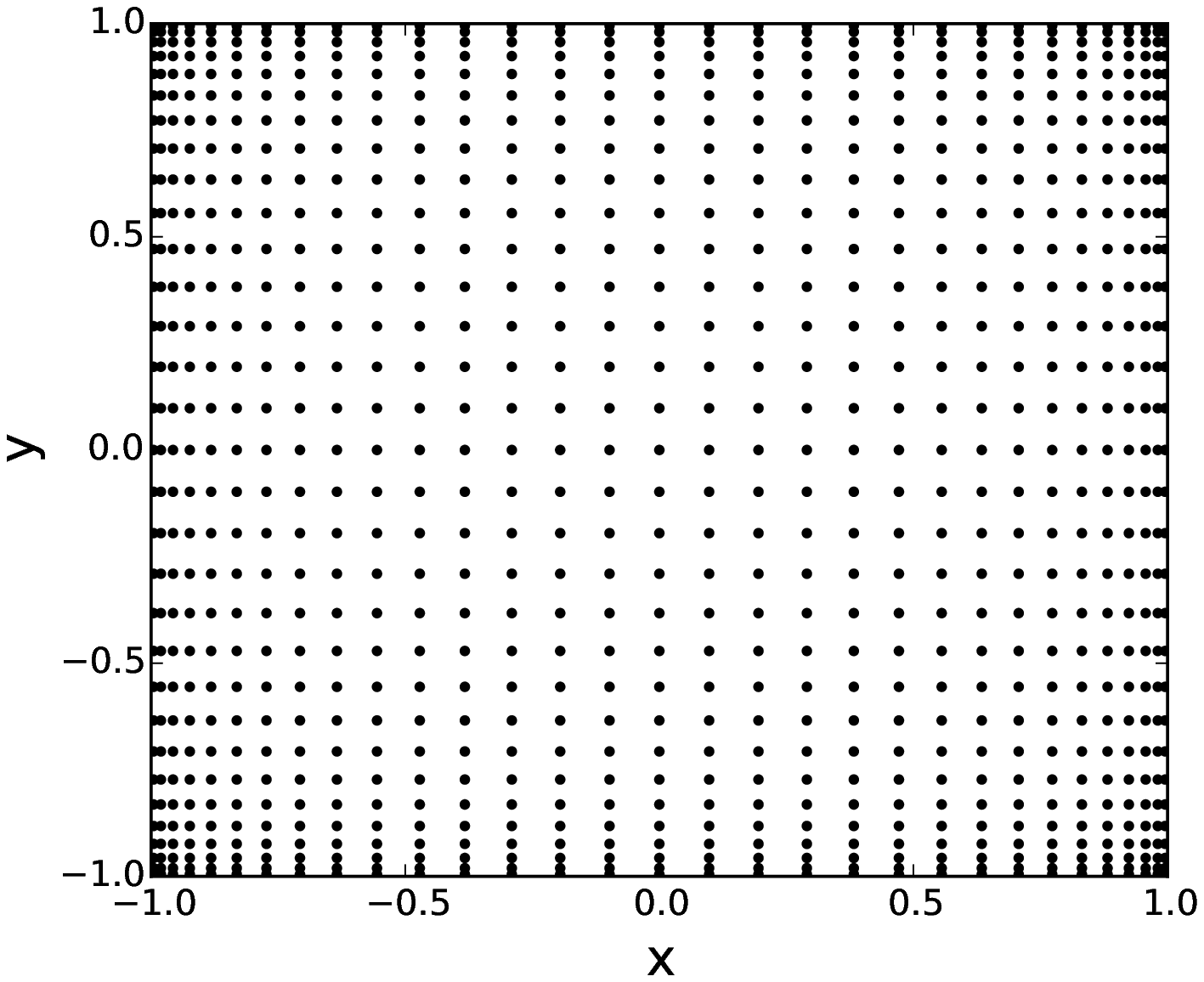}}
	\subfloat[]{\label{fig:UQ:FEJSparseGrid}\includegraphics[width=0.32\textwidth]{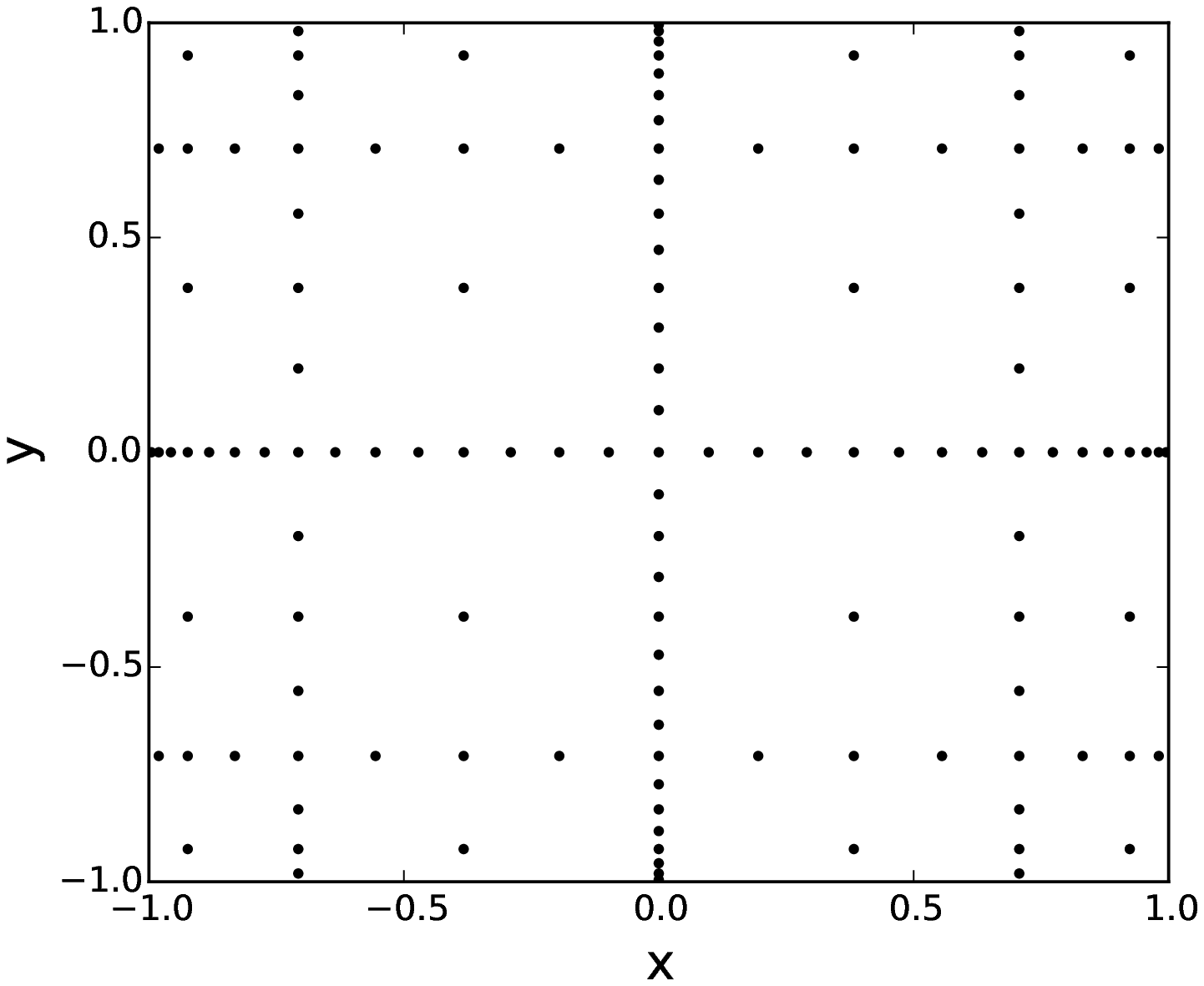}}
	\subfloat[]{\label{fig:UQ:FEJSG_Npoints}\includegraphics[width=0.32\textwidth]{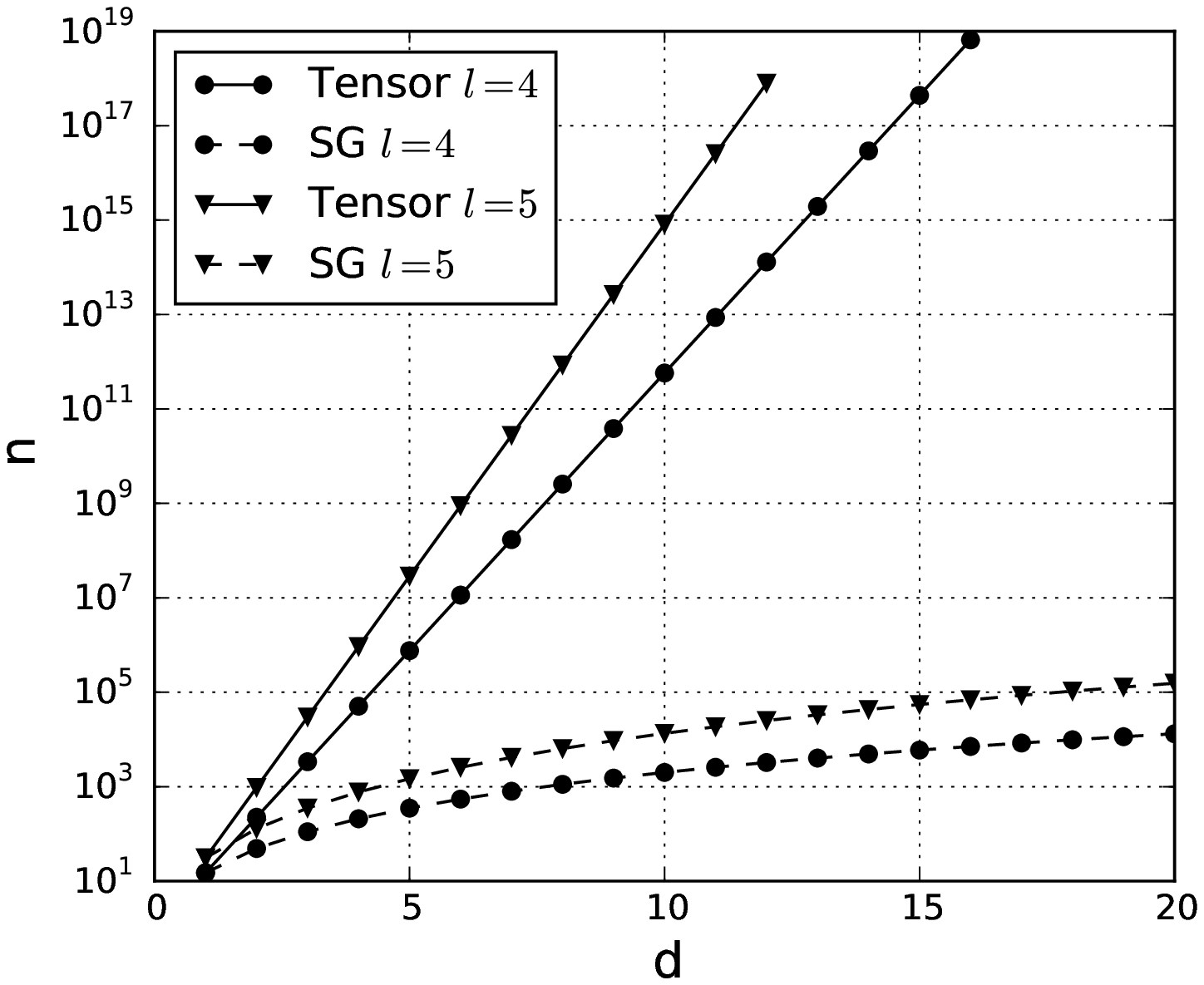}}
	\caption{A tensor grid -- fig. \protect\subref{fig:UQ:FEJTensorGrid} -- and a sparse grid -- fig. \protect\subref{fig:UQ:FEJSparseGrid} -- of order $l=5$ based on Fej\'{e}r's quadrature rule \cite{Fejer1933,Waldvogel2006}. Fig. \ref{fig:UQ:FEJTensorVSsg} shows how the number of quadrature points scale with respect to $d$ for $l=5$.}
	\label{fig:UQ:FEJTensorVSsg}
\end{figure}

Before addressing the curse of dimensionality, we need first to observe that quadrature rules based on the zeros of orthogonal polynomials are not \textit{nested} in general, meaning that the quadrature points $\Theta_l$ based on the polynomial of order $l$ are not in $\Theta_{l'}$, with $l' \geq l$. This property is important in practical calculations in case one would like to increase the accuracy without wasting results already computed. Common choices of nested quadrature rules are the Clenshaw-Curtis and Fej\'er's \cite{Clenshaw1960,Fejer1933,Waldvogel2006}, that uses the maxima of the Chebyshev polynomials as quadrature points, and the Kronrod-Patterson rules \cite{Kronrod1965}. With appropriate scaling, this quadrature rule works on the bounded interval $[0,1]$ with a probability density function $\rho(z)=1$, corresponding to the uniform distribution. In general we will have to compute integrals as in  \eqref{eq:UQ:SCM}, where $\rho_z$ does not need to be the uniform density function. However, using the fact that the CDF $F_z$ is bijective, we can use a variable transformation s.t.
\begin{align}
	\int_\mathbb{R} f(z) \rho_z(z) \dif z = \int_0^1 g(x) \dif x, \qquad g(x)\equiv f(F_z^{-1}(x)).
\end{align}

Using these nested rules, we can attempt to alleviate the curse of dimensionality. One particularly successful approach is given by \textit{Sparse grid} (see \cite{Petras2003,Conrad2013} for details). The idea is not to take the complete tensor product of the 1-dimensional grids, but only products up to the desired order for each stochastic dimension, very much alike the construction of the set of basis $\left\lbrace \Phi_\mathbf{i} \right\rbrace_{ \vert \mathbf{i} \vert = 0}^P $. This procedure assumes a certain level of separability of the function, meaning that the cross-contribution of the parameters is lower than the contribution of the parameters considered separately. Figure \ref{fig:UQ:FEJTensorVSsg} shows a comparison of tensor grids and sparse grid. From figure \ref{fig:UQ:FEJSG_Npoints} we can see that the gain given by sparse grid over tensor grids increases with the stochastic dimension $d$. 

\subsection{Sensitivity analysis using the method of Sobol'}\label{sec:sensitivityAnalysis}
Sensitivity analysis is aimed at the quantification and the ranking of the influence of input uncertainties on QoIs. This analysis allows then to refine the model, disregarding input uncertainties that have negligible effects on the QoIs. \rvnote*{\#4-7}{In many engineering fields the sensitivity of a QoI to its parameters is interpreted as its gradient with respect to them \cite{Maly1996,Errico1997,Cao2003,Ulbrich2007,Herzog2010}, and a ranking of the parameters is obtained observing the absolute value of the partial derivatives of the QoI. This kind of sensitivity analysis is termed ``local'', in order to stress the locality nature of the gradient. From the UQ perspective the uncertainty in the input parameters of a QoI is not necessarily ``local'', but it has often a non negligible variance. For this reason we will focus on the variance based ``global'' sensitivity analysis which expresses the sensitivity of a QoI to its parameters by the amount of variance in the QoI distribution that is caused by such parameters.}

For the sake of simplicity, we let the QoI be described by the scalar function $f:\mathbb{R}^d\rightarrow \mathbb{R}$ of the inputs. The influence of an input uncertainty on a QoI is quantified by the variance of the QoI that is due to the input uncertainty and its combination with any other group of uncertainties. This definition is based on the concept of ANOVA decomposition
\begin{equation}
  \label{eq:ANOVAdecomposition}
  {\bf Var}[f] = \sum_i D_i + \sum_{i < j} D_{ij} + \cdots + D_{1,2,\ldots,d}\;,
\end{equation}
where $D_i$ are the contributions to the variance due to input $i$ alone, $D_{ij}$ are the contributions to the variance due to inputs $i$ and $j$, and so on. The \rvnote*{\#5-6}{Main Sensitivity $S_{\bf i}$} and Total Sensitivity $TS(i)$ of the QoI on input $i$ is then defined as
\begin{equation}
  \label{eq:TotSensitivity}
  TS(i) = 1-S_{\neg i}\;,\qquad S_{i_1,\ldots,i_l} = \frac{D_{i_1,\ldots,i_l}}{D}\;,
\end{equation}
where $S_{\neg i}$ is the sum of all $S_{\bf i}$ for which $i\notin {\bf i}$. Note that the $TS(i)$ indices do not sum to one, because for $i\neq j$ there are many multi-indices ${\bf i}$ \rvnote*{\#2-3}{that} contain both of them. Even for mild dimensions $d$, the number of terms in the decomposition \eqref{eq:ANOVAdecomposition} grows exponentially, thus it needs to be truncated considering a limited number of cross-interactions between inputs. A good heuristic for this truncation is to use the fact that by \eqref{eq:ANOVAdecomposition} the components $D_{\bf i}$ must sum up to the total variance. Then, for $0 < q \leq 1$ one seek $0\leq L \leq d$ such that
\begin{equation}
  \label{eq:effectiveDimensionality}
  \sum_{0 < \# {\bf i} \leq L} D_{\bf i} \geq q {\bf Var}[f] \;,
\end{equation}
where $\# {\bf i}$ is the number of indices contained in the multi-index ${\bf i}$. The number $L$ is called the \textit{effective dimension} of $f$.

\begin{figure}
  \centering
  \subfloat[]{\label{fig:P1:Ch5:HDMR:Points:FullCubature3D}    \includegraphics[width=0.48\textwidth]{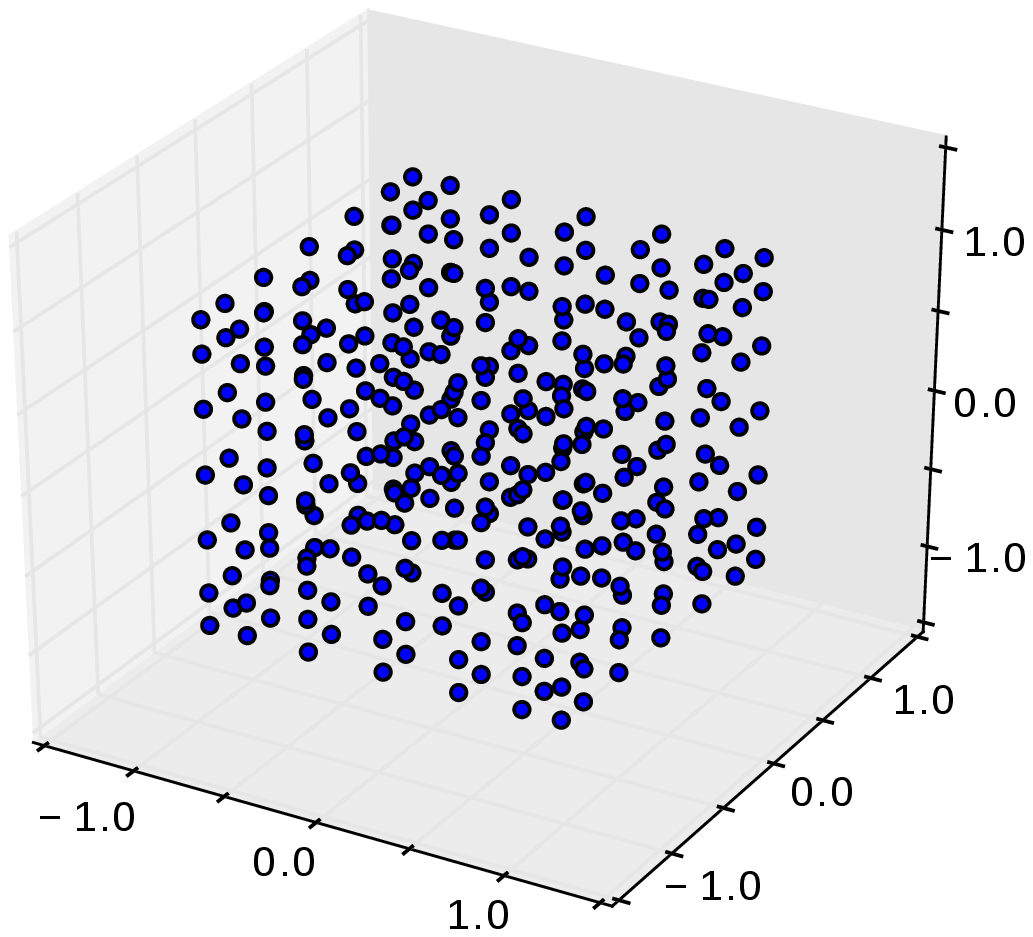}}
  \hspace{5pt}
  \subfloat[]{\label{fig:P1:Ch5:HDMR:Points:CutHDMR}\includegraphics[width=0.48\textwidth]{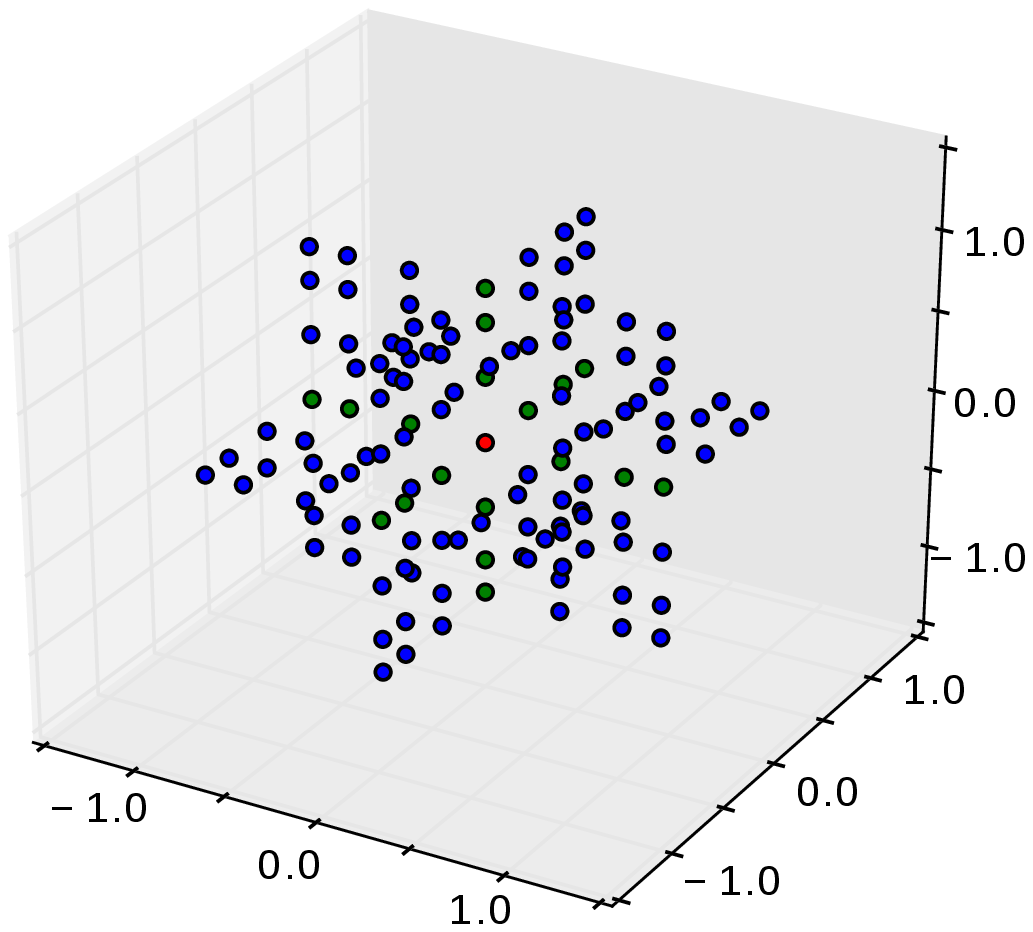}}
  \caption{Full cubature rule -- fig. \protect\subref{fig:P1:Ch5:HDMR:Points:FullCubature3D} -- and cubature for the construction of the cut-HDMR including 2-nd order interactions, for $d=3$ -- fig. \protect\subref{fig:P1:Ch5:HDMR:Points:CutHDMR}.}
  \label{fig:P1:Ch5:HDMR:Points}
\end{figure}

The calculation of the factors $D_{\bf i}$ is based on the construction of projection operators onto mutually orthogonal spaces -- see \cite{Sobol1993,Rabitz2000,Chan2000}. These projection operators involve high dimensional integrals which can be evaluated through random sampling or through the introduction of a surrogate function
\begin{equation}
  \label{eq:cut-HDMR}
  f({\bf x}) \simeq f^C({\bf x}) = \underbrace{f^C({\bf x}_0) + \sum_i f^C_i({\bf x}_i) + \sum_{i<j} f_{ij}^C({\bf x}_i,{\bf x}_j) + \cdots}_{\text{$L$-th order interactions}} \;,
\end{equation}
called the cut (anchored) High Dimensional Model Representation (cut-HDMR), where ${\bf x}_0$ is a chosen anchor point in the parameter space and $L$ is commonly chosen to be the effective dimension of $f$. The functions $f^C_{\bf i}$ are approximated through PC approximation, which result in approximations along lines, planes and hyper-cubes \cite{Gao2011}. \rvnote*{\#4-9}{As an example of the discretization of the cut-HDMR approximation \eqref{eq:cut-HDMR},} figure \ref{fig:P1:Ch5:HDMR:Points} shows the PC based Gauss quadrature points of order $N=13$ for the approximation of the functions $f^C_{\bf i}$ including interactions of order $L=2$. \rvnote*{\#4-8}{The introduction of the cut-HDMR approximation \eqref{eq:cut-HDMR} and its PC discretization result in an efficient and accurate approximation of the sensitivity indices \eqref{eq:TotSensitivity}.}

\section{Uncertainty Quantification in Water Wave Simulations}\label{sec:UQonWaterWaves}

We now use the formulation with random input introduced in  \eqref{FreeSurfaceEqsStochastic} to describe the uncertainty of the evolution of water waves, in terms of free surface position $\zeta(\mathbf{x},t,\mathbf{Z})$ and velocity potential $\tilde{\phi}(\mathbf{x},t,\mathbf{Z})$. For both the random sampling approaches and the non-intrusive polynomial chaos approaches, we need to solve  \eqref{FreeSurfaceEqsStochastic} at a set of points $\left\lbrace \mathbf{z}^{(i)} \right\rbrace_{i=1}^N$, producing the ensembles of solutions
\begin{align}
	\left\lbrace \zeta\left(\mathbf{x},t,\mathbf{z}^{(i)}\right) \right\rbrace_{i=1}^N, \text{ and } \left\lbrace \tilde{\phi}\left(\mathbf{x},t,\mathbf{z}^{(i)}\right) \right\rbrace_{i=1}^N \; .
\end{align}
The sampling strategy depends on the particular method chosen. Furthermore, the PC method constructs the surrogate functions
\begin{subequations}
\begin{align}
	P_N \zeta(\mathbf{x},t,\mathbf{Z}) = \sum_{\vert \mathbf{i} \vert \leq N} \hat{\zeta}_\mathbf{i}(\mathbf{x},t) \Phi_\mathbf{i}(\mathbf{Z}) \approx \zeta(\mathbf{x},t,\mathbf{Z})\;,\\
	P_N \tilde{\phi}(\mathbf{x},t,\mathbf{Z}) = \sum_{\vert \mathbf{i} \vert \leq N} \hat{\tilde{\phi}}_\mathbf{i}(\mathbf{x},t) \Phi_\mathbf{i}(\mathbf{Z}) \approx \tilde{\phi}(\mathbf{x},t,\mathbf{Z})  \;, \label{eq:gPC:freesurf}
\end{align}
\end{subequations}
that provide an easy way to compute statistics and to reproduce the PDFs of the solution variables.

We revisit two classical benchmarks to illustrate how the forward propagation of uncertainties can be done efficiently and we construct a synthetic experiment where we analyze the sensitivity of the load on an off-shore structure to a number of input uncertainties. The presentation of PC methods have been preferred over the random sampling methods whenever the dimension of the problem allowed their application, i.e. when this resulted in reduced CPU time. However, random sampling methods have been used in all the presented cases in order to obtain reference solutions.

% and we propose new stochastic benchmarks. \fixme{If we propose benchmarks, probably we should deliver some results that can be used in direct comparisons??}

\subsection{Harmonic generation over a submerged bar}
\label{sec:submergerbar}

We now consider an experimental setting originally proposed by Beji and Battjes (1994) \cite{BB94}. In the experiment a \rvnote*{\#1-1}{nonlinear} wave propagates across a submerged bar. In the process the bound wave harmonics are decomposed into free harmonics which are released on the lee side of the bar and causes a transformation of the initial wave profile as described in \cite{BL09}. 
The measurements of the experiment are commonly used to validate dispersive and \rvnote*{\#1-1}{nonlinear} wave models such as \eqref{FreeSurfaceEqs}.
% It is generally accepted that the experiment can be reproduced within engineering accuracy by a verified wave model such as \eqref{FreeSurfaceEqs}, which describe both the \rvnote*{\#1-1}{nonlinear} and dispersive effects accurately. 
However, calibration details and measurement errors are not included in the public report by Beji and Battjes. Therefore, in the following, we will assume uncertainties and detail how these can be \rvnote*{\#5-1}{accounted for}.

To analyze the wave evolution we use the bottom topography of the experiments shown in figure \ref{fig:BarTest:BottomProfiles}. We consider the setup corresponding to Case A in the original experiments \cite{BL09}, where the input wave signal is defined by a wave period $T=2.02$s and a wave height $H=2$cm. In the numerical solver the input waves are generated using Stokes second order theory.

The shape of the bar and the shape of the incoming wave influence the spectrum of the waves that reach the right end of the domain as analyzed in \cite{BL09}. In the following different sources of uncertainties are considered and the results are compared with deterministic results often presented in existing literature as well as to the experimental measurements due to Luth {\em et al.} \cite{LKK94}.

\subsubsection{Deterministic results}
\begin{figure}[t]
  \centering
  \subfloat[]{\label{fig:BarTest:Deterministic:0}\includegraphics[width=0.48\textwidth]{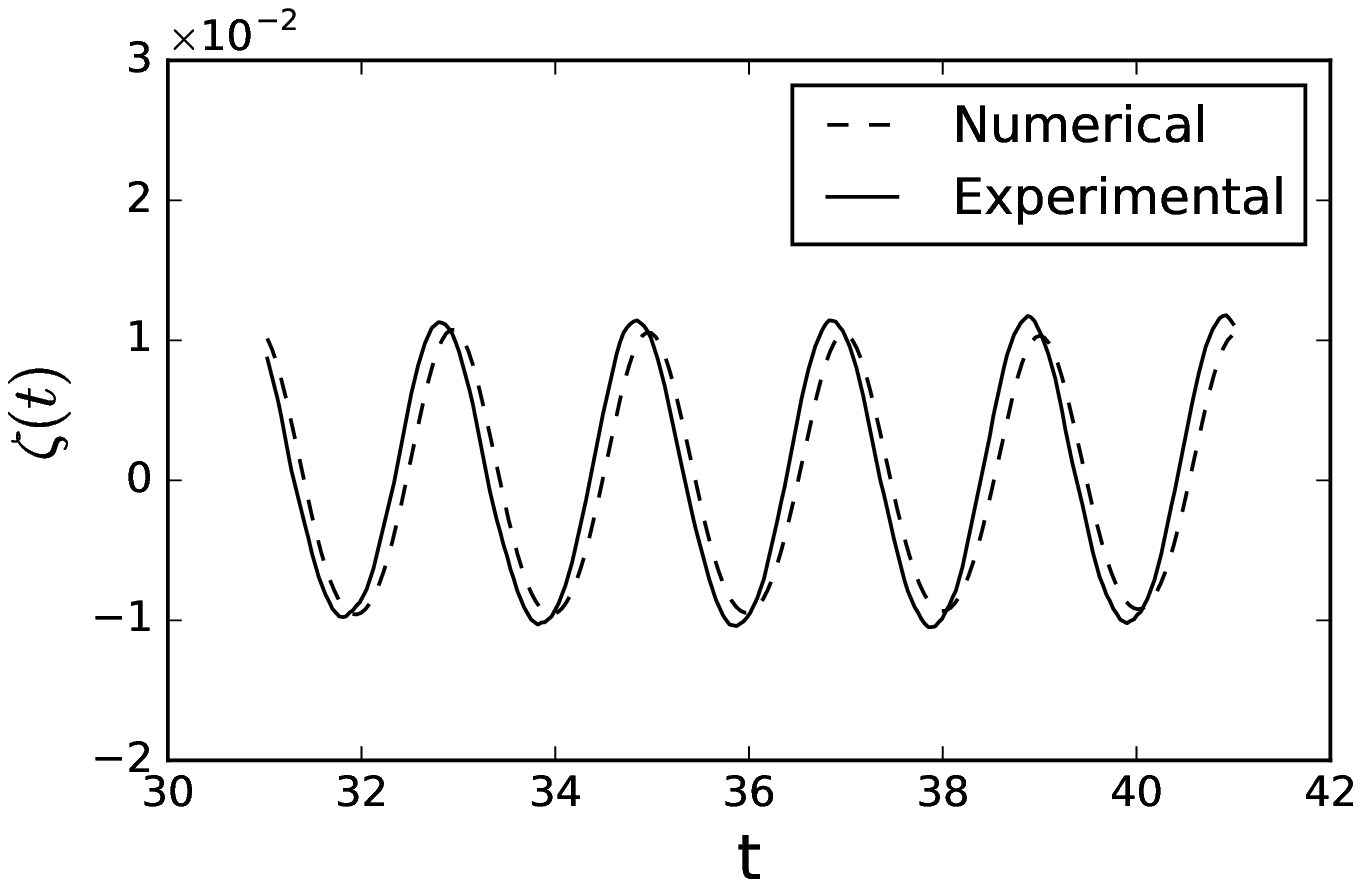}}
  \hspace{3pt}
  \subfloat[]{\label{fig:BarTest:Deterministic:1}\includegraphics[width=0.48\textwidth]{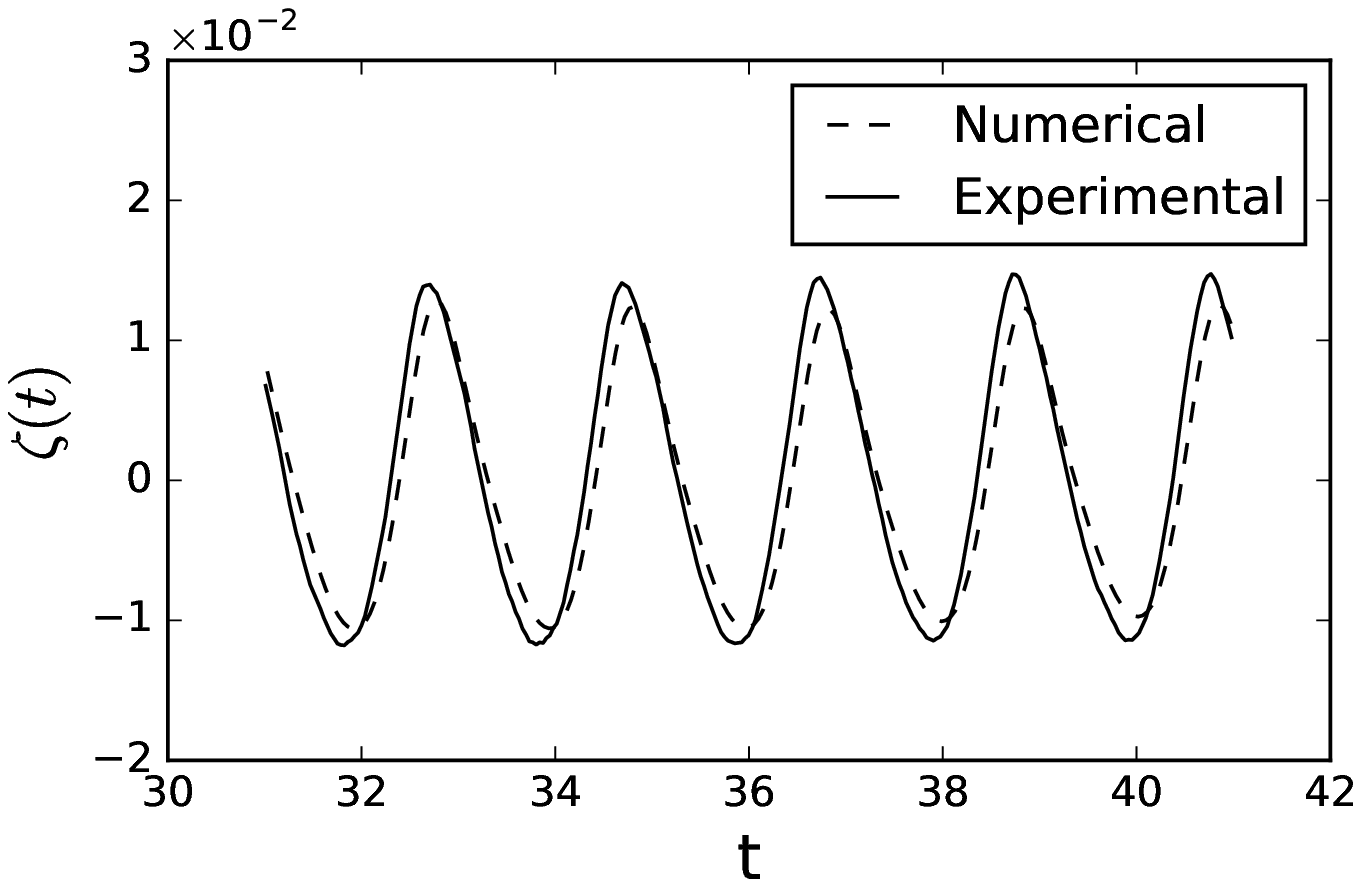}}\\
  \subfloat[]{\label{fig:BarTest:Deterministic:2}\includegraphics[width=0.48\textwidth]{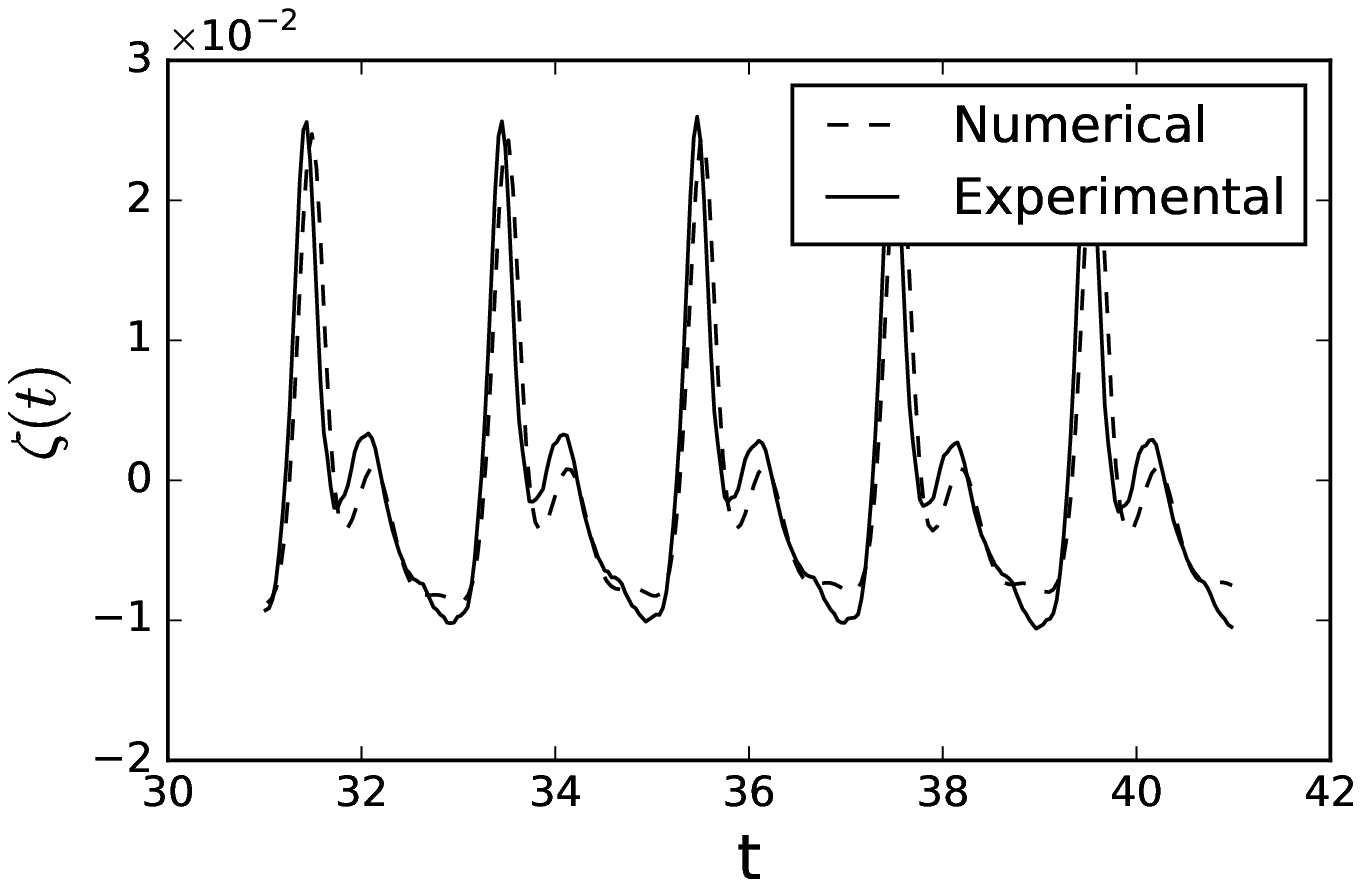}}
  \hspace{3pt}
  \subfloat[]{\label{fig:BarTest:Deterministic:3}\includegraphics[width=0.48\textwidth]{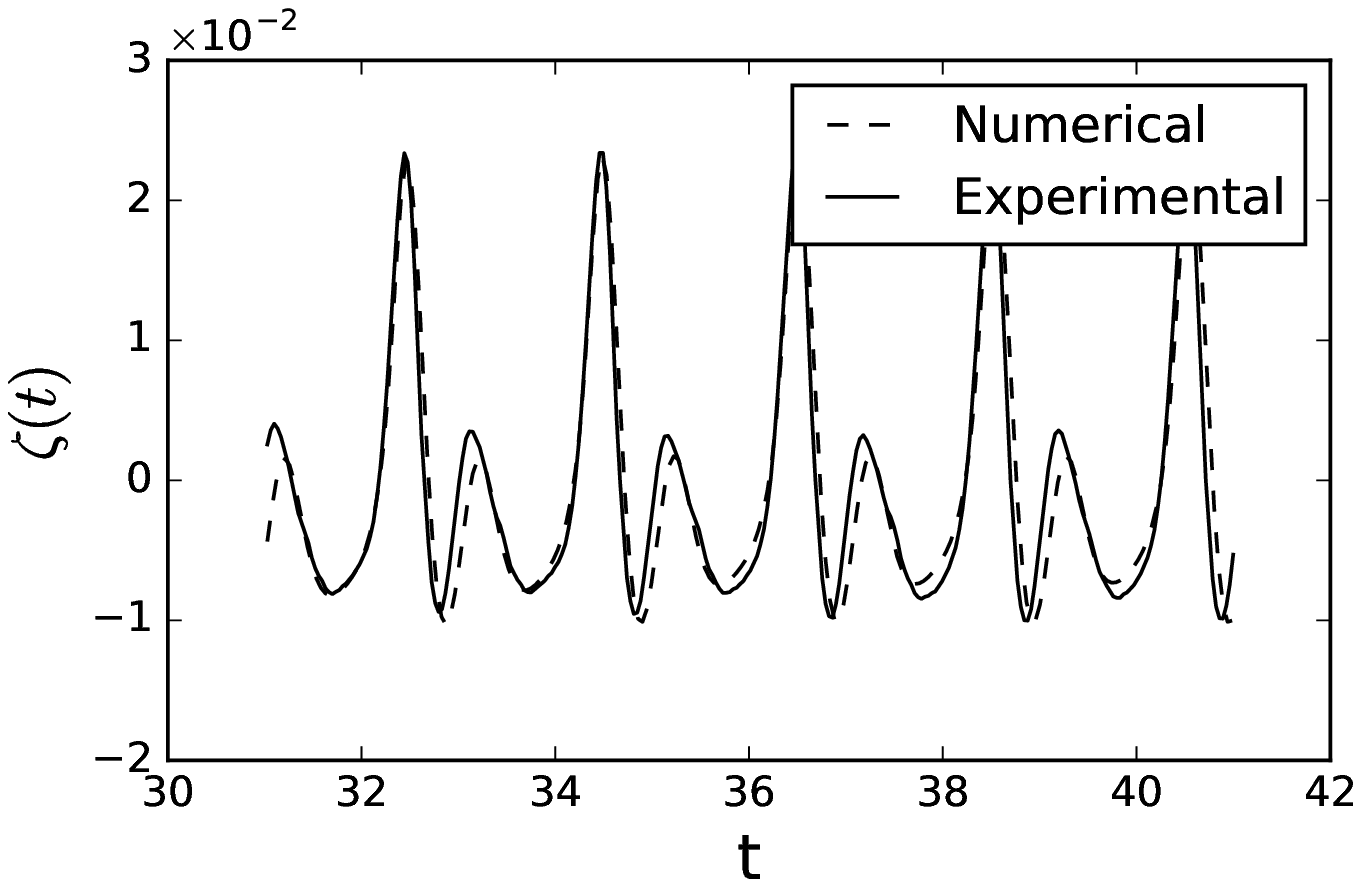}}\\
  \subfloat[]{\label{fig:BarTest:Deterministic:4}\includegraphics[width=0.48\textwidth]{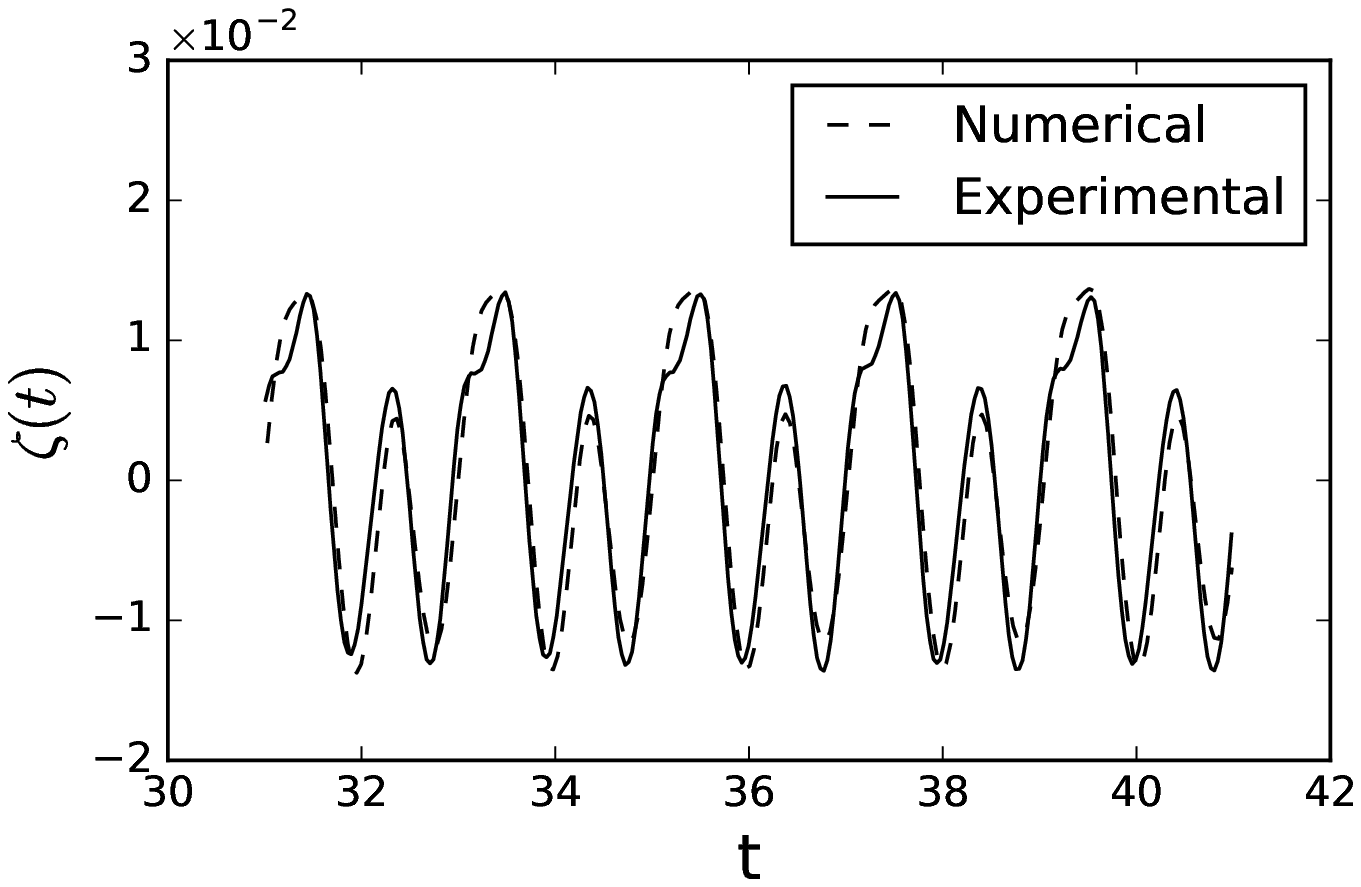}}
  \hspace{3pt}
  \subfloat[]{\label{fig:BarTest:Deterministic:5}\includegraphics[width=0.48\textwidth]{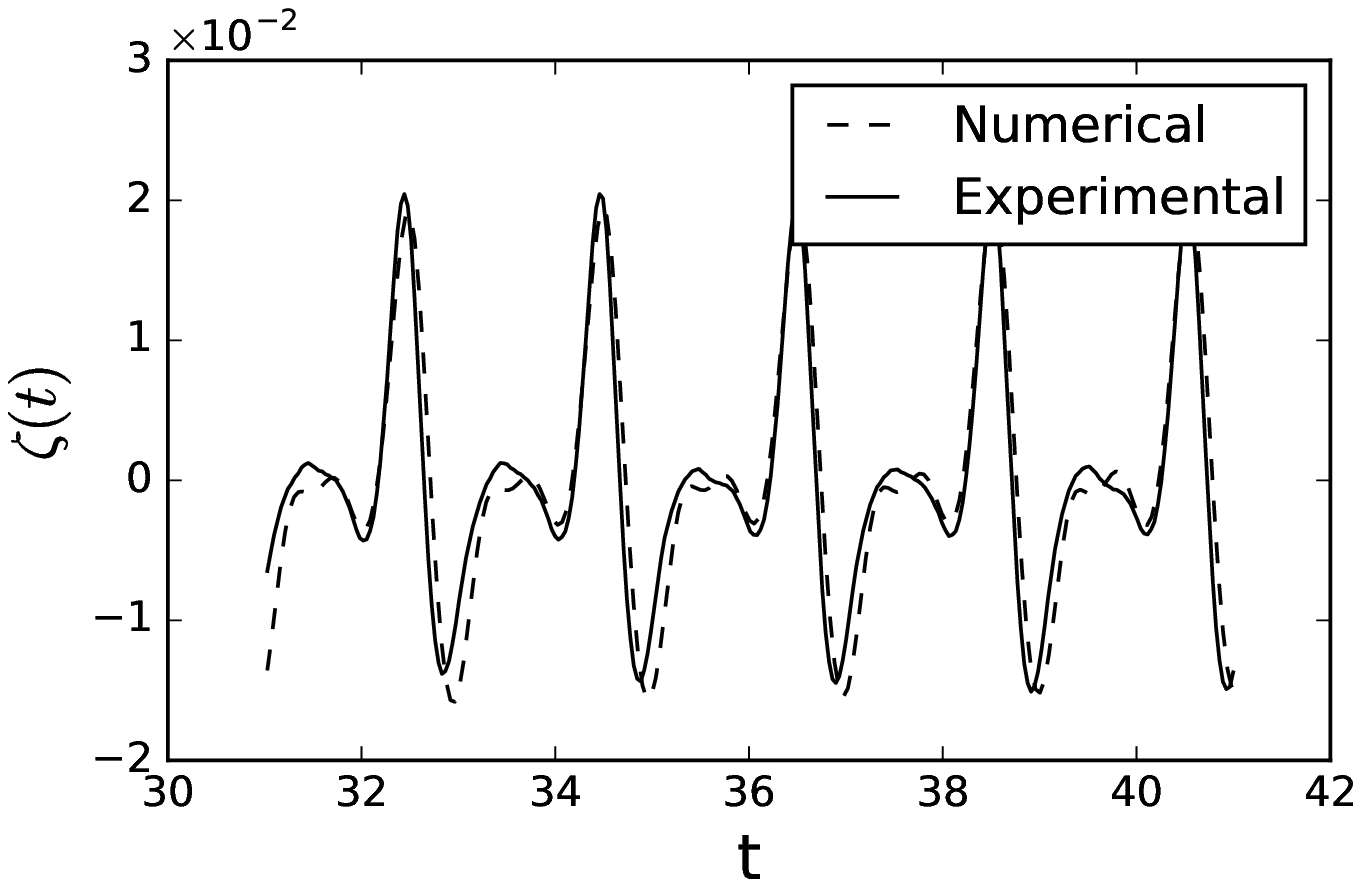}}\\
  \subfloat[]{\label{fig:BarTest:Deterministic:6}\includegraphics[width=0.48\textwidth]{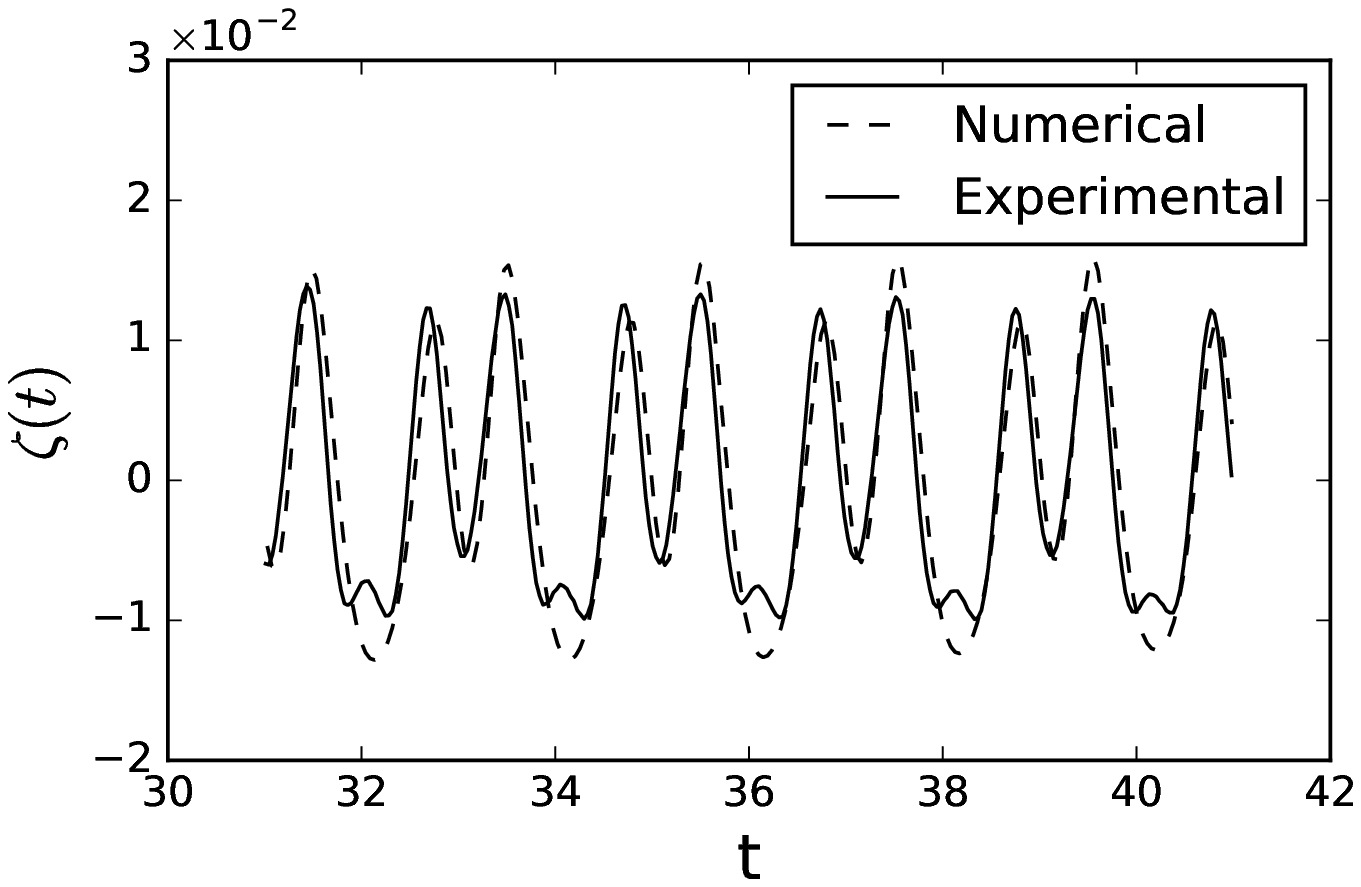}}
  \hspace{3pt}
  \subfloat[]{\label{fig:BarTest:Deterministic:7}\includegraphics[width=0.48\textwidth]{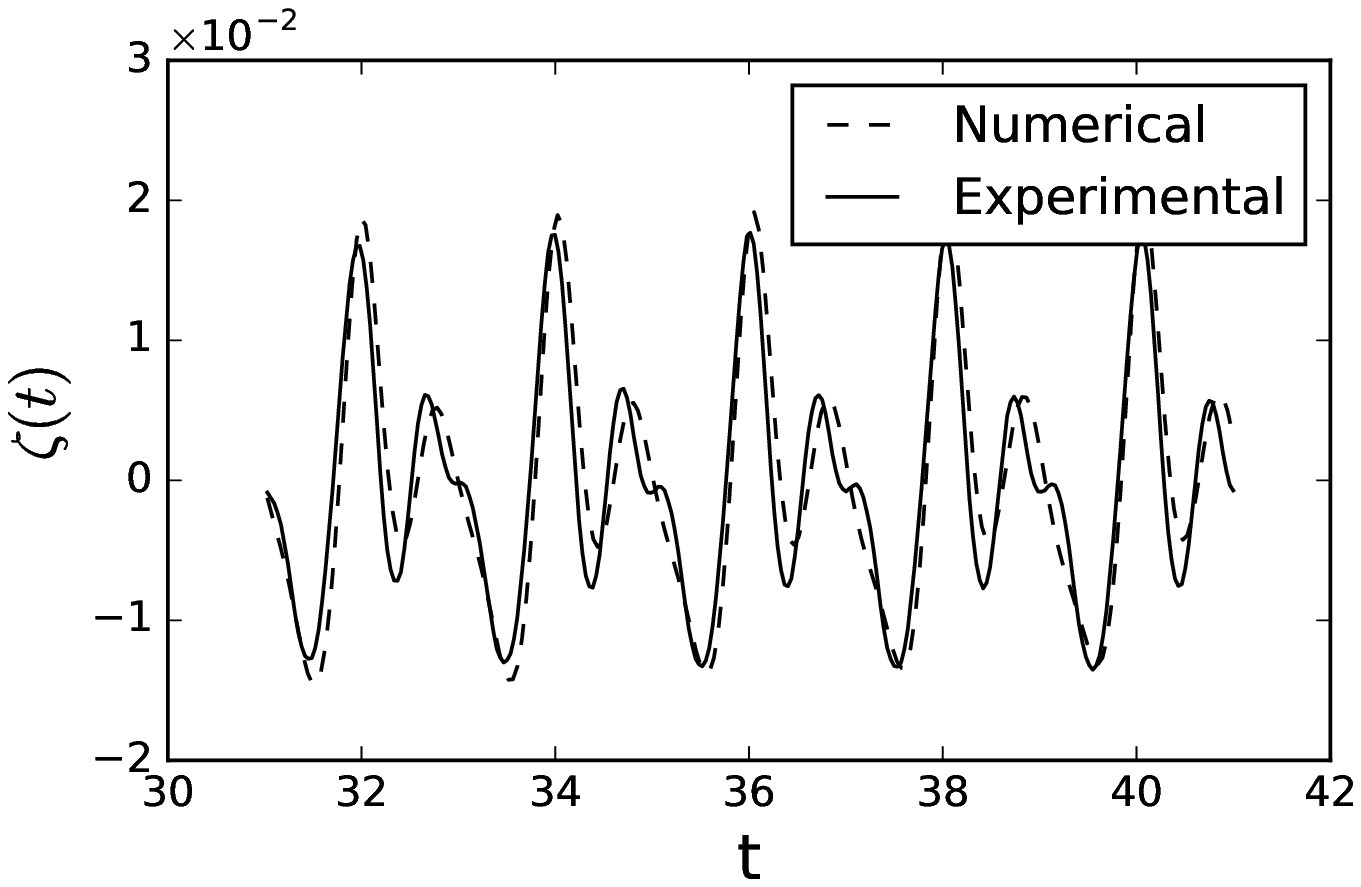}}
  \caption{Deterministic solution of the submerged bar experiment at the eight different gauge locations listed in table \ref{tab:RES:NominalVals}. The experimental data are due to Luth {\em et al.} \cite{LKK94}.}
  \label{fig:BarTest:Deterministic}
\end{figure}
As a conventional \rvnote*{\#5-2}{tool} for validation of the numerical wave model, we compare with the experimental measurements at eight gauges positioned in the wave tank. The results of this comparison are presented in figure \ref{fig:BarTest:Deterministic}, where the bathymetry used is the exact bathymetry illustrated in figure \ref{fig:BarTest:BottomProfiles}.
The results have been computed with the parameters of the experiment given in table \ref{tab:RES:NominalVals}.
\begin{table}[t]
	\centering
	\begin{tabular}{lcc}	
		\hline
		Description 				& Variable 			& Value \\ \hline\hline
		Bar height from bottom 	& $h_\text{bar}$ 	& $0.3m$ \\
		Bottom floor 			& $h_b$ 				& $-0.4m$ \\
		Entering wave period		& $T$				& $2.02s$ \\
                Basin Length & & $29.0m$\\
                Gauges positions & \multicolumn{2}{c}{$\{4.0, 10.5, 13.5, 14.5, 15.7, 17.3, 19.0, 21.0\}$} \\
		\hline
	\end{tabular}
	\caption{Nominal values and experimental settings used for the deterministic solution of the water wave problem.}
	\label{tab:RES:NominalVals}
\end{table}
These parameters will be changed in the following to reflect single realizations of uncertain parameters. Clearly, the computation and the experiments match qualitatively very well, however there are noticeable discrepancies between the numerical calculations and the experimental data. For example, the wave heights and phases are not exactly reproduced at the first and second measurement stations. Discrepancies in the wave signal are observed at the high peaks in measurements from stations 5, 7 and 9, and in the low and intermediate peaks of station 6.

The numerical calculations are done using a high-order accurate numerical method \cite{EngsigKarupEtAl2008} with sufficient resolution to accurately resolve the dispersion and \rvnote*{\#1-1}{nonlinear} wave effects, and are therefore assumed to be converged to a grid-independent solution. The absorption zone introduced behind the bar has been defined so that minimum wave reflections occur. % \dbnote*{Allan: We need some ref for this statement}{The results compare well to other published results.}

The solution up to time $T=41s$ is obtained in approximately 13s on an Intel$^\circledR$~Core\texttrademark ~i7-2640M CPU {@} 2.80GHz. In the following it can be assumed that the influence of the choice of the parameters on the computational cost of solving the system is negligible, and thus the total computational cost scales proportionally with the number of solutions computed.

\subsubsection{Uncertainty on wave period and water height}
\begin{figure}[h]
  \centering
  \subfloat[]{\label{fig:BarTest:UQWaveLengthBottomNormalDistr:0}\includegraphics[width=0.48\textwidth]{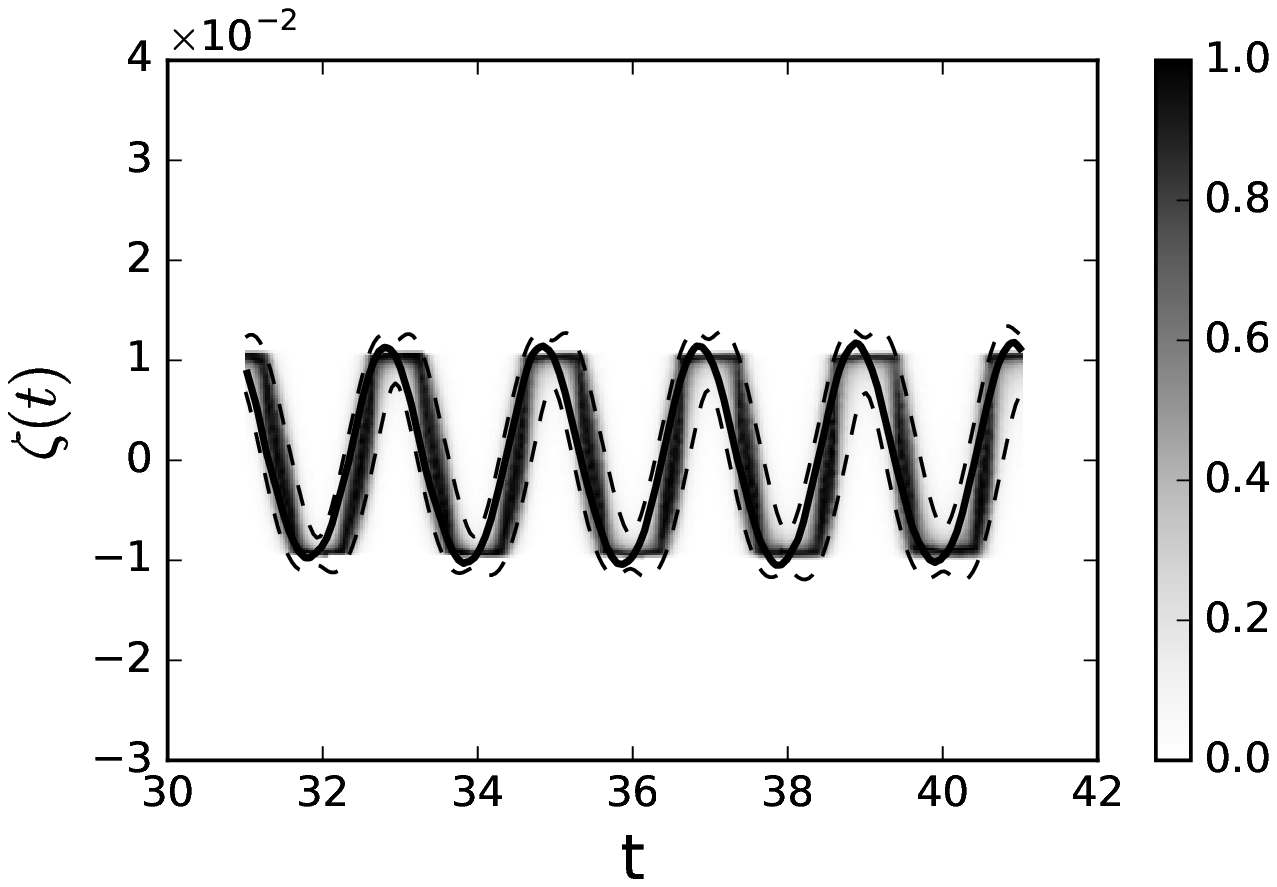}}
  \hspace{3pt}
  \subfloat[]{\label{fig:BarTest:UQWaveLengthBottomNormalDistr:1}\includegraphics[width=0.48\textwidth]{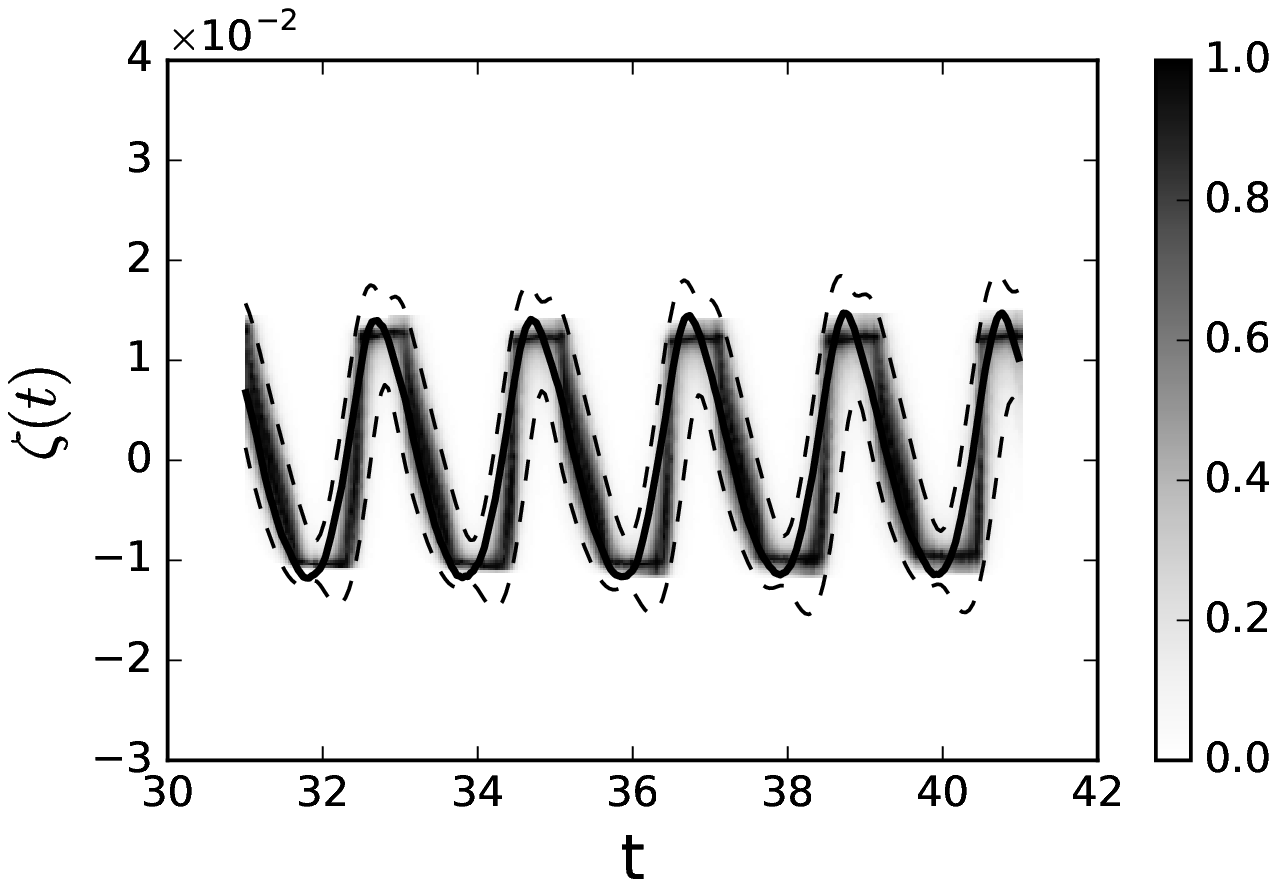}}\\
  \subfloat[]{\label{fig:BarTest:UQWaveLengthBottomNormalDistr:2}\includegraphics[width=0.48\textwidth]{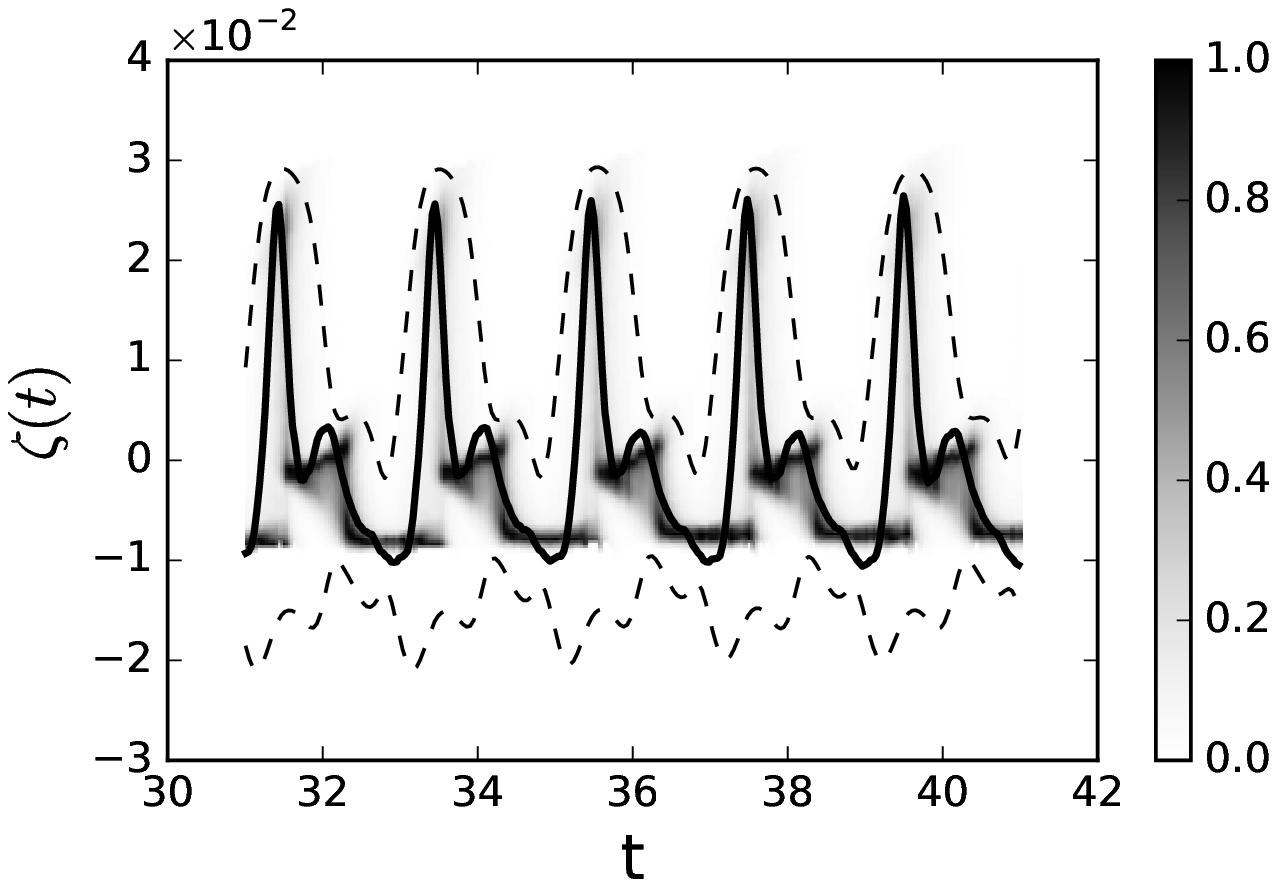}}
  \hspace{3pt}
  \subfloat[]{\label{fig:BarTest:UQWaveLengthBottomNormalDistr:3}\includegraphics[width=0.48\textwidth]{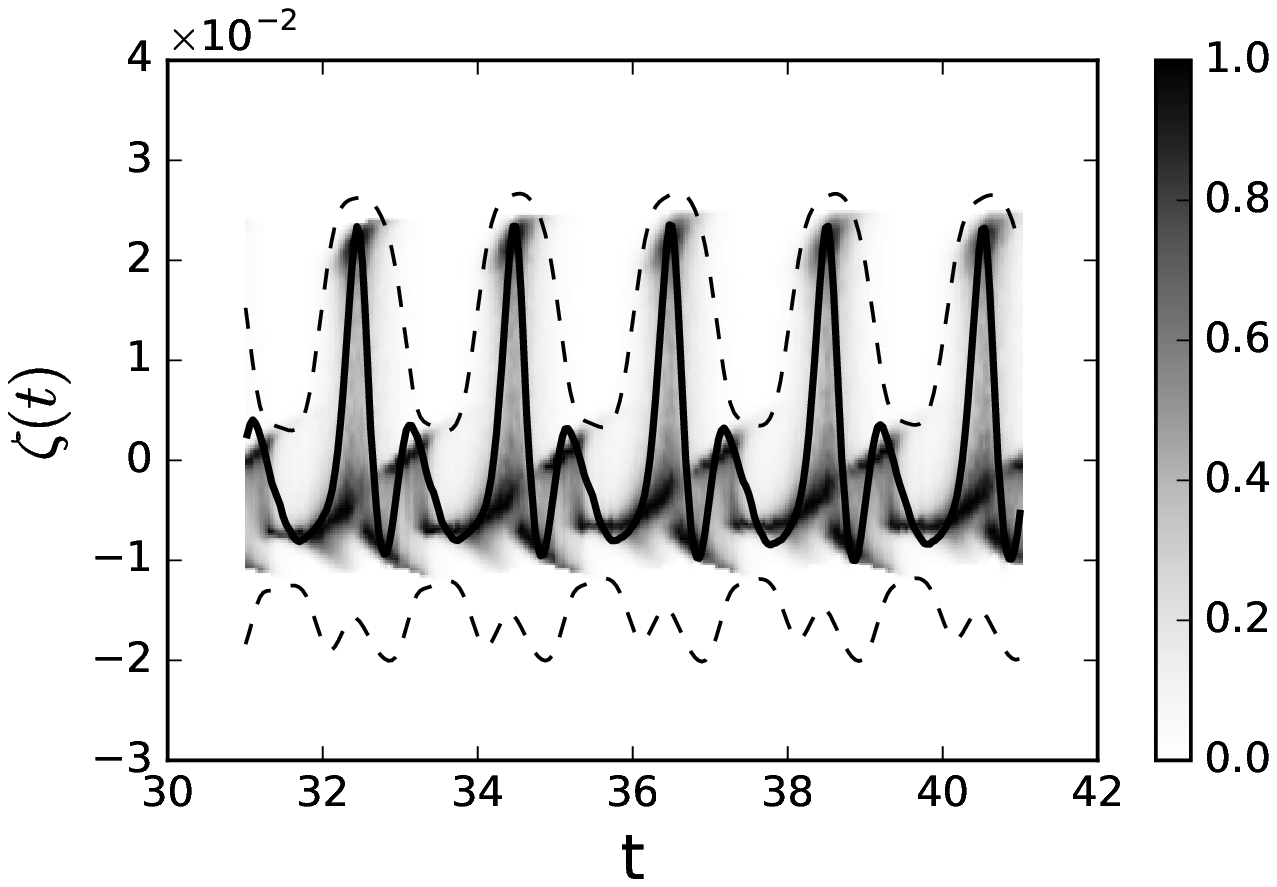}}\\
  \subfloat[]{\label{fig:BarTest:UQWaveLengthBottomNormalDistr:4}\includegraphics[width=0.48\textwidth]{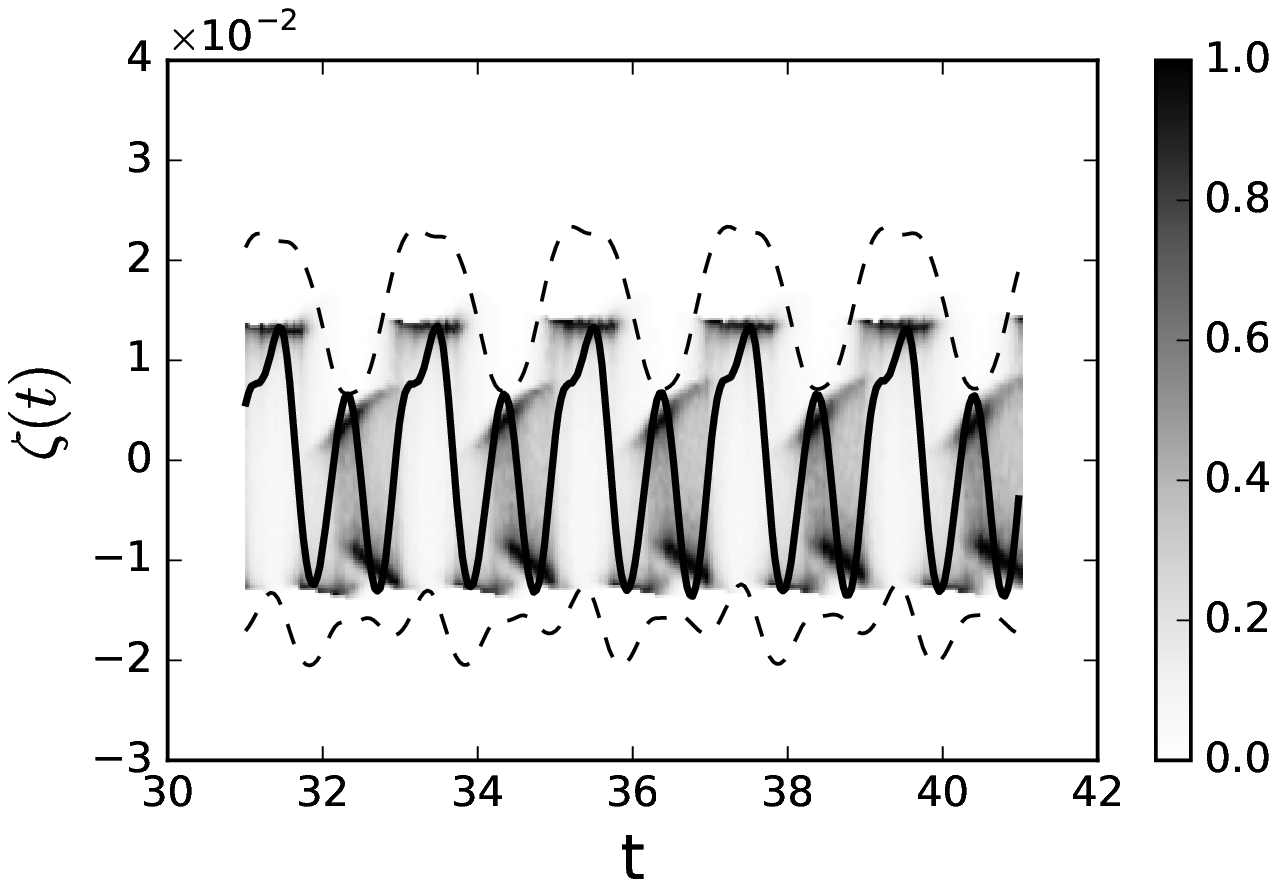}}
  \hspace{3pt}
  \subfloat[]{\label{fig:BarTest:UQWaveLengthBottomNormalDistr:5}\includegraphics[width=0.48\textwidth]{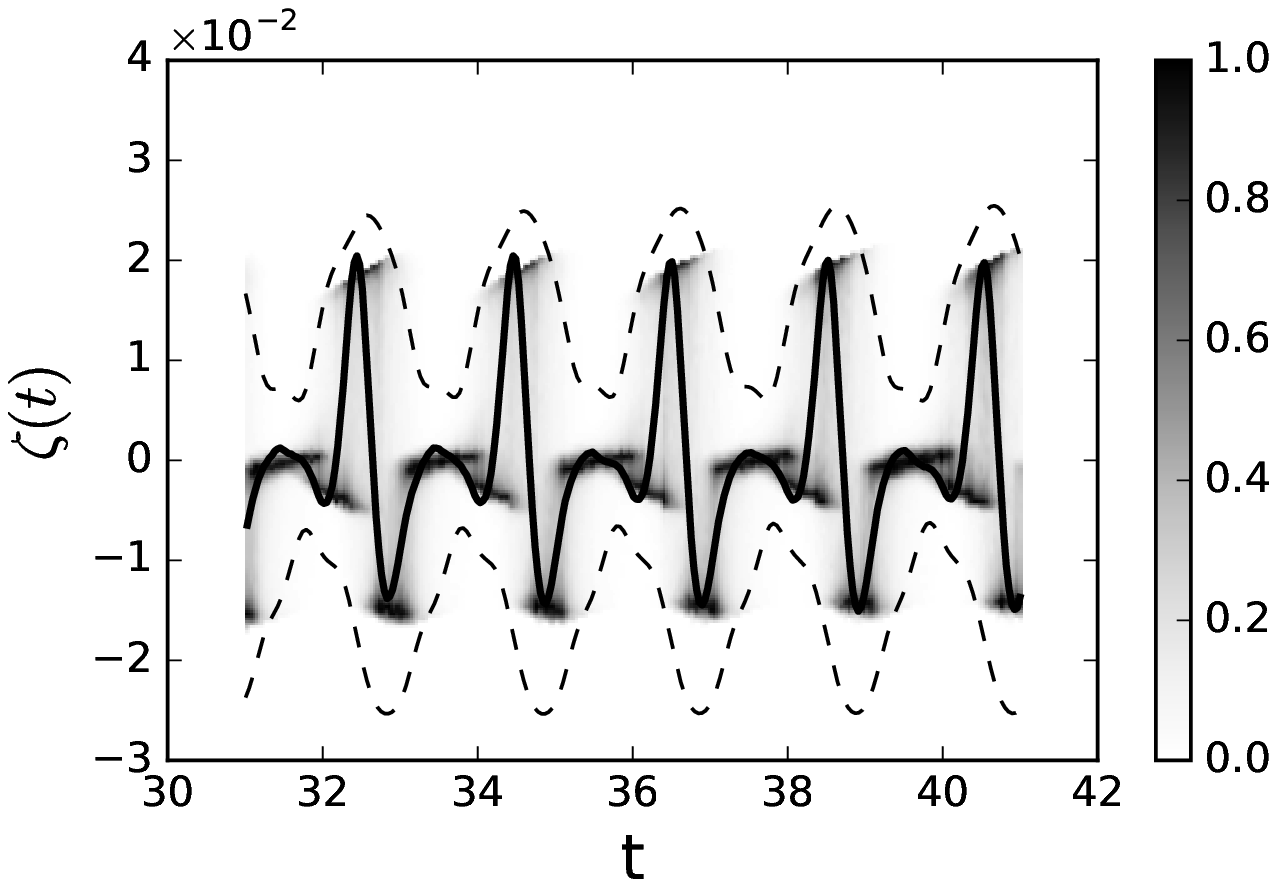}}\\
  \subfloat[]{\label{fig:BarTest:UQWaveLengthBottomNormalDistr:6}\includegraphics[width=0.48\textwidth]{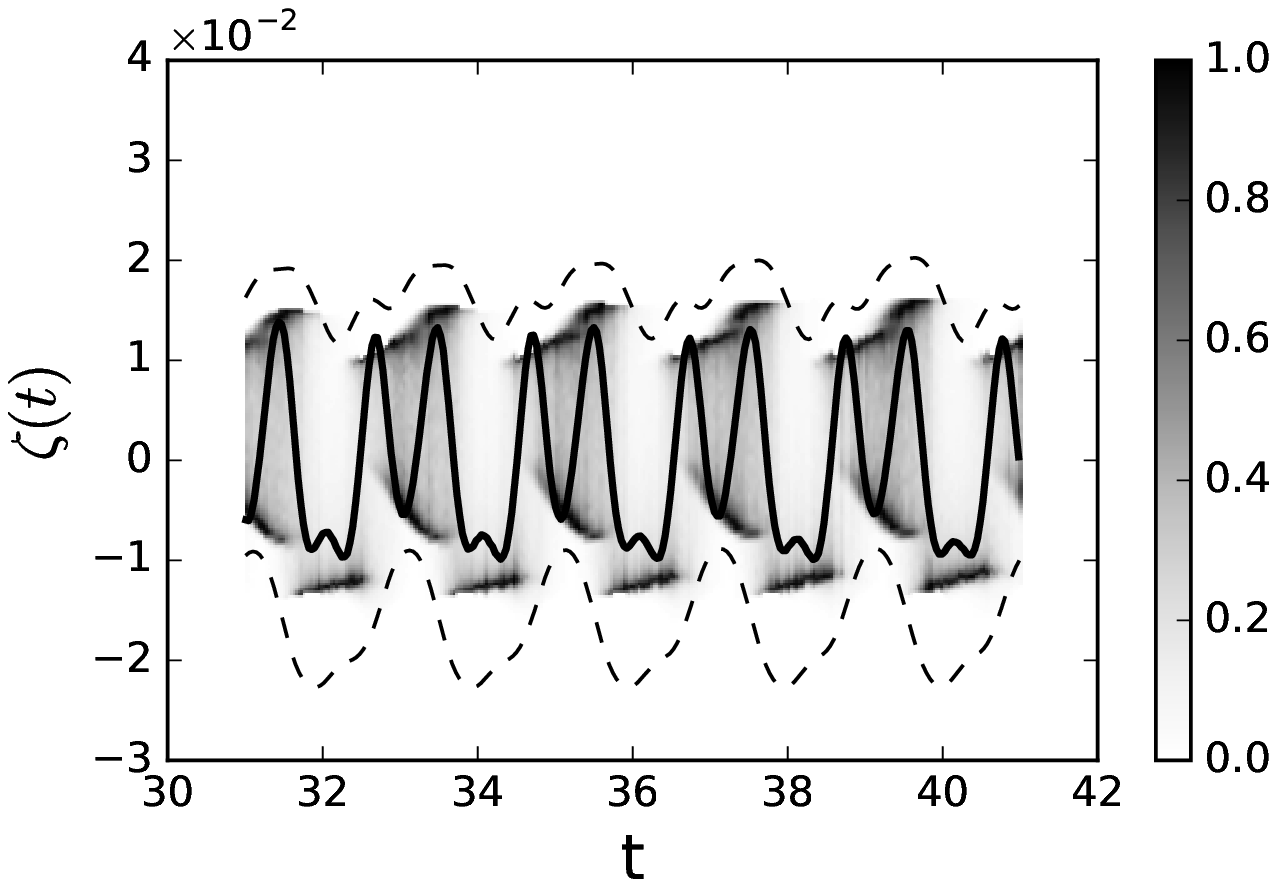}}
  \hspace{3pt}
  \subfloat[]{\label{fig:BarTest:UQWaveLengthBottomNormalDistr:7}\includegraphics[width=0.48\textwidth]{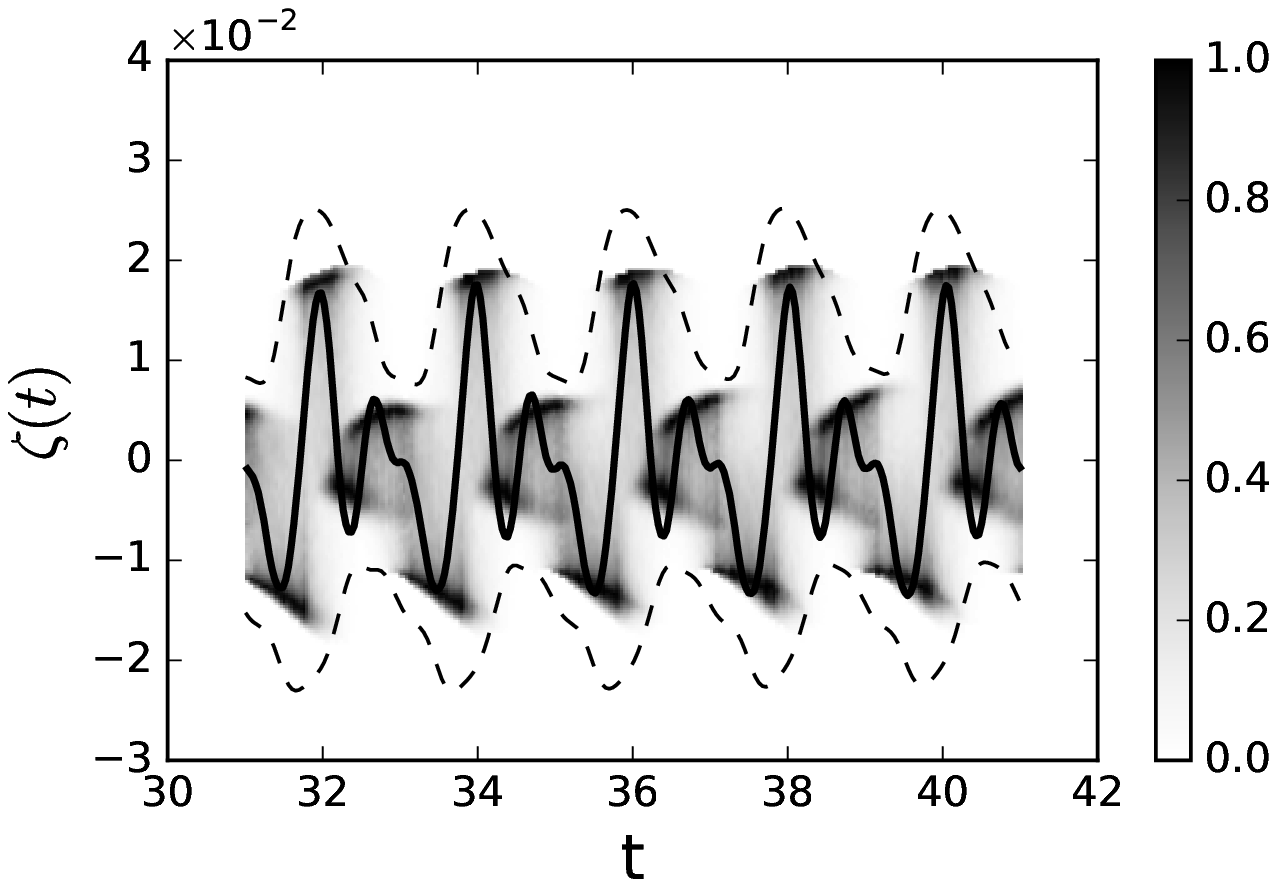}}\\
  \caption{Probability distributions of the time-varying free surface elevation in the submerged bar experiment with uncertain still water height $h_b$ and wave period $T$, at the measurement locations listed in table \ref{tab:RES:NominalVals}. The thick black lines show the experimental results at the different gauges, while the dashed lines show the $95\%$ tolerance intervals.}
  \label{fig:BarTest:UQWaveLengthBottomNormalDistr}
\end{figure}

\rvnote*{\#5-3}{Experiments in manufactured basins are sensitive to fluctuations, evaporation and spill of water as well as to inaccuracies in the mechanical generation of the waves.}
% Very difficult parameters to be controlled when experiments in a manufactured basin are performed, are the exact height of the still water and the input wave period. The accuracy of the measured height is sensitive to fluctuations, evaporation and spill of water. 
Here we use the truncated normal distribution 
\begin{equation}
  \label{eq:WaterHeightDistr}
  h_b \sim \text{tr}\mathcal{N}(0.3m,0.0125^2m^2,[0.375m,0.425m])
\end{equation}
to represent the fact that large defects in the water height can be detected and corrected. An accurate representation of the wave signal requires that the wave maker displacement and the wave amplitudes are matched. This can be difficult to achieve in practice, especially for \rvnote*{\#1-1}{nonlinear} wave signals, and may lead to unintentional harmonic generation. To illustrate how such uncertainty in the signal can be accounted for, we use 
\begin{equation}
  \label{eq:WaveLengthDistr}
  T \sim \mathcal{N}(2.02s,0.01^2s^2)
\end{equation}
to represent the uncertainty due to the generation of the input waves. 

The SC method  with order 10 was used to produce figure \ref{fig:BarTest:UQWaveLengthBottomNormalDistr}, which shows the time dependent distribution function of the solution, its mean and the 95\% tolerance interval. For the expansion of the uncertainty in the wave signal, the Hermite polynomials were used because they form an orthogonal basis with respect to the Gaussian distribution. \rvnote*{\#5-4}{For the expansion of the uncertainty in the water height, a set of polynomials were constructed through the Stieltjes procedure\footnote{The package \texttt{ORTHPOL} \cite{Gautschi1994} is used within the Python modules \texttt{UQToolbox} and \texttt{SpectralToolbox}, for the construction of polynomials orthogonal with respect to an arbitrary measure.} such that they fulfilled the orthogonality condition with respect to the truncated Normal \eqref{eq:WaterHeightDistr}.} The obtained uncertainty is not merely the superposition of the uncertainties that one would obtain in the one dimensional cases where inputs are considered separately, but is the result of the \rvnote*{\#1-1}{nonlinear} interaction between the two. This effect will be evident through the observation of the coefficients in the gPC expansion \eqref{eq:gPC:freesurf}.

\begin{figure}
  \centering
  \subfloat[]{\label{fig:2d-convergence-convplot}\includegraphics[width=0.48\textwidth]{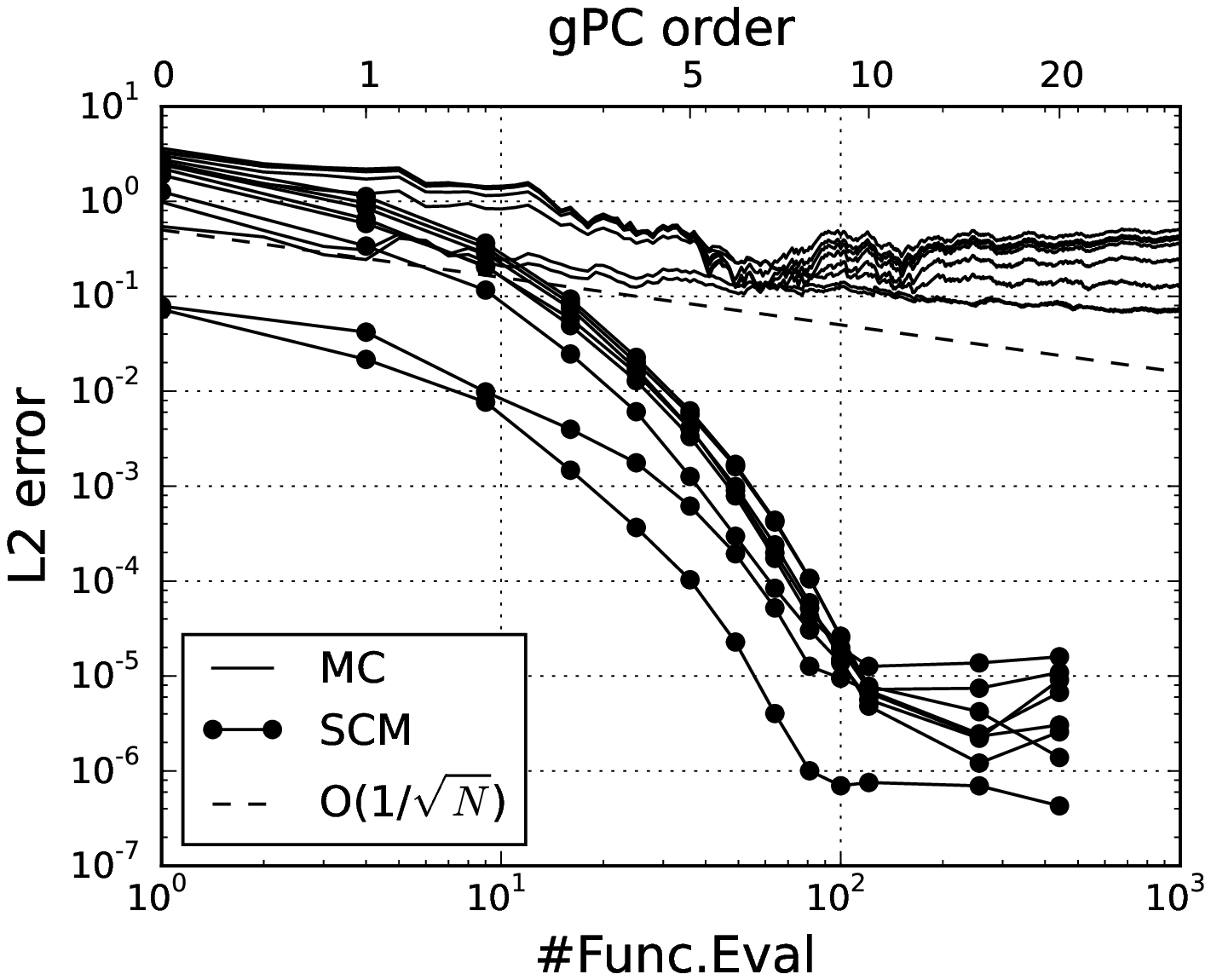}}
  \hspace{3pt}
  \subfloat[]{\label{fig:2d-convergence-pdf}\includegraphics[width=0.48\textwidth]{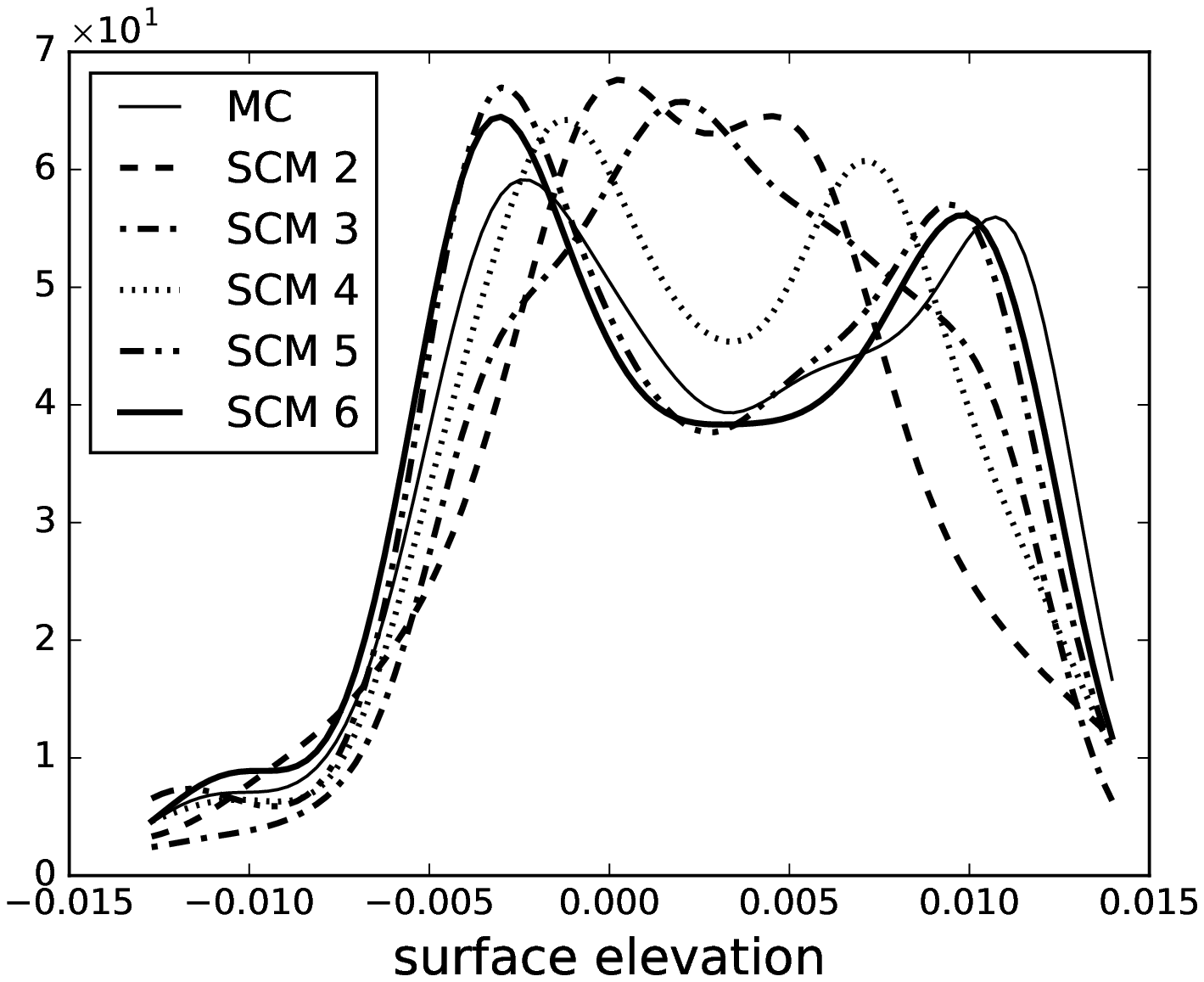}}
  \caption{Convergence rate of the MC and the SC methods. Fig. \protect\subref{fig:2d-convergence-convplot}: decay of the $L^2$ error in the approximation of the mean of 10s of simulation. The convergence is computed with respect to an highly accurate reference solution obtained using the SC method of order 25. The different lines belong to different gauges. Fig. \protect\subref{fig:2d-convergence-pdf}: Comparison of the PDFs of the solution at gauge number 7 at time $t=41s$, obtained with the MC and the SC methods with increasing orders.}
  \label{fig:2d-convergence}
\end{figure}
In order to check the convergence of the method, the SC method with order 25 was used to generate a reference solution, to which lower order SC methods and the MC method were compared. Figure \ref{fig:2d-convergence-convplot} shows the $L^2$ error in the approximation the mean of 10s of simulation as one uses an increasing number of simulations both for the MC method and the SC method. The plateau appearing above order 10 marks the maximum accuracy achievable with the selected tolerance set for the approximate solution of the Laplace problem \eqref{eq:laplaceproblem} ($\varepsilon=10^6$). Using gPC approximations of higher order will result in over-fitting the under-resolved part of the solution. Figure \ref{fig:2d-convergence-pdf} shows the convergence of the PDF of the solution at gauge 7 at time $t=41$, as an increasing number of simulations is used in the SC method, and the approximation to the PDF obtained using the MC method\footnote{The method of Kernel Density Estimation \cite{Friedman2001} with a Gaussian kernel was used here in order to obtain the approximations of the PDFs from samples of the solution.}.

\begin{figure}[h]
  \centering
  \subfloat[]{\label{fig:BarTest:UQWaveLengthBottomNormalCoeffLast:0}\includegraphics[width=0.48\textwidth]{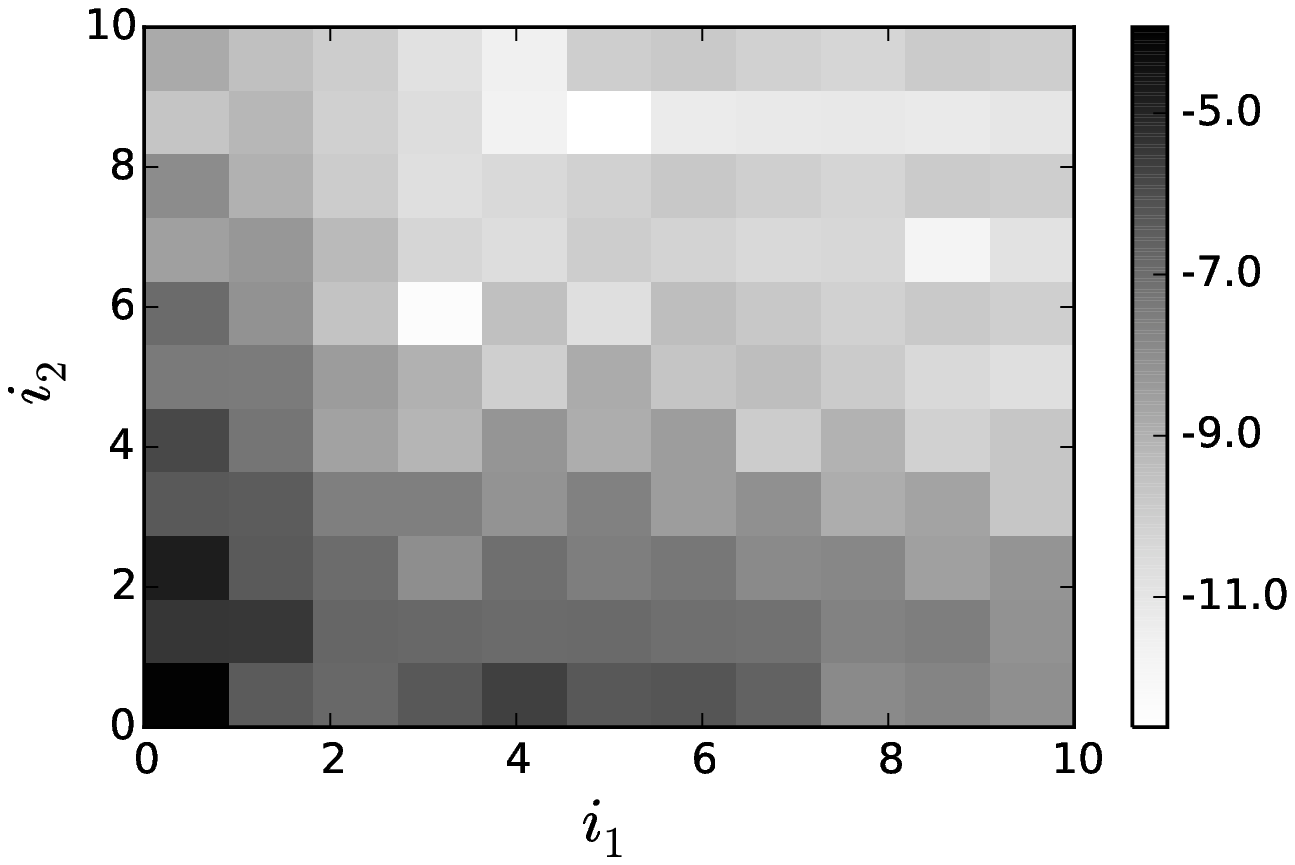}}
  \hspace{3pt}
  \subfloat[]{\label{fig:BarTest:UQWaveLengthBottomNormalCoeffLast:1}\includegraphics[width=0.48\textwidth]{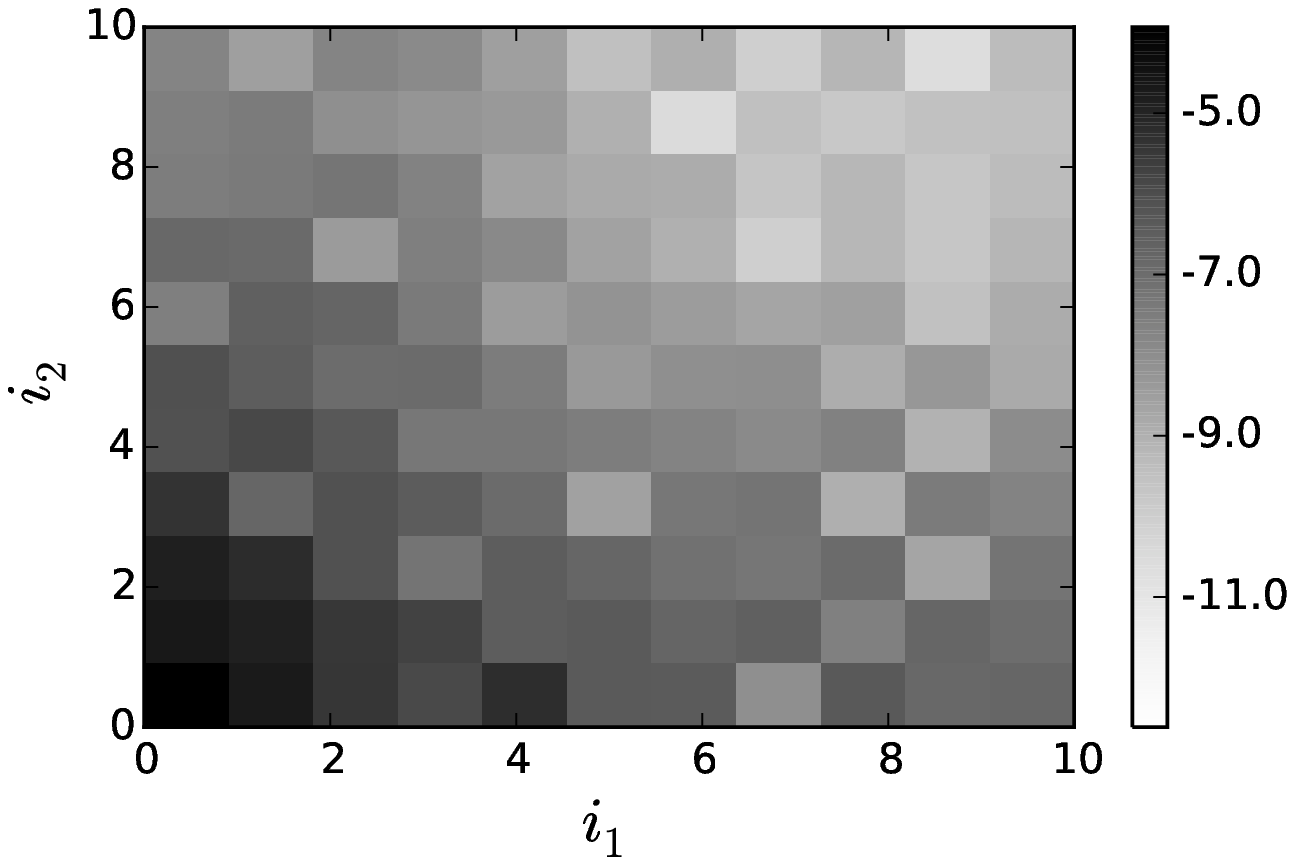}}\\
  \subfloat[]{\label{fig:BarTest:UQWaveLengthBottomNormalCoeffLast:2}\includegraphics[width=0.48\textwidth]{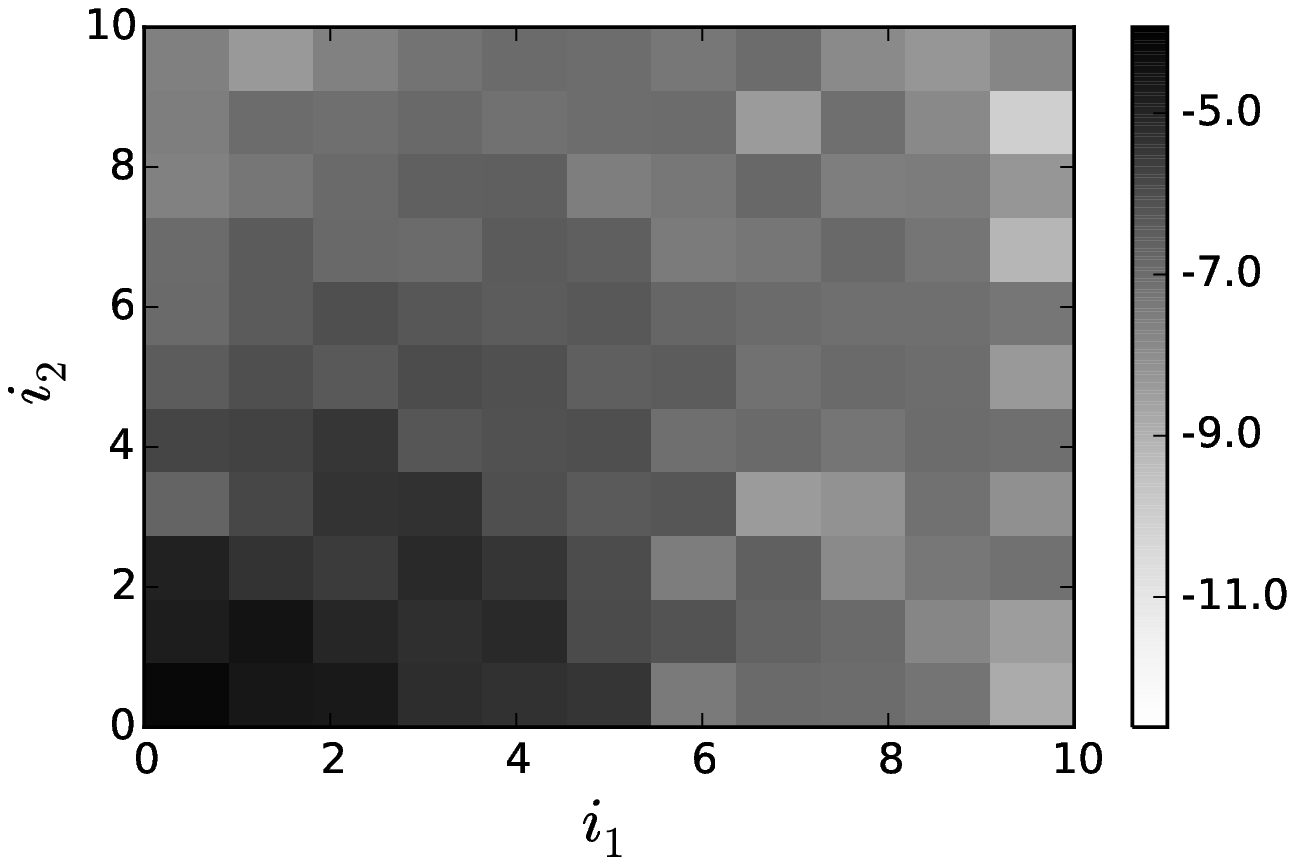}}
  \hspace{3pt}
  \subfloat[]{\label{fig:BarTest:UQWaveLengthBottomNormalCoeffLast:3}\includegraphics[width=0.48\textwidth]{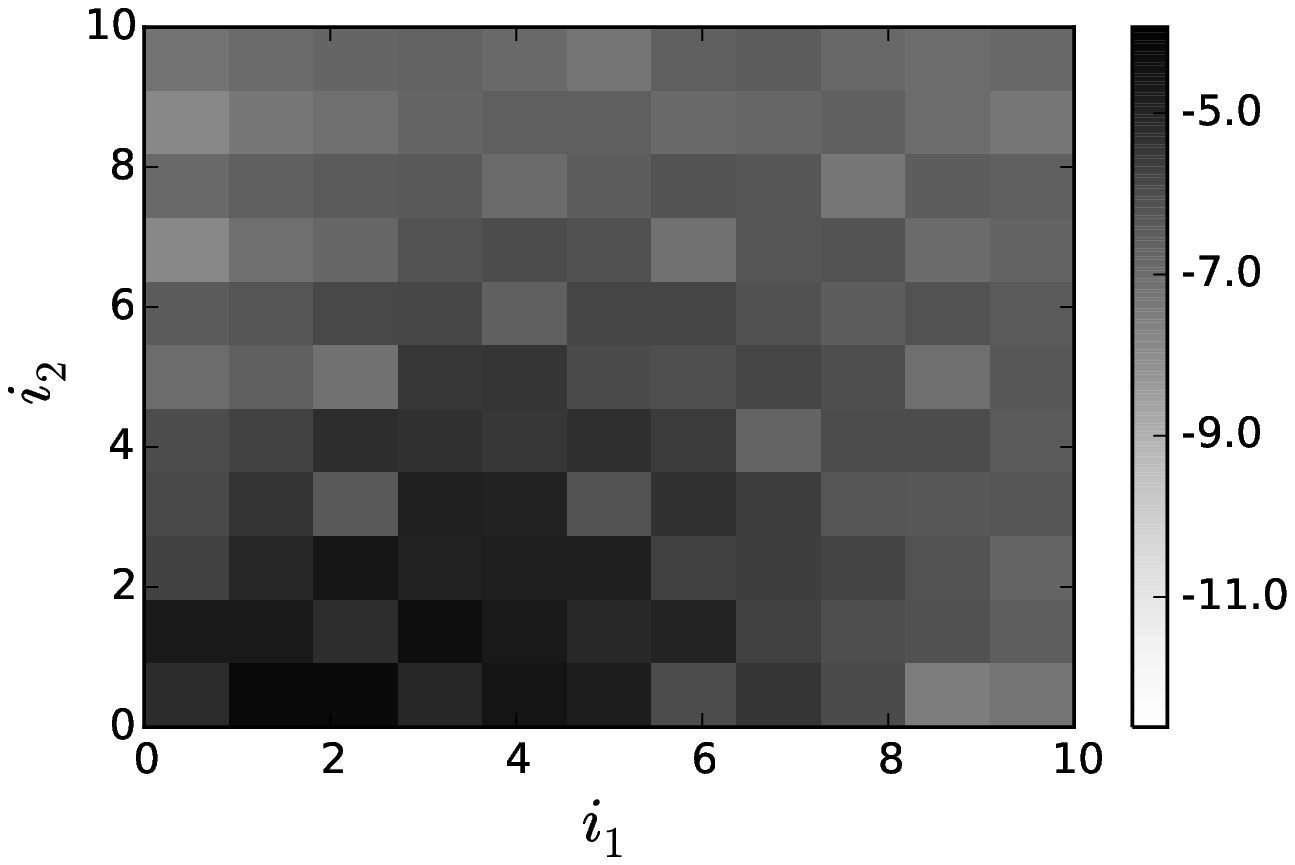}}\\
  \subfloat[]{\label{fig:BarTest:UQWaveLengthBottomNormalCoeffLast:4}\includegraphics[width=0.48\textwidth]{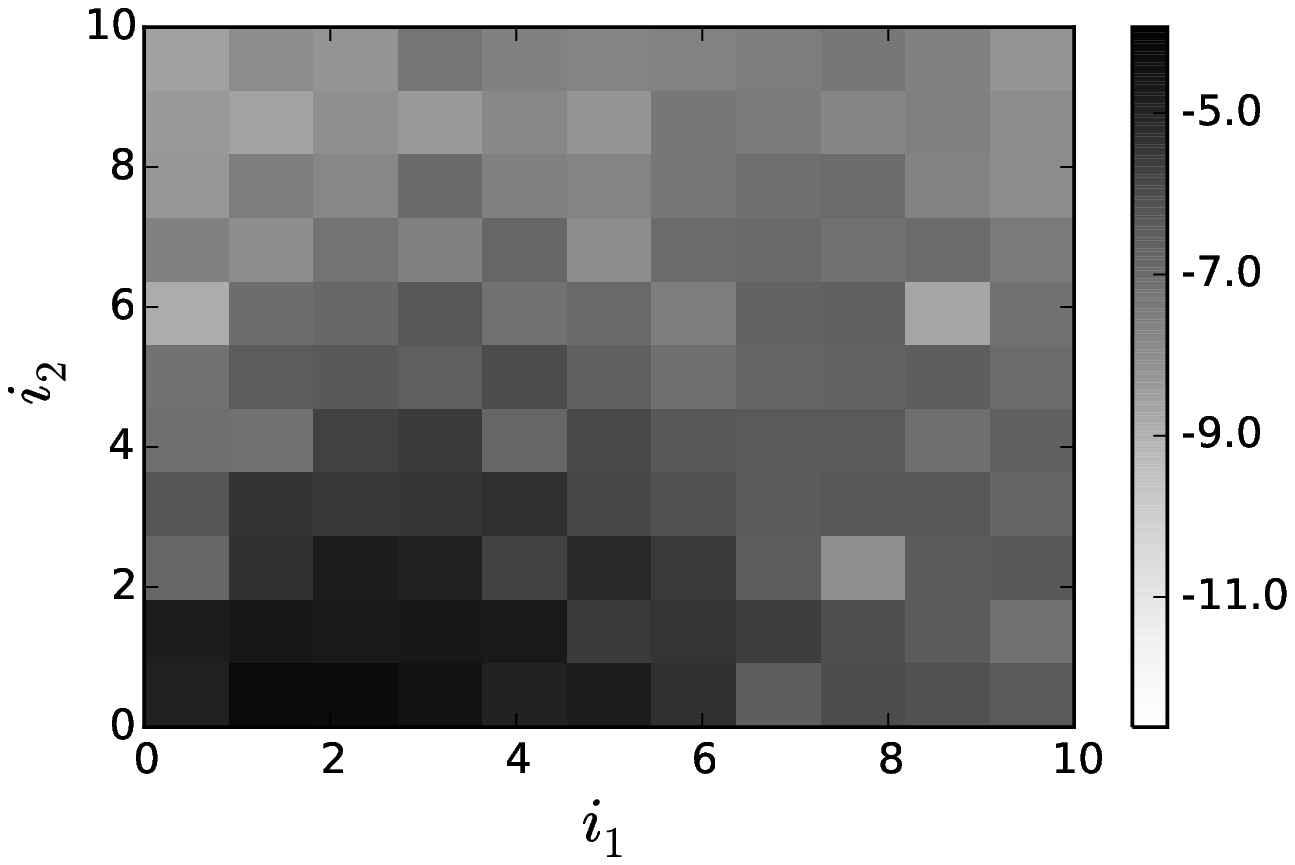}}
  \hspace{3pt}
  \subfloat[]{\label{fig:BarTest:UQWaveLengthBottomNormalCoeffLast:5}\includegraphics[width=0.48\textwidth]{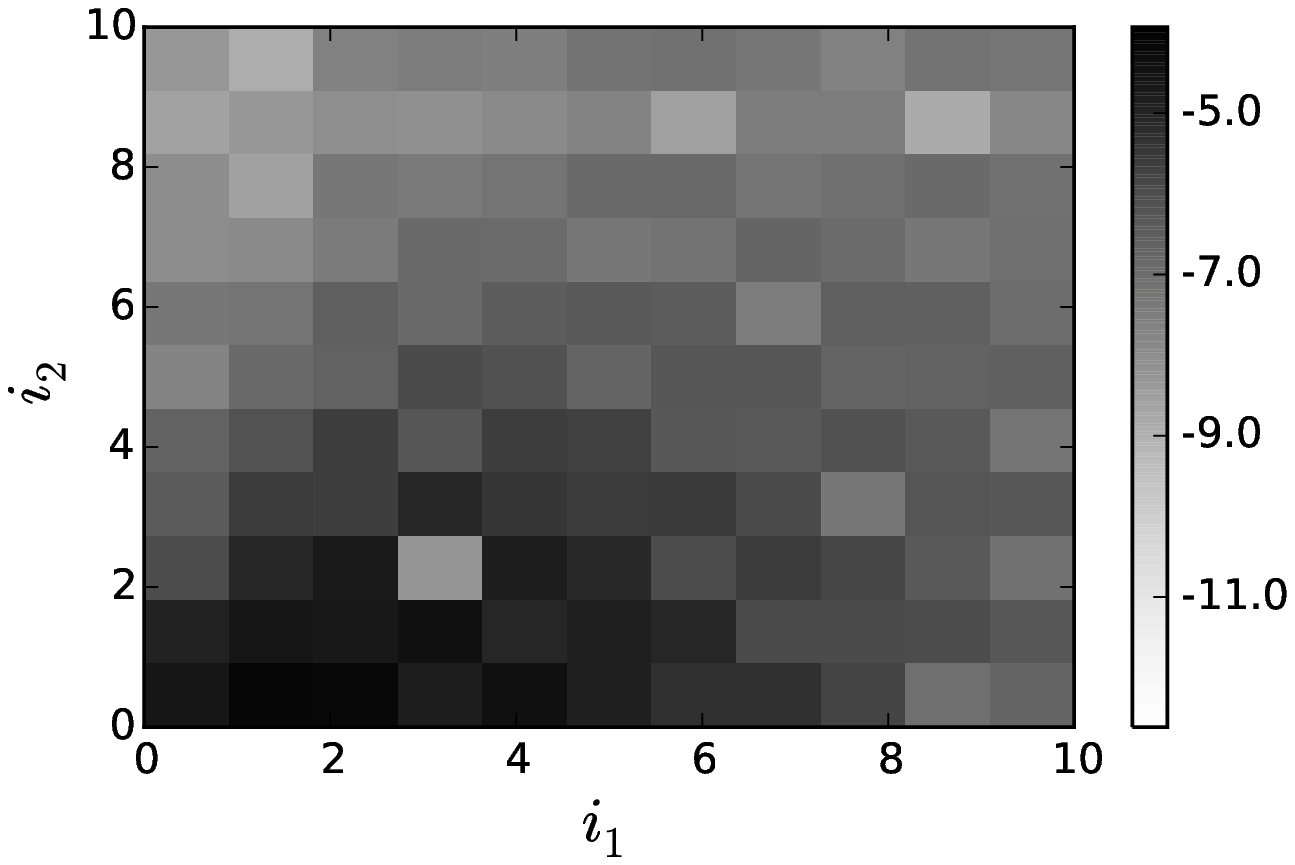}}\\
  \subfloat[]{\label{fig:BarTest:UQWaveLengthBottomNormalCoeffLast:6}\includegraphics[width=0.48\textwidth]{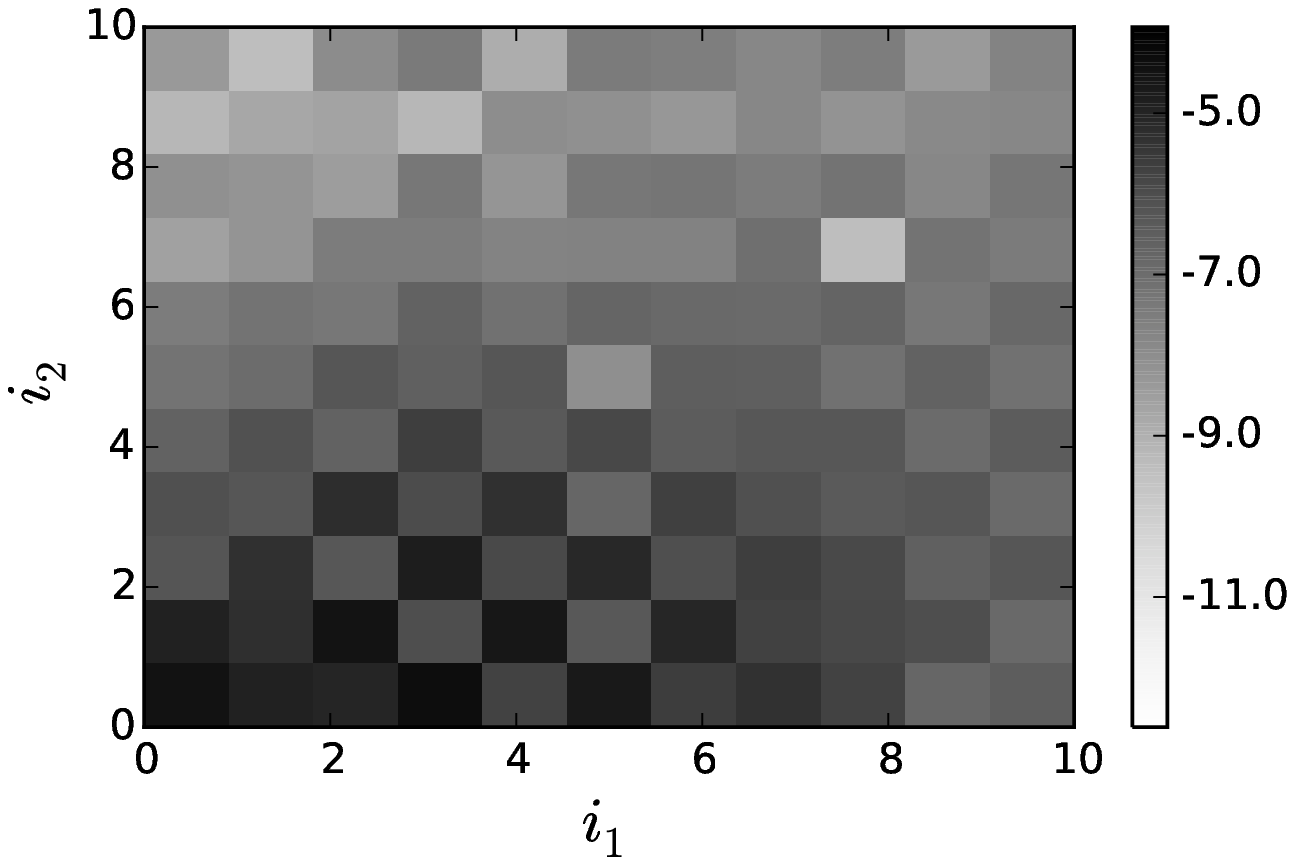}}
  \hspace{3pt}
  \subfloat[]{\label{fig:BarTest:UQWaveLengthBottomNormalCoeffLast:7}\includegraphics[width=0.48\textwidth]{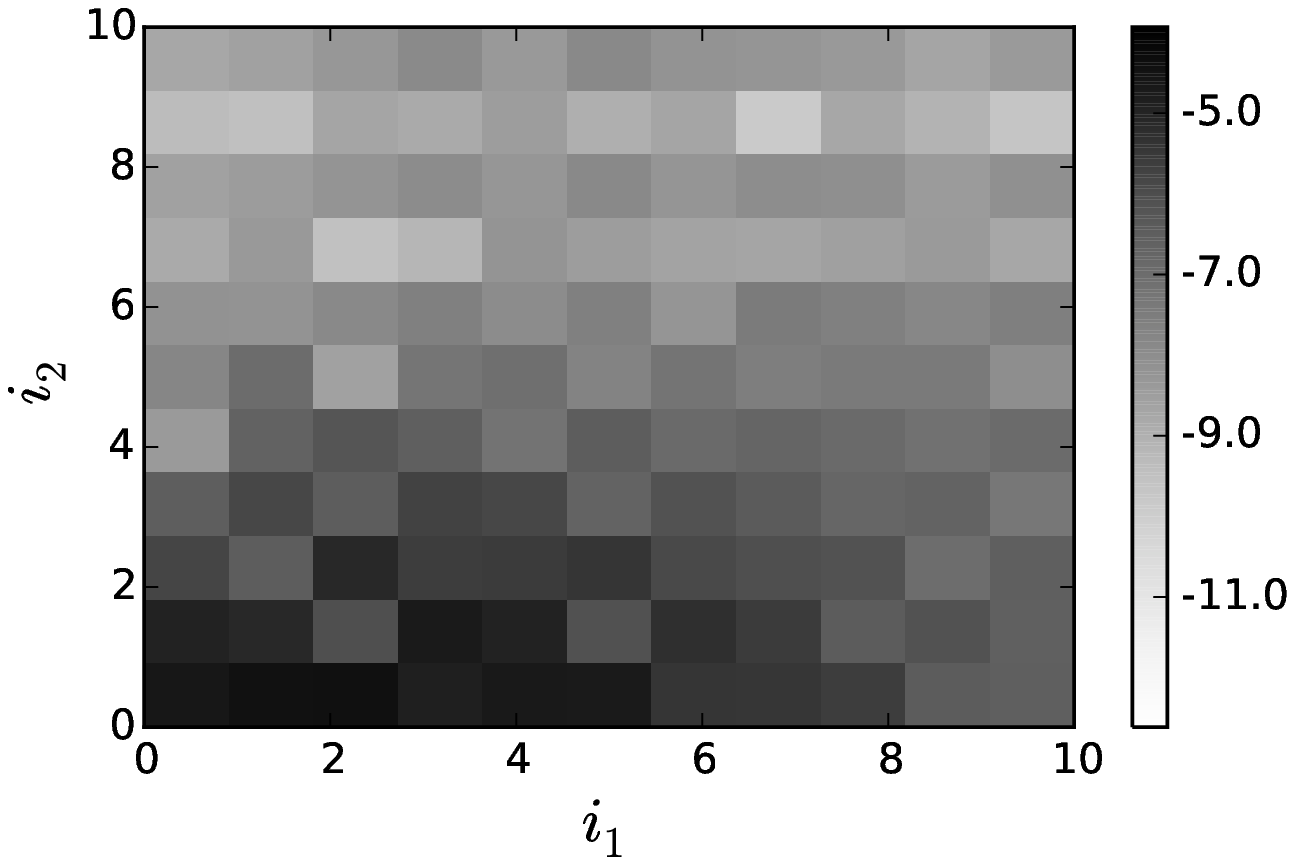}}\\
  \caption{Decay of the 2-dimensional gPC-expansion coefficients $\hat{\zeta}_\mathbf{i}$ in  \eqref{eq:gPC:freesurf} for the last integration time at the measurement locations listed in table \ref{tab:RES:NominalVals}. The two axes span the multi-index ${\bf i}$. The values of the coefficients are shown in the $\log_{10}$ scale.}
  \label{fig:BarTest:UQWaveLengthBottomNormalCoeffLast}
\end{figure}

Figure \ref{fig:BarTest:UQWaveLengthBottomNormalCoeffLast} shows the decay of the projection coefficients in relation to both the input uncertainties. A total independence of the two parameters in the influence of the system would produce an expansion \eqref{eq:UQ:gPCexpansion} where all the non-zero coefficients $\hat{f}_\mathbf{i}$ are the ones with $\mathbf{i}=(i,0)$ or $\mathbf{i}=(0,j)$, $i,j=0,\dots,20$. This corresponds to have decays similar to the one shown in the first upper-left plot in figure \ref{fig:BarTest:UQWaveLengthBottomNormalCoeffLast}. The next plots, however, show that the two input uncertainties act on the solution in non-trivial ways when combined. This means that the results of the UQ analysis on the two separate sources, cannot be trivially superposed, but they need to be considered together in a unique UQ analysis. 
% Furthermore, the application of Sparse grid on this case would suffer from this property which corresponds to the lack of separability of the function of interest. Methods which allow a sparse sampling in these situations are still lacking in the scientific literature.

%\subsubsection{Uncertain position of wave gauges}

\subsubsection{Uncertain bottom topography}\label{sec:Bartest:Bottom}
\begin{figure}
  \centering
  \subfloat[]{\label{fig:BarTest:Field-a30:0}\includegraphics[width=0.48\textwidth]{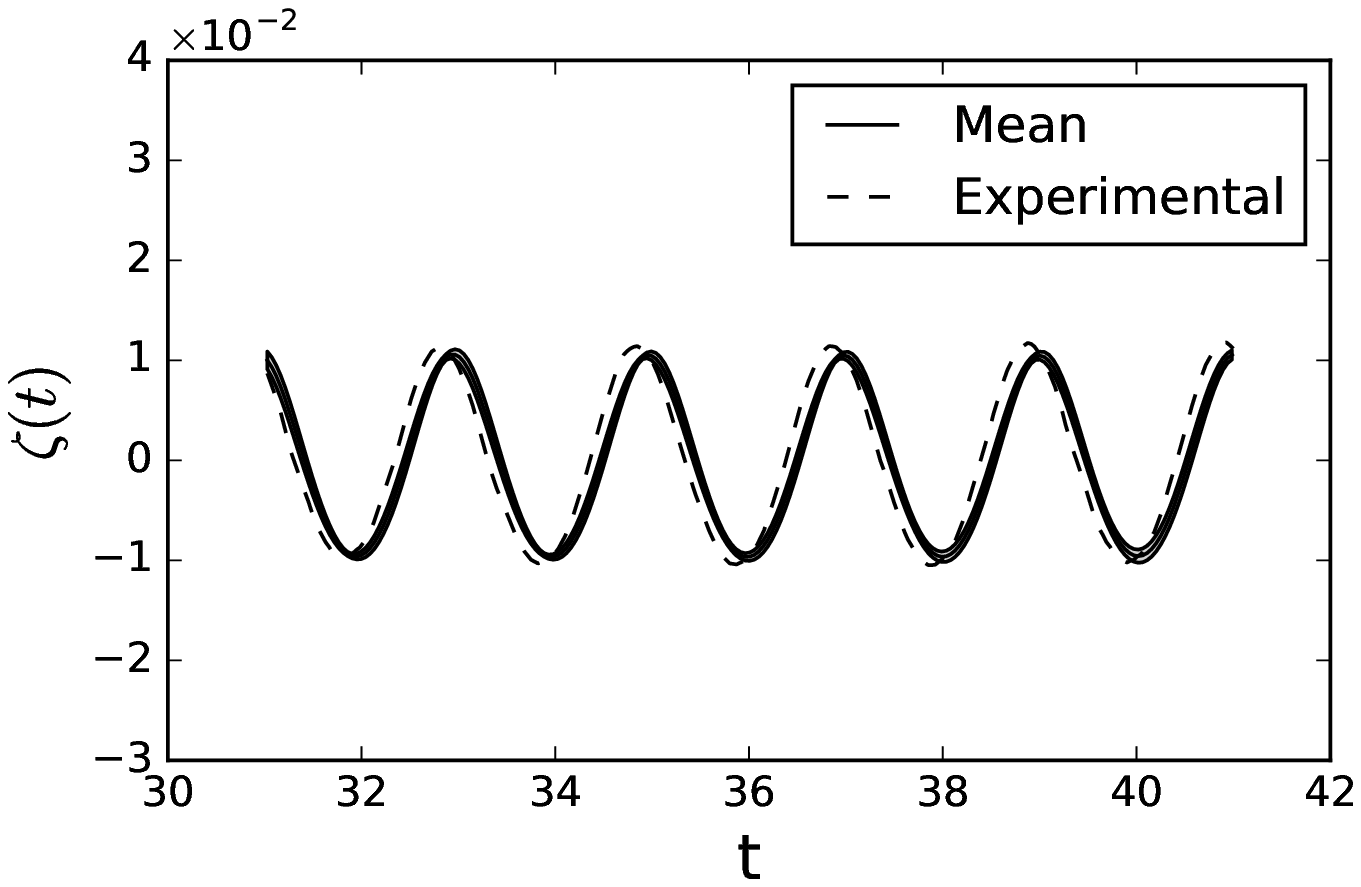}}
  \hspace{3pt}
  \subfloat[]{\label{fig:BarTest:Field-a30:1}\includegraphics[width=0.48\textwidth]{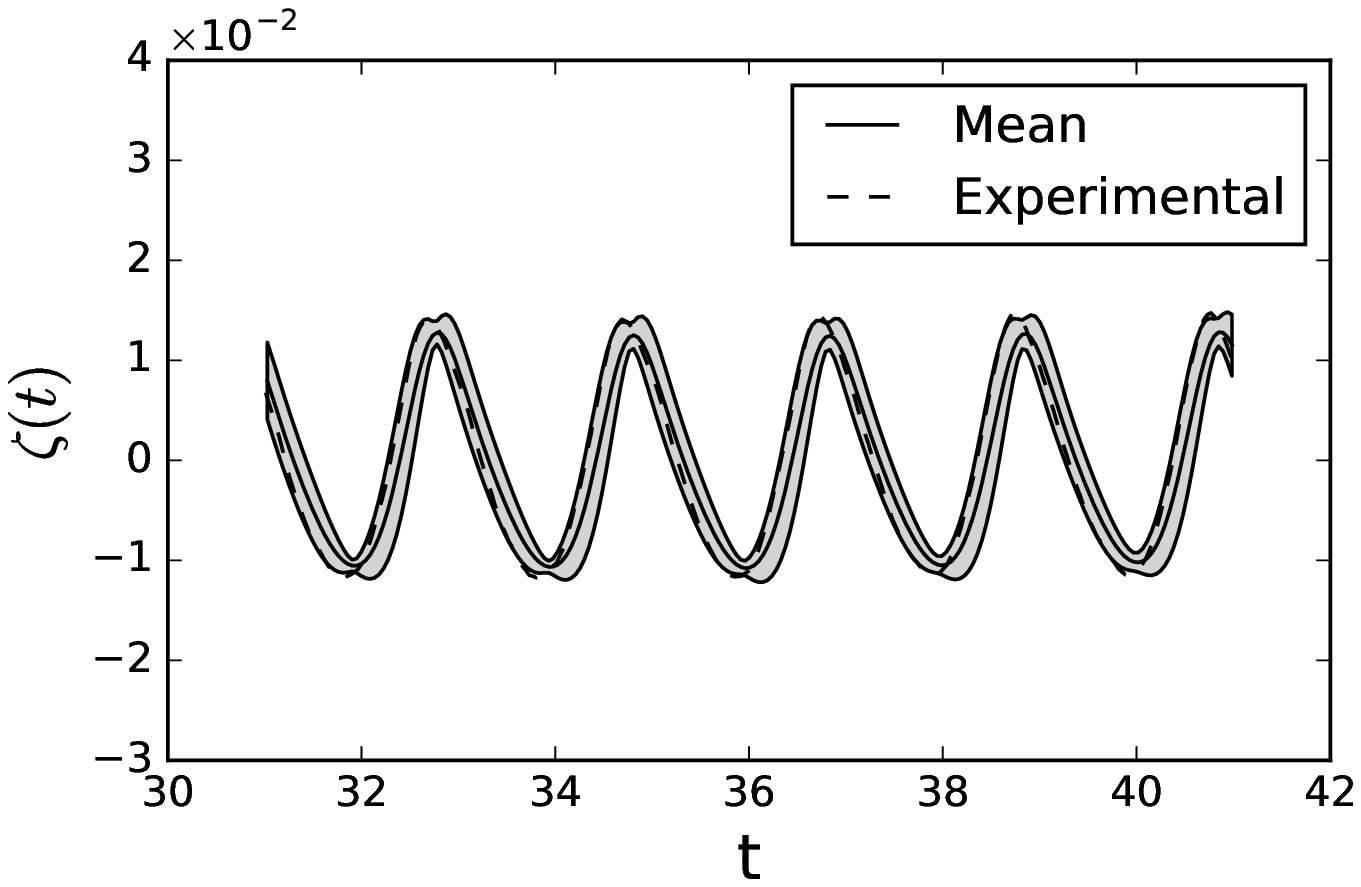}}\\
  \subfloat[]{\label{fig:BarTest:Field-a30:2}\includegraphics[width=0.48\textwidth]{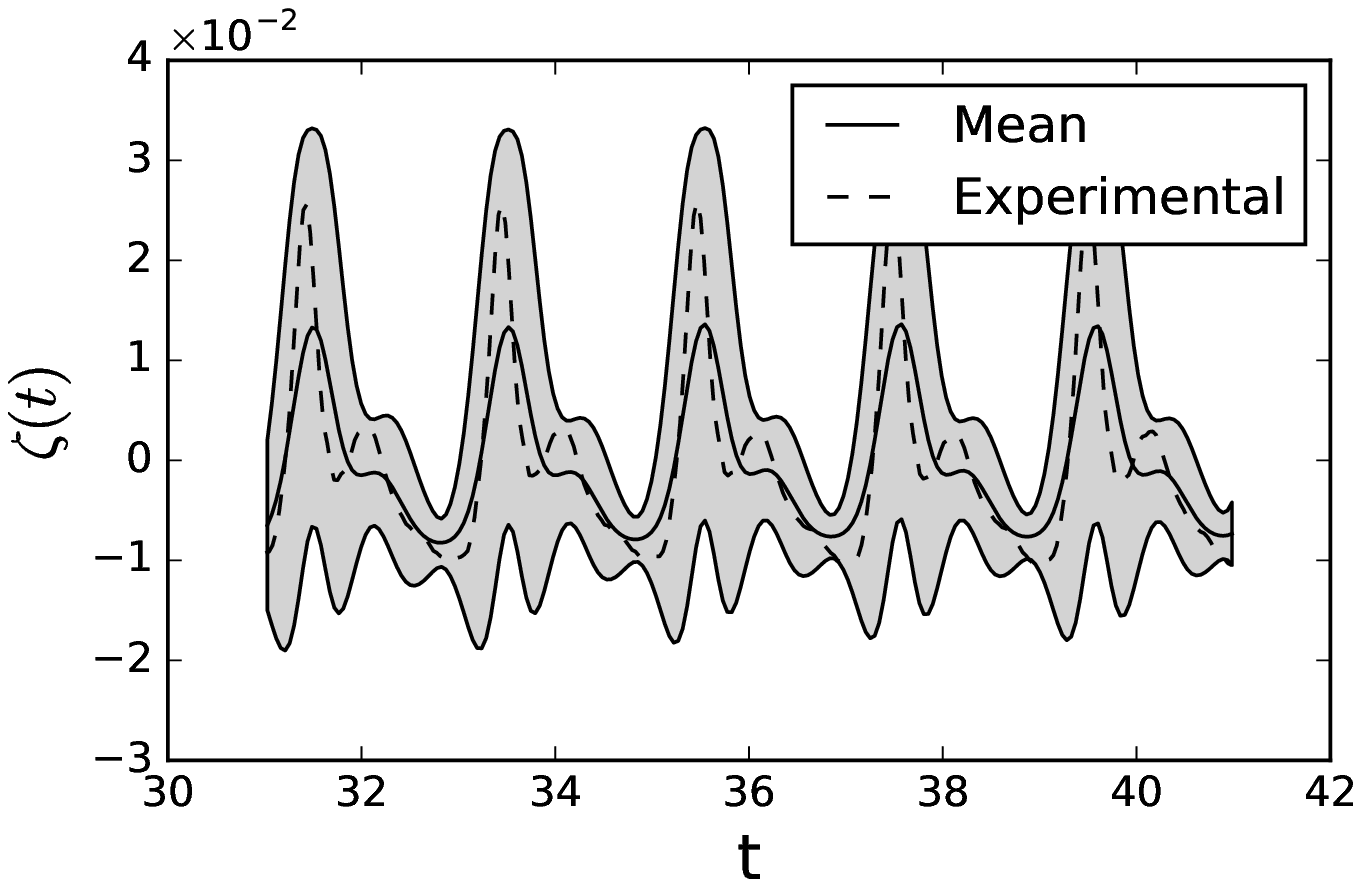}}
  \hspace{3pt}
  \subfloat[]{\label{fig:BarTest:Field-a30:3}\includegraphics[width=0.48\textwidth]{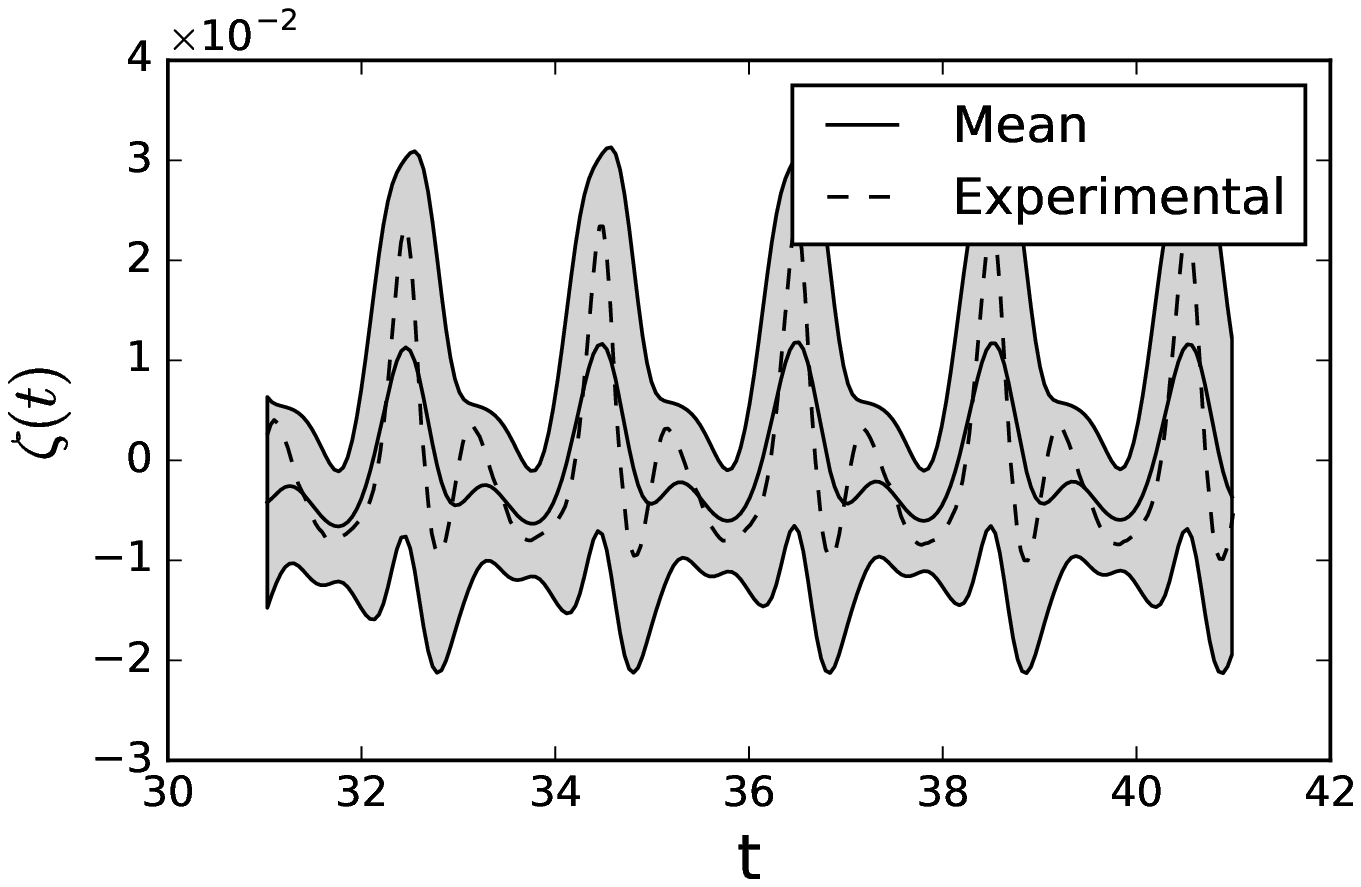}}\\
  \subfloat[]{\label{fig:BarTest:Field-a30:4}\includegraphics[width=0.48\textwidth]{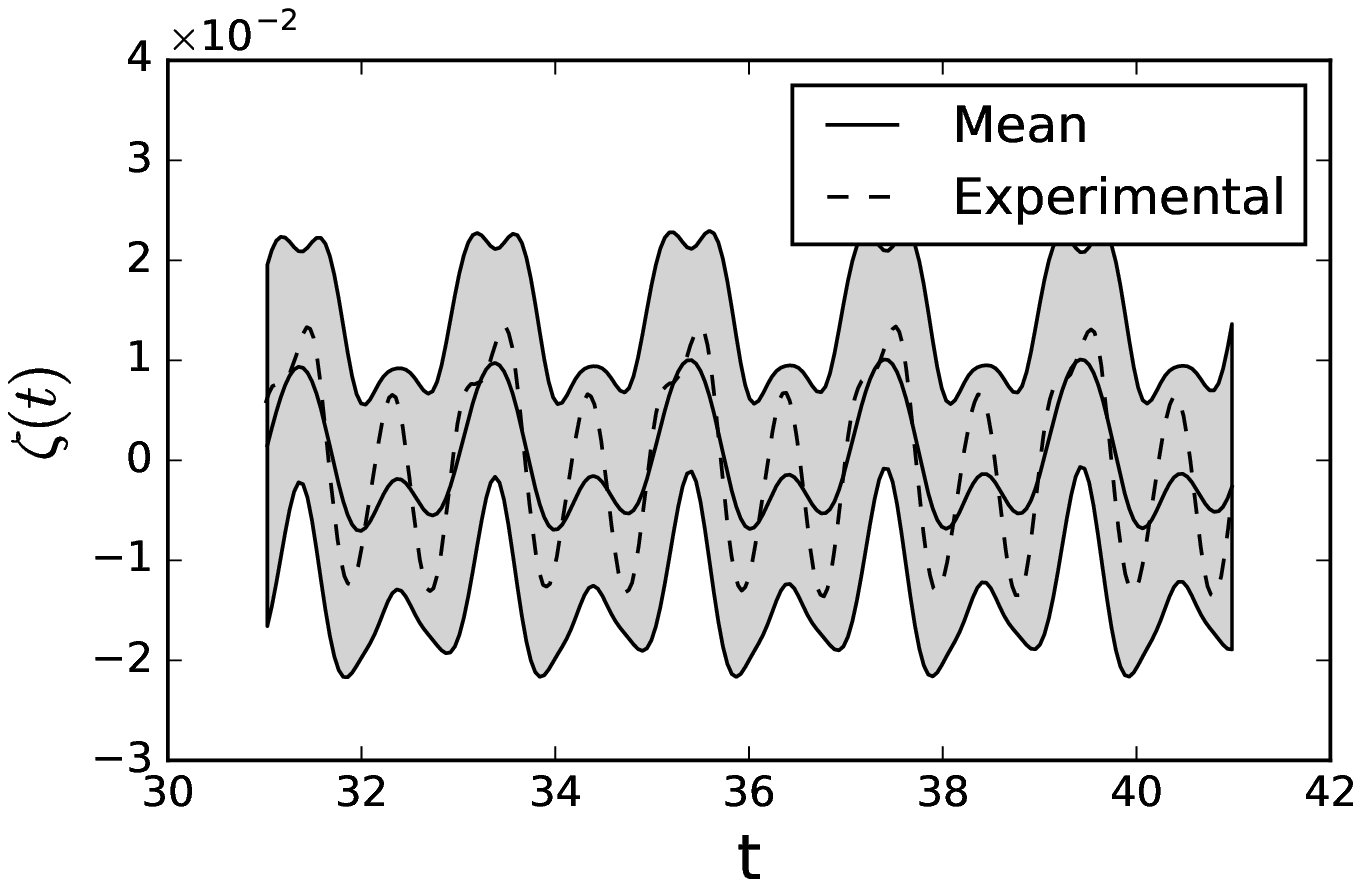}}
  \hspace{3pt}
  \subfloat[]{\label{fig:BarTest:Field-a30:5}\includegraphics[width=0.48\textwidth]{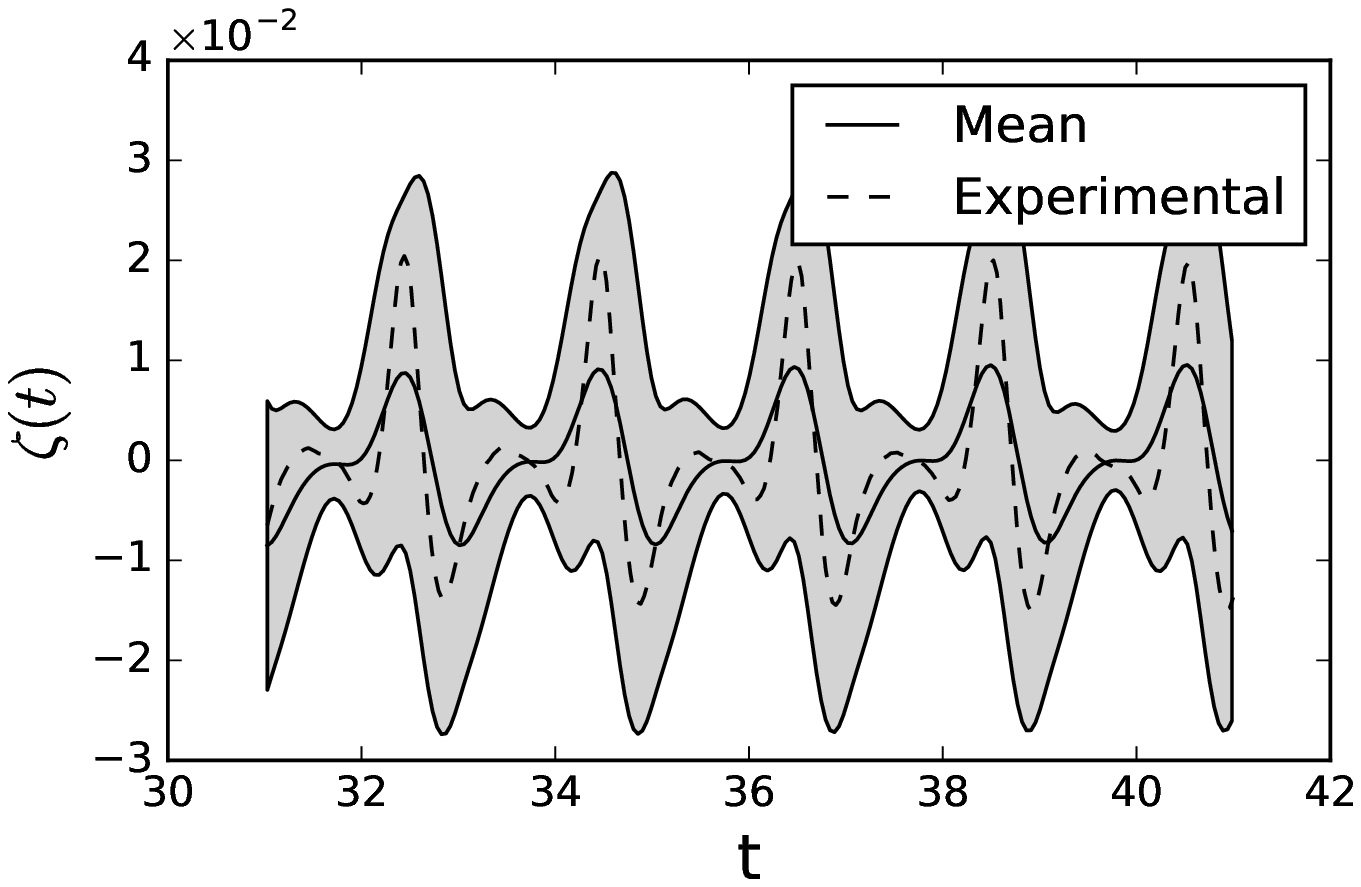}}\\
  \subfloat[]{\label{fig:BarTest:Field-a30:6}\includegraphics[width=0.48\textwidth]{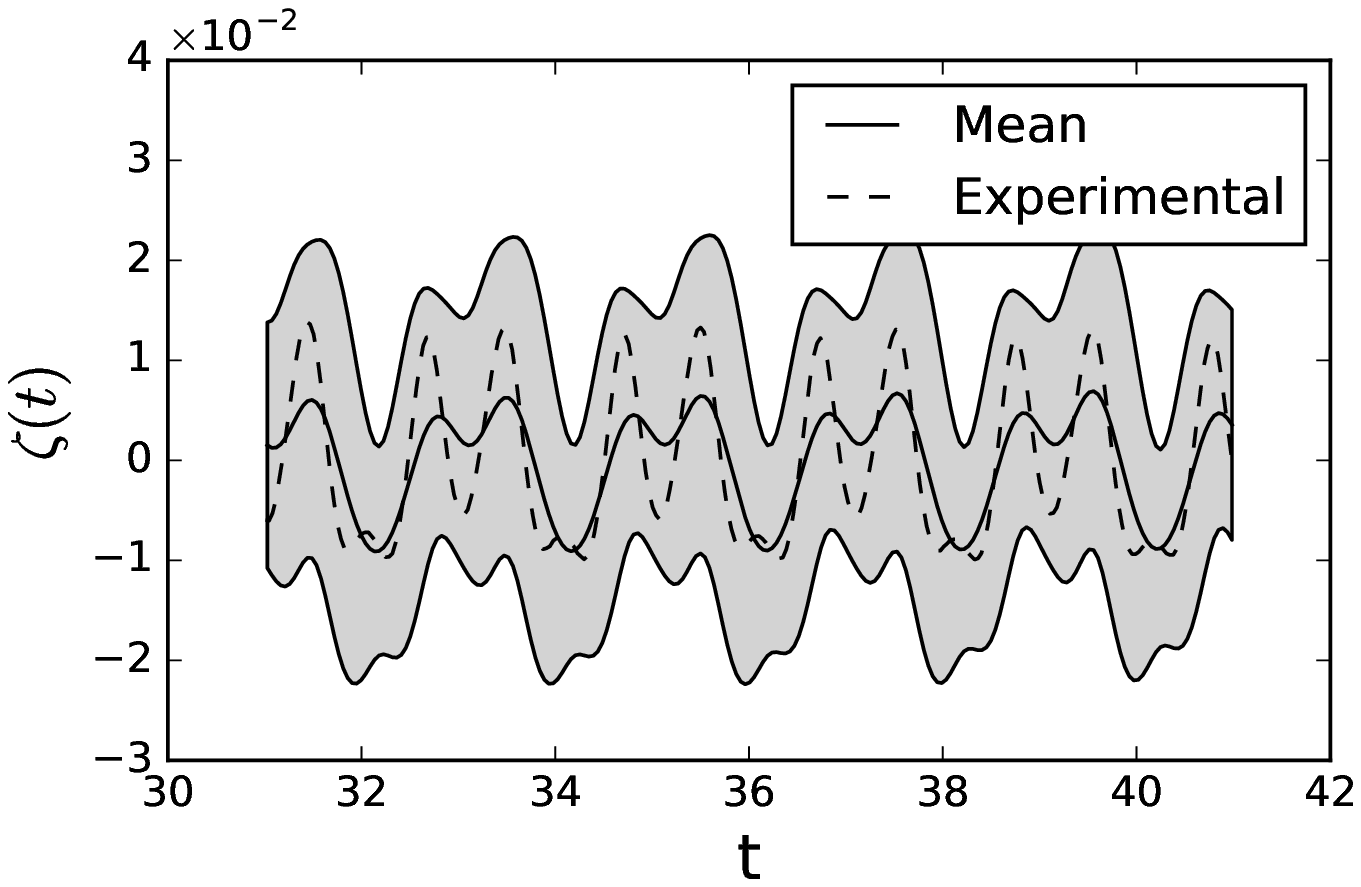}}
  \hspace{3pt}
  \subfloat[]{\label{fig:BarTest:Field-a30:7}\includegraphics[width=0.48\textwidth]{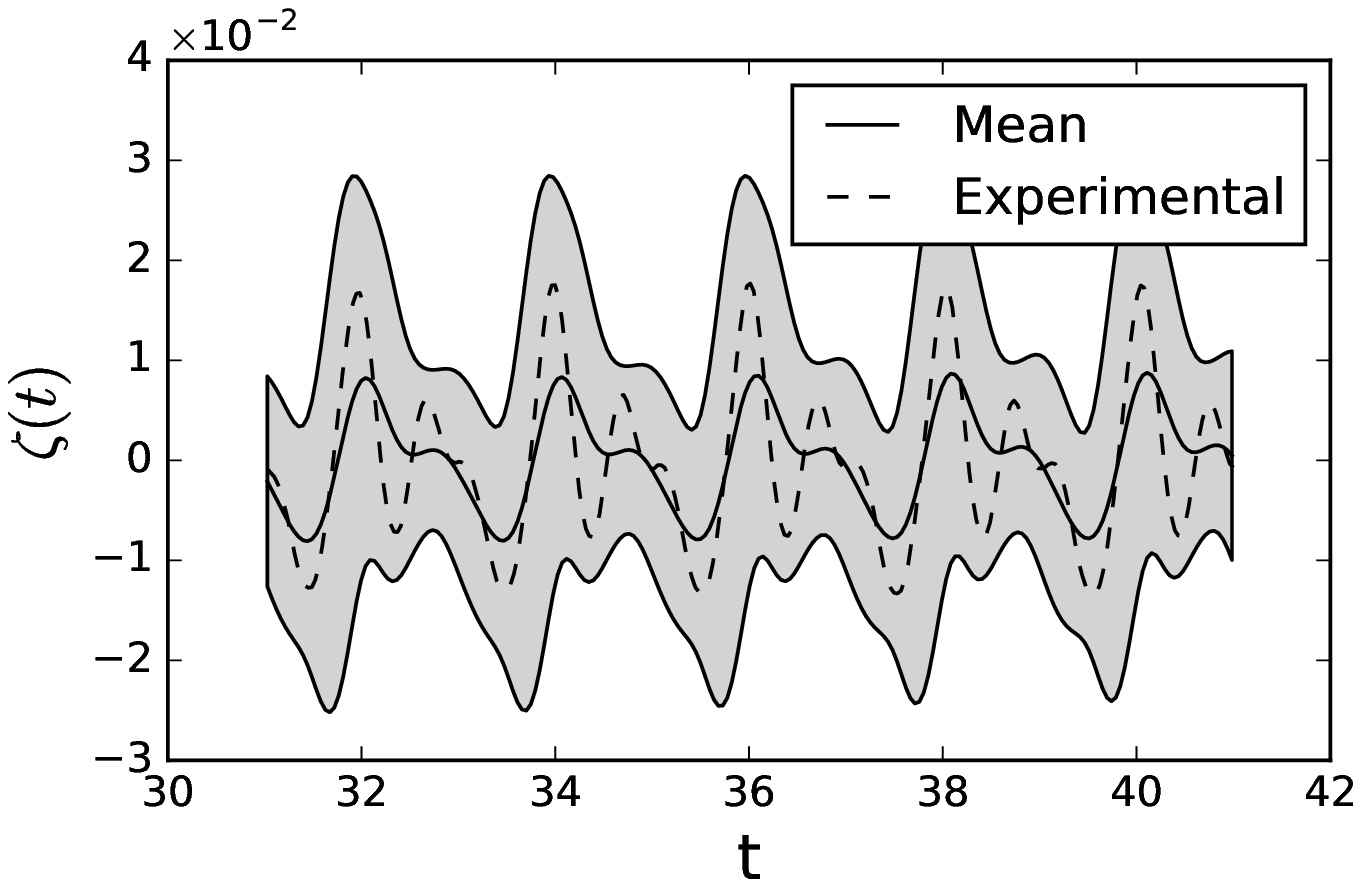}}\\
  \caption{Mean (solid line) and 95\% tolerance interval (shaded) of time-dependent free surface elevation at the eight gauge locations in the submerged bar experiment. Here the bottom topography is described by a Gaussian random field with Ornstein-Uhlenbeck \cite{Uhlenbeck1930} covariance function and correlation length $a=30.0m$. The experimental data (dashed line) is also shown.}
  \label{fig:BarTest:Field-a30}
\end{figure}
The topography of the basin is often precise in experimental settings, but rarely for real sites. Discrepancies with respect to the ideal/real design can be present. We will model these discrepancies using Gaussian random fields added on top of the deterministic basin, as shown in figure \ref{fig:BarTest:BottomProfiles}. 

We consider three Gaussian random fields with exponential covariance \eqref{eq:KL-exp:exp-cov-kernel} and with correlation lengths $a=(30.0, 10.0, 3.0)$. The mean of the fields is set to be the nominal bottom topography and the total variance of the fields is set to $\sigma^2=0.01^2$. Two realizations of such random fields are shown in figure \ref{fig:BarTest:BottomProfiles}. The random fields are expanded using the KL-expansion \eqref{eq:KL-expansion}, capturing $95\%$ of the total variance of the fields. This results in truncated KL-expansions with $5$, $13$ and $40$ terms respectively.

\begin{figure}
  \centering
  \subfloat[]{\label{fig:BarTest:FieldComparison-a10}\includegraphics[width=0.48\textwidth]{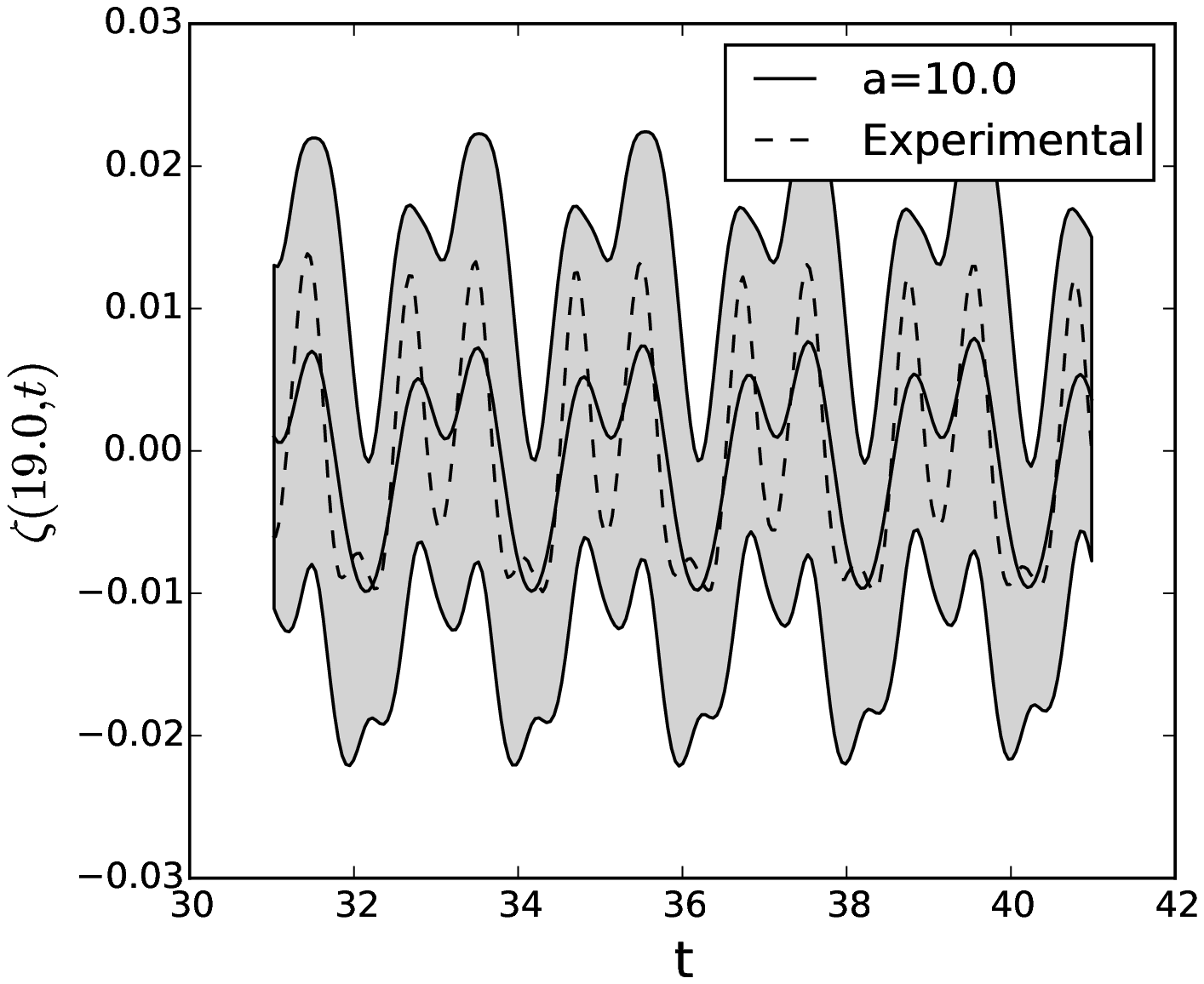}}
  \hspace{3pt}
  \subfloat[]{\label{fig:BarTest:FieldComparison-a3}\includegraphics[width=0.48\textwidth]{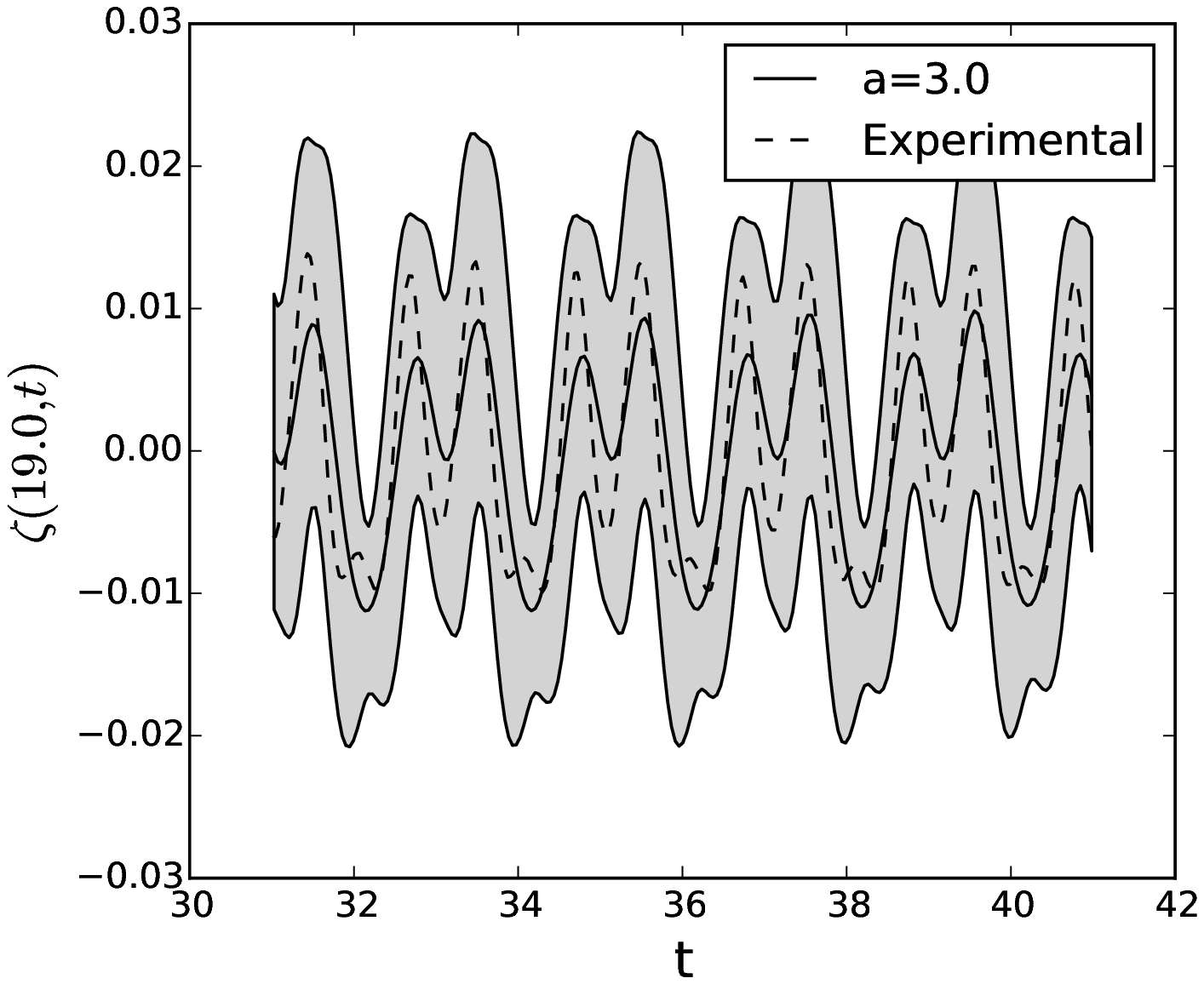}}
  \caption{Comparison of the mean and the 95\% tolerance interval of the time-dependent free surface elevation at the gauge number 7 for topographies described by two random fields with correlation lengths $a=10.0$ -- fig. \protect\subref{fig:BarTest:FieldComparison-a10} -- and $a=3.0$ -- fig. \protect\subref{fig:BarTest:FieldComparison-a3}.}
  \label{fig:BarTest:FieldComparison}
\end{figure}

The LHS method is used in the three cases checking that the number of samples ($n=5000$) is sufficient to estimate the mean up to the second digit of accuracy\footnote{The theoretical error of MC method can be expressed by the standard deviation of the MC estimator \eqref{eq:MC}, which is $\sigma/\sqrt{n}$. This can be approximated inserting the estimated variance $\bar{\sigma}$ and then used to test a convergence criteria for the method. Note that the estimate of LHS is often better than the MC estimate. In the example at hand both mean and variance are time and space dependent, i.e. $\mu({\bf x},t),\sigma^2({\bf x},t)$. For the gauge locations $\{{\bf x}_i\}_{i=1}^8$, we defined the convergence criteria by $\max_i\left(\frac{\Vert \sigma({\bf x}_i,t) \Vert_\infty / \sqrt{n}}{\Vert \mu({\bf x}_i,t) \Vert_\infty}\right)\leq 10^{-2}$.}.
Figure \ref{fig:BarTest:Field-a30} shows the results obtained using 5000 realizations of the deterministic model. We can see that the uncertain bottom topography considered plays an important role in the wave transformation downstream of the bar, even if the random field considered has a relatively long correlation length and small variance. Figures \ref{fig:BarTest:FieldComparison-a10} and \ref{fig:BarTest:FieldComparison-a3} show the mean and the 95\% tolerance interval for $a=10.0$ and $a=3.0$ respectively. One can notice that in the case with $a=3.0$, i.e. shorter correlation length and thus rougher topographies, there is a clear decrease in variance of the solution with respect to $a=30.0$ and $a=10.0$. The reasons behind this phenomena are subject of ongoing investigations

\subsection{Load of shoaling waves on off-shore structures}
A relevant part of the research on water waves is devoted to the estimation of loads -- see sec. \ref{sec:calculationLoads} -- on off-shore and coastal structures. In this context engineers are often interested in estimating the effect of extreme conditions or accumulated fatigue damage. To this end we will investigate the influence of uncertainties on the maximum load experienced by a structure positioned on the top of a shoaling bathymetry, after the system has reached its periodic solution.

\begin{figure}
  \centering
  \subfloat[]{\label{fig:LoadShoalin:ExperimentalSetting}\includegraphics[width=0.48\textwidth]{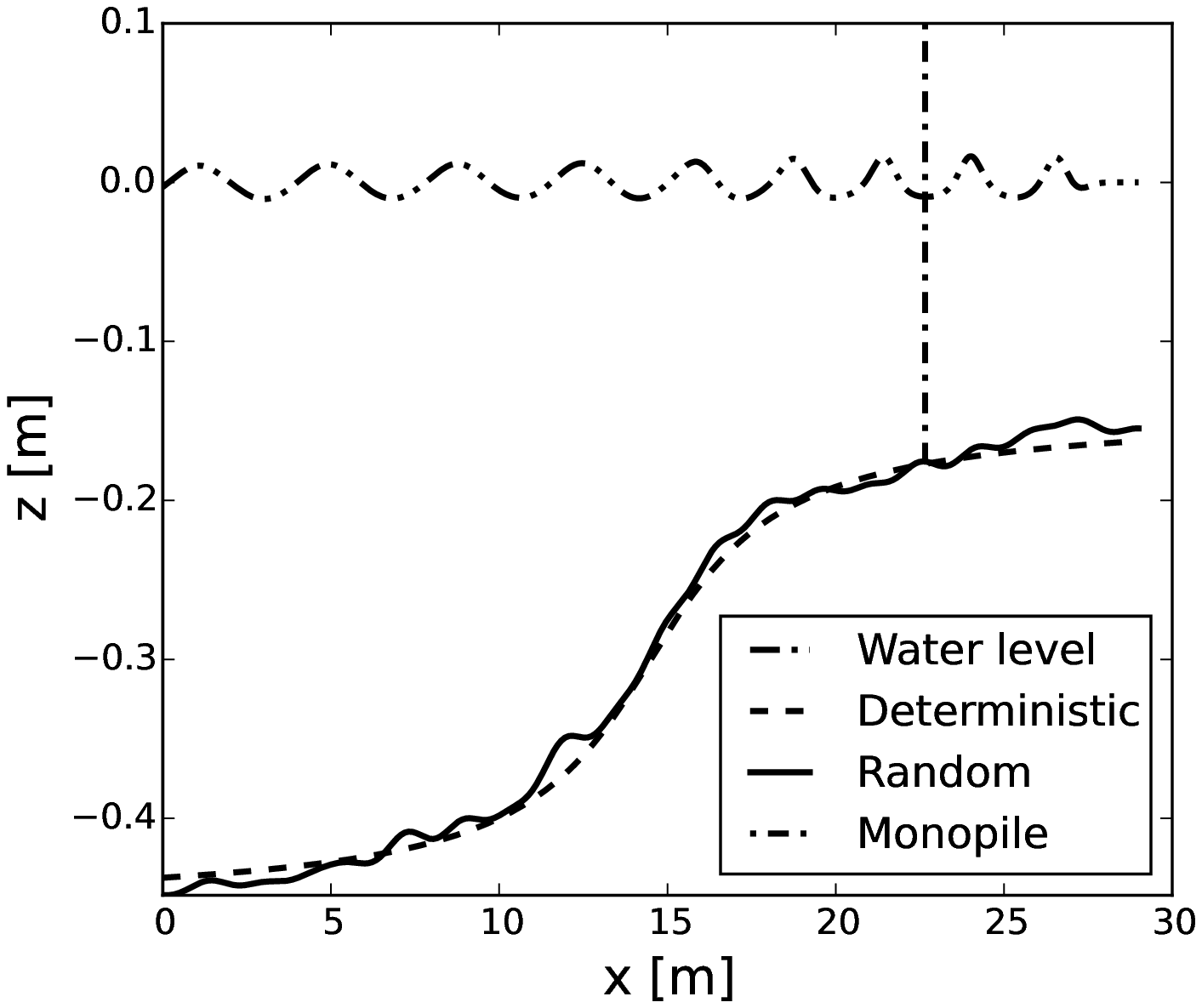}}
  \hspace{3pt}
  \subfloat[]{\label{fig:LoadShoalin:SolutionRealization}\includegraphics[width=0.48\textwidth]{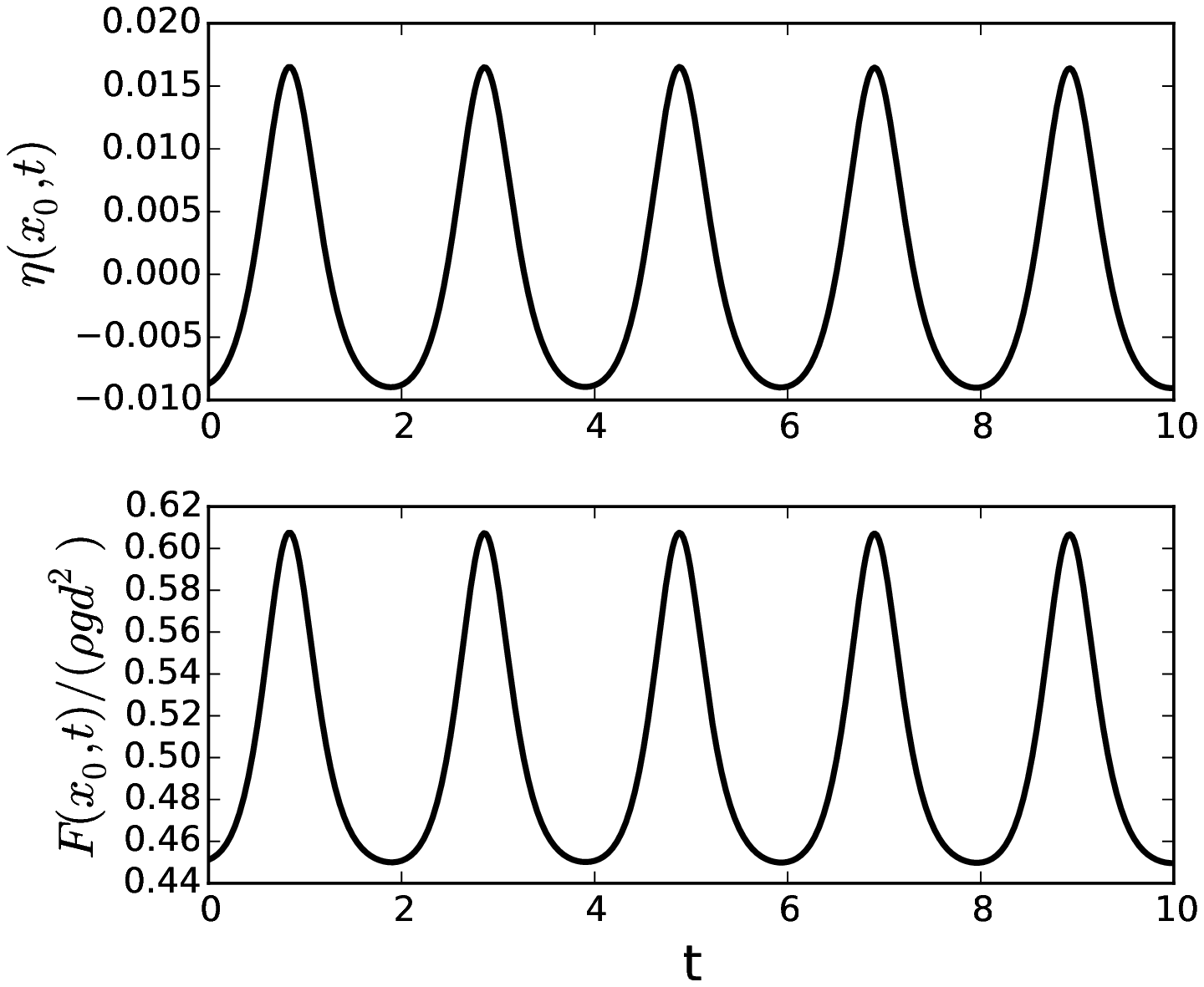}}
  \caption{Load of shoaling waves on an off-shore structure. Fig. \protect\subref{fig:LoadShoalin:ExperimentalSetting}: waves are propagated left to right over a uncertain shoaling bathymetry. The load is estimated at $x_0$, where a vertical structure is to be built. Fig. \protect\subref{fig:LoadShoalin:SolutionRealization}: time-dependent free surface elevation and load at the measuring point $x_0$.}
  \label{fig:LoadShoalin:SettingAndRealization}
\end{figure}

The experimental setting is shown in figure \ref{fig:LoadShoalin:ExperimentalSetting}, where waves are propagated left to right and shoal over the sloping bathymetry. It is assumed that a structure is to be built in the shallower part of the bathymetry at a fixed position $x_0$. Wave loads are then calculated at $x_0$ and their maximum is taken when the system has reached its periodic solution. The experiment does not account for wave-structure interactions \rvnote*{\#2-3}{that} would occur in the case a real structure was put in place. In such cases wave loads are often  estimated using Morison's equation. Instead, we estimate loads based directly on the estimate pressure distribution in the water column of an  assumed position of a cylinder. Figure \ref{fig:LoadShoalin:SolutionRealization} shows a possible realization of the solution when an uncertain bathymetry is considered (see sec. \ref{sec:LoadShoaling:UQ_bottom}).

\subsubsection{Uncertainty on wave height and wave period}\label{sec:LoadShoaling:UQ_H_T}
\begin{figure}
  \centering
  \subfloat[]{\label{fig:LoadShoalin:UQinputWaves}\includegraphics[width=0.48\textwidth]{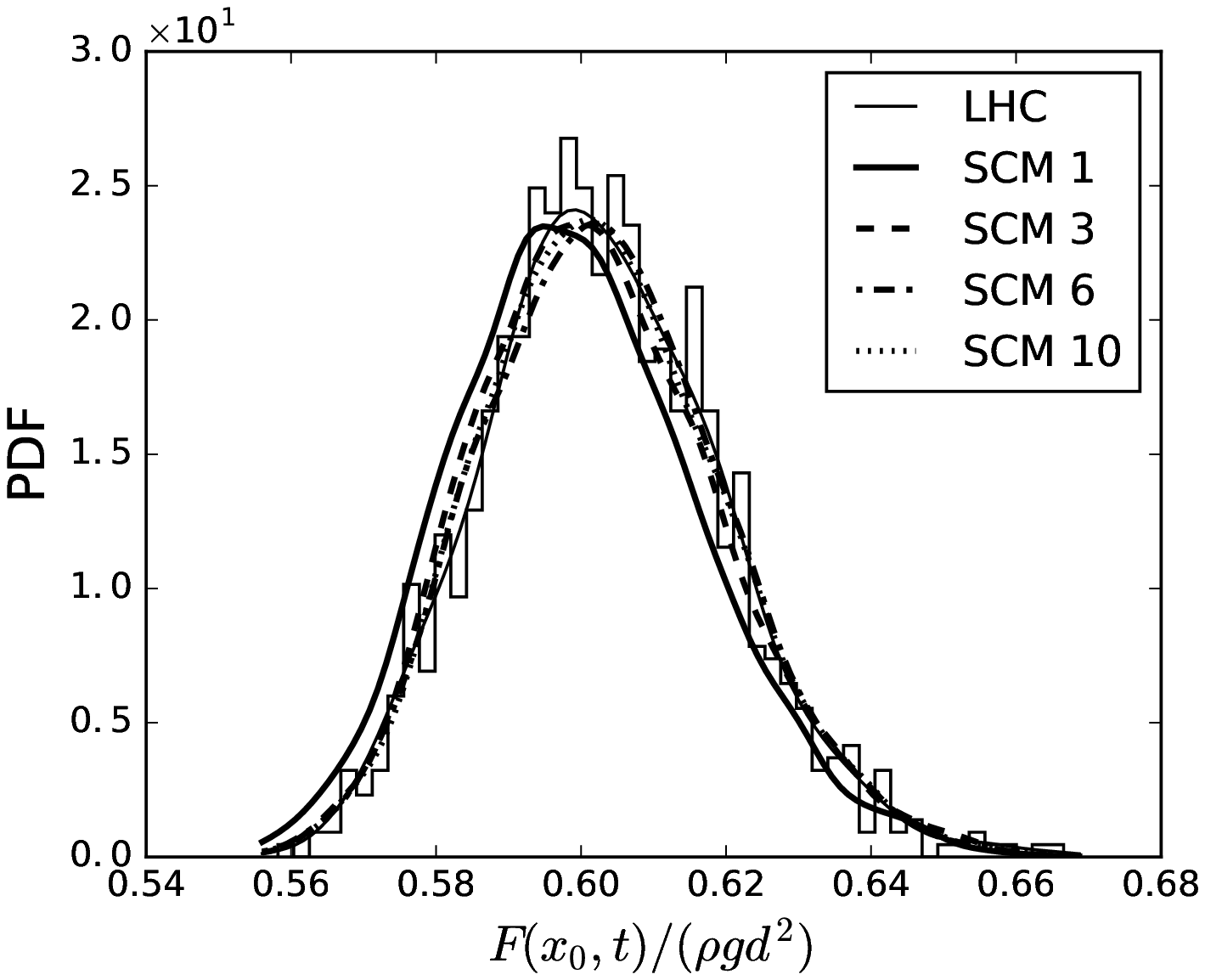}}
  \hspace{3pt}
  \subfloat[]{\label{fig:LoadShoalin:UQbottom}\includegraphics[width=0.48\textwidth]{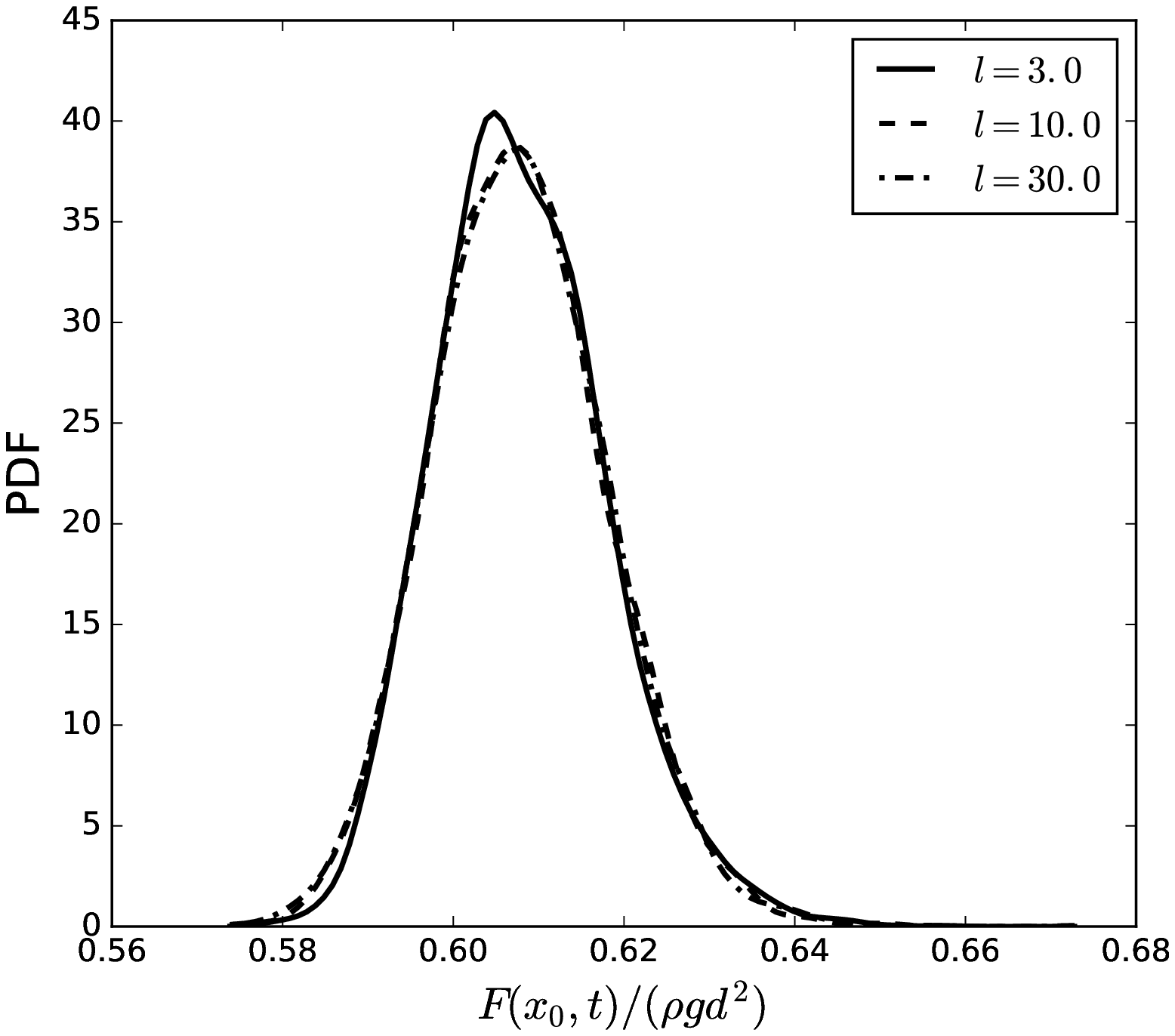}}
  \caption{PDF of the load of shoaling waves on an off-shore structure. Fig. \protect\subref{fig:LoadShoalin:UQinputWaves}: PDF of the load under uncertain input wave characteristics. The results obtained using SC method of orders 1,3,6 and 10 are compared to the one obtained using $10^3$ LHS samples. Fig. \protect\subref{fig:LoadShoalin:UQbottom}: PDF of the load under uncertain bathymetry for different correlation lengths.}
  \label{fig:LoadShoalin:SettingAndRealization}
  \rvnote{\#4-2}
\end{figure}
We will first focus on the uncertainties affecting the input wave characteristics. The input waves are regular stream function waves due to Dean \cite{Dean1965} and are defined by their height $H$ and period $T$, from which the wave speed is derived such that it satisfies the dispersion relation at the inlet of the domain. We describe the assumed uncertainties by Gaussian distributions with standard deviations of 10\% of their nominal values:
\begin{subequations}
  \label{eq:LaodShoalinWaveUQ}
  \begin{align}
    H \sim \mathcal{N}(0.02,0.002^2) \label{eq:LaodShoalinWaveH}\\
    T \sim \mathcal{N}(2.02,0.202^2) \label{eq:LaodShoalinWaveT}
  \end{align}
\end{subequations}
Figure \ref{fig:LoadShoalin:UQinputWaves} shows the PDF obtained using $10^3$ samples of the \rvnote*{\#4-2}{LHS} method and the PDF obtained using the gPC approximation constructed through the SC method with orders $P=1, 3, 6, 10$, using $(P+1)^2$ function evaluations. The results agree very closely even for low orders, suggesting that the underlying dependency of the load on the parameters is not significantly complex.

\subsubsection{Uncertain bottom topography}\label{sec:LoadShoaling:UQ_bottom}
We study now the influence of the uncertainty in the bottom topography to the maximum load predicted. The uncertainty is modeled through a Gaussian random field with Ornstein-Uhlenbeck \cite{Uhlenbeck1930} covariance and correlation lengths $a=30.0, 10.0, 3.0$. The mean of the field is set to the nominal bathymetry while the total variance is set to $\sigma^2=0.01^2$. A realization of such field is shown in figure \ref{fig:LoadShoalin:ExperimentalSetting}. The fields are parametrized by a KL-expansion \eqref{eq:KL-expansion} preserving 95\% of the total variance, resulting in truncations with $5$, $13$ and $40$ terms respectively.

Figure \ref{fig:LoadShoalin:UQbottom} shows the PDFs of the load with respect to the three uncertain bathymetries. The results are obtained using the \rvnote*{\#4-2}{LHS} method with $5000$ samples, which produce estimates of the mean accurate to its third digit. The PDFs obtained are very similar for the three cases, suggesting that the load is mostly sensitive to slowly varying perturbations of the nominal bathymetry.

\subsubsection{Sensitivity of loads to input uncertainties}

We now consider all the uncertainties studied in section \ref{sec:LoadShoaling:UQ_H_T} and \ref{sec:LoadShoaling:UQ_bottom} at the same time and we quantify the influence of each of these inputs through the variance based sensitivity analysis provided by the method of Sobol' (see section \ref{sec:sensitivityAnalysis}). Drawing from the experience acquired in section \ref{sec:LoadShoaling:UQ_bottom} we know that the three correlation lengths considered give very similar results. Then we consider an uncertain bottom topography modeled by a Gaussian random field with correlation length $a=30.0$, resulting in a KL-expansion with $5$ terms. The total number of uncertain parameters in the system is thus $7$.

\begin{table}
  \centering
  \begin{tabular}{c|c|ccccc}
     & \multirow{2}{*}{T.S.} & \multicolumn{5}{c}{Sparse grid} \\ \cline{3-7}
     & & $l=1$ & $l=2$ & $l=3$ & $l=4$ & $l=5$ \\ \hline
     \# $f$ eval.                & 3000 & 1    & 15   & 99   & 407  & 1317 \\ 
     Mean ($\times 10^{-1}$)      & 6.04 & 6.08 & 6.05 & 6.05 & 6.05 & 6.04 \\
     Std. dev. ($\times 10^{-2}$) & 2.03 & 0.0  & 1.90 & 1.92 & 1.97 & 2.03 \\
  \end{tabular}
  \caption{Sparse grid and LHS applied to estimate the mean $\mu$ and total standard deviation $\sigma$ in the load of shoaling waves on off-shore structures due to uncertainties in the input wave characteristics and in the bottom topography. The Kronrod-Patterson \cite{Kronrod1965} quadrature rule of different levels $l$ is used to construct the sparse grid approximation.}
  \label{tab:LoadShoalin:totvariance}
  \rvnote{\#4-2}
\end{table}

A necessary prerequisite to the computation of the sensitivities is the calculation of the effective dimension of the load function. This is done in terms of the amount of total variance expressed by the ANOVA-decomposition used in the method of Sobol'. This means one needs to compute the total variance first. We do this using the \rvnote*{\#4-2}{LHS} method and the sparse grid method of increasing levels. Table \ref{tab:LoadShoalin:totvariance} shows the convergence of the sparse grid method in terms of total variance as the number of evaluations is increased.

\begin{table}
  \centering
  \begin{tabular}{|c|c|c|c|c|c|c|c|c|c|c|}
    \hline
    \multicolumn{2}{|c|}{cut order} & \multicolumn{3}{c|}{1} & \multicolumn{3}{c|}{2} & \multicolumn{3}{c|}{3} \\ \hline
    \multicolumn{2}{|c|}{poly. order} & 2 & 4 & 6 & 2 & 4 & 6 & 2 & 4 & 6 \\ \hline
    \multicolumn{2}{|c|}{\# $f$ eval.} & 15 & 29 & 43 & 99 & 365 & 799 & 379 & 2605 & 8359 \\ \hline
    \multicolumn{2}{|c|}{Variance (\%)} & 83 & 89 & 93 & 87 & 93 & 98 & 87 & 96 & 98 \\ \hline
    \multirow{7}{*}{M.S.} 
     & $H$          & 0.50 & 0.50 & 0.50 & 0.49   & 0.48   & 0.47   & 0.49   & 0.49 & 0.48 \\ \cline{2-11}
     & $T$          & 0.10 & 0.15 & 0.19 & 0.10   & 0.14   & 0.19   & 0.10   & 0.15 & 0.18 \\ \cline{2-11}
     & bot.       & 0.24 & 0.24 & 0.24 & 0.26   & 0.29   & 0.29   & 0.26   & 0.30 & 0.28 \\ \cline{2-11}
     & $H$-$T$      &      &      &      & 4.8e-3 & 6.1e-3 & 7.8e-3 & 5.1e-3 & 6.8e-3 & 7.8e-3 \\ \cline{2-11}
     & $H$-bot.     &      &      &      & 8.7e-3 & 8.8e-3 & 8.9e-3 & 9.0e-3 & 1.0e-2 & 1.0e-2 \\ \cline{2-11}
     & $T$-bot.     &      &      &      & 7.4e-3 & 1.2e-2 & 1.5e-2 & 7.6e-3 & 1.3e-2 & 1.5e-2 \\ \cline{2-11}
     & $H$-$T$-bot. &      &      &      &        &        &        & 1.9e-4 & 2.3e-4 & 3.0e-4 \\ \hline
    \multirow{3}{*}{T.S.} & $H$ & 0.50 & 0.50 & 0.50 & 0.50 & 0.49 & 0.48 & 0.50 & 0.50 & 0.50 \\ \cline{2-11}
     & $T$ & 0.10 & 0.15 & 0.19 & 0.11 & 0.16 & 0.21 & 0.11  & 0.17 & 0.20 \\ \cline{2-11}
     & bot. & 0.24 & 0.24 & 0.24 & 0.28 & 0.31 & 0.32 & 0.28 & 0.32 & 0.31 \\ \hline
  \end{tabular}
  \caption{Main sensitivities (M.S.) $S_{\bf i}$ and total sensitivities (T.S.) $TS(i)$, as defined in \eqref{eq:TotSensitivity}, of the load of shoaling waves on an off-shore structure to input wave characteristics ($H$ and $T$) and uncertain bathymetry (bot.). Different analysis are shown, obtained using different cut orders (number of interactions) and polynomial orders. The number of function evaluations increase with both of these orders, but it leads also to an improved exploration of the total variance.}
  \label{tab:LoadShoalin:totsens}
  \rvnote{\#5-6}
\end{table}

The sensitivity analysis is then performed for increasing truncation orders in the number of interactions -- see \eqref{eq:cut-HDMR} -- and increasing PC orders. Table \ref{tab:LoadShoalin:totsens} shows that increasing both the number of interactions (cut order) and the PC order increases the percentage of total variance represented by the ANOVA decomposition. If one sets $q=0.95$ in the effective dimension criteria \eqref{eq:effectiveDimensionality}, then the analysis shows that the load function can be represented by a function including up to second order interactions. However, we can see that a sufficiently high PC order is required in order for the criteria to be fulfilled. Table \ref{tab:LoadShoalin:totsens} shows also the total sensitivities obtained. The sensitivities due to all the parameters characterizing the bottom topography are grouped in one unique entry. Under the assumed distributions for the input uncertainties, the height of the generated wave appears to be the most influential input on the load. Nevertheless, the other two inputs are not negligible either, meaning that none of them can be disregarded in the UQ analysis.

\subsection{Harmonic generation over a semi-circular shoal}
\begin{figure}
  \centering
  % \subfloat[Bottom Topography]{\label{fig:WhailnShoal3D:DeterministicBottomTopography}\includegraphics[width=0.48\textwidth]{DeterministicBottomTopography-bw}}
  \subfloat[]{\label{fig:WhailnShoal3D:DeterministicBottomTopography}\includegraphics[width=0.48\textwidth]{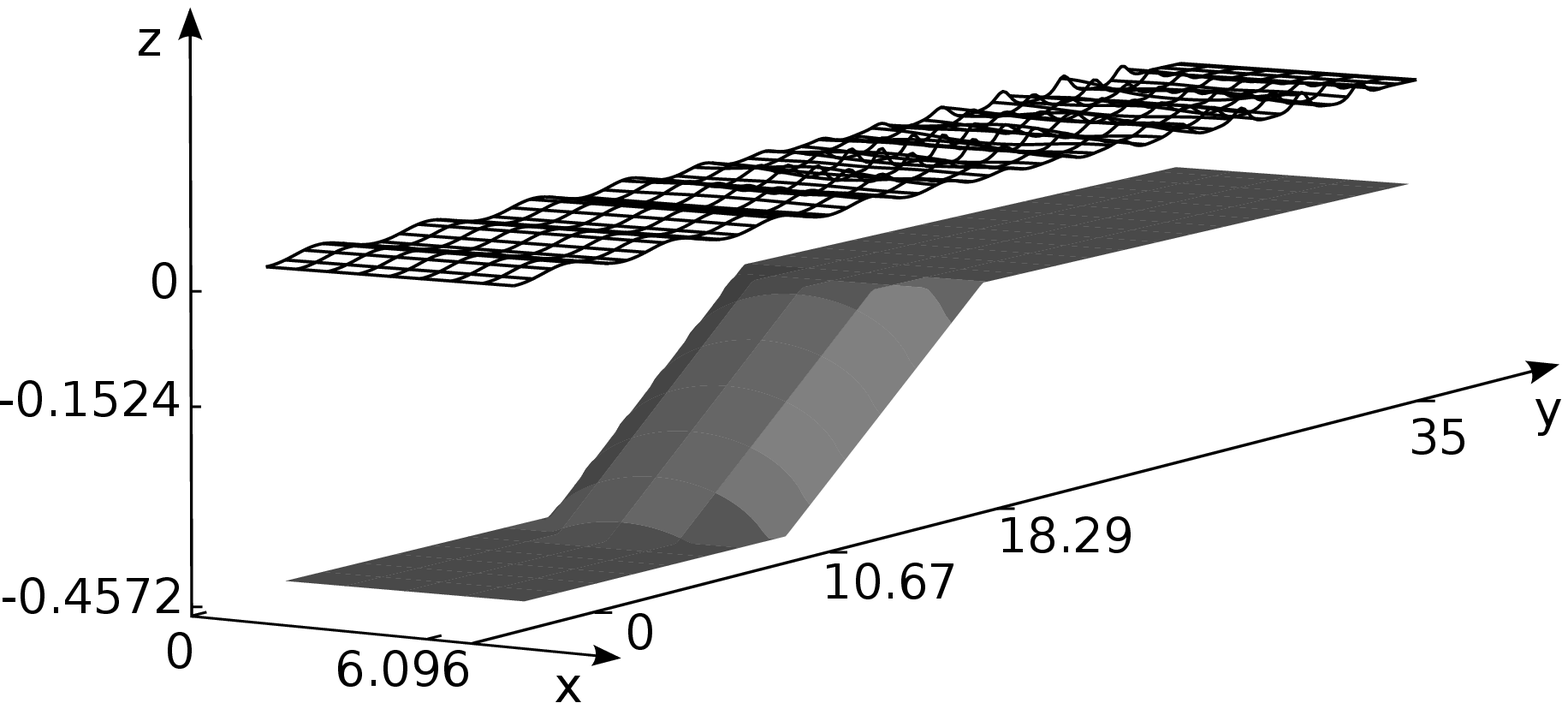}}
  \hspace*{5pt}
  \subfloat[]{\label{fig:WhailnShoal3D:DeterministicSolution}\includegraphics[width=0.48\textwidth]{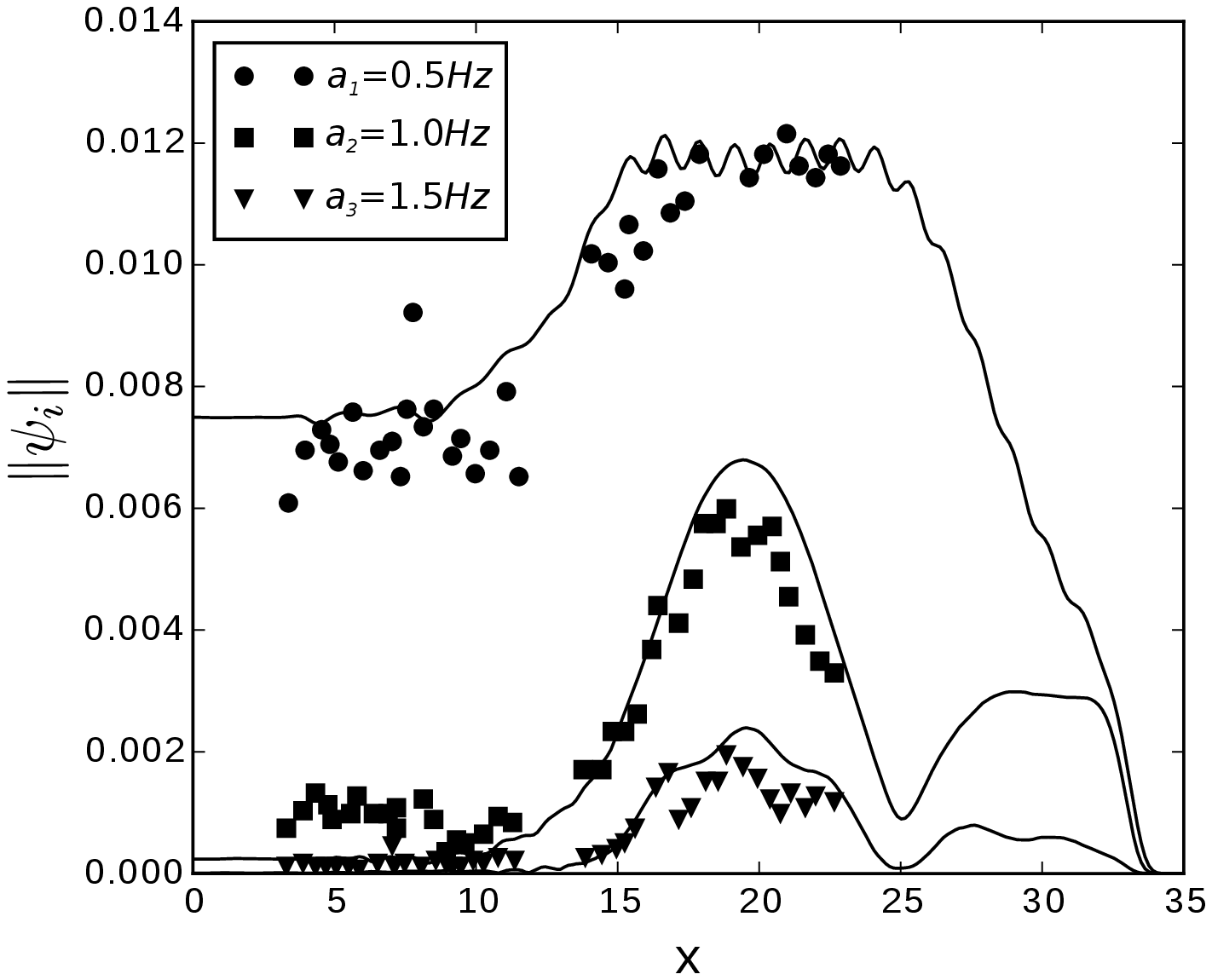}}
  \caption{Experimental setting and deterministic solution of the wave propagation in three dimensions over a semi-circular shoal. Fig. \protect\subref{fig:WhailnShoal3D:DeterministicBottomTopography}: deterministic bottom topography and free surface elevation at fixed time. Fig. \protect\subref{fig:WhailnShoal3D:DeterministicSolution}: the first three harmonics of the numerical solution (full lines) for the center-line are compared with the corresponding experimental measurements at different longitudinal locations in the basin (dots).}
  \label{fig:WhalinShoal3D:Deterministic}
\end{figure}

Extending the analysis to the full three dimensional problem we will proceed to the experiments of Whalin \cite{Whalin1971}. The experiments consist of a regular wave propagating over a semi-circular shoal, see figure \ref{fig:WhailnShoal3D:DeterministicBottomTopography}. The shoaling process transfers energy between the bound harmonics but, in contrast to the submerged bar case, the harmonics remain bounded and refraction adds complexity to the solution. The Whalin experiments have become standard benchmarks for dispersive wave models regardless of a rather substantial scatter  present in the experimental data. We will look into the case of of incoming waves with height $H=0.015$m and period $T=2$s. For this case most numerical models tend to over predict the amplitude of the second harmonic. % As the present model is able to accurately capture all the major phenomena taking place in the experiments we are interested to see what level of uncertainty this corresponds to in the experimental values.

% \begin{figure}
%   \centering
%   \subfloat[Wave Height]{\label{fig:WhailnShoal3D:UQWaveHeightMeanVar}\includegraphics[width=0.45\textwidth]{./Figures/WhalinShoal3D/WaveHeight/UQ_WaveHeight_Normal_000075_SCM_10-MeanVarExp-bw}}
%   \hspace*{5pt}
%   \subfloat[Wave Period]{\label{fig:WhailnShoal3D:UQWavePeriodMeanVar}\includegraphics[width=0.45\textwidth]{./Figures/WhalinShoal3D/WavePeriod/UQ_WavePeriod_Normal_01_SCM_20-MeanVarExp-bw}}
%   \caption{Uncertainty quantification of the solution of the Whalin test with respect to the wave height and the wave period. The numerical means are plotted as full lines. The shaded areas represent one standard deviation from the mean. The full dots are experimental measurements.}
%   \label{fig:WhalinShoal3D:UQ1DMeanVar}
% \end{figure}

The deterministic numerical solution is computed for $t\in[0,50]$, to ensure the reaching of the periodic solution, and then is compared with the experimental measurements of the magnitude of the first three harmonics at different measurement locations through the center line of the domain. Due to the symmetry of the domain along its center line, the solution is computed only on half of the domain. Figure \ref{fig:WhailnShoal3D:DeterministicSolution} shows the fitting of the deterministic numerical solution to the measurement data. 

For the deterministic case, the solution is obtained in approximately 4min on an Intel$^\circledR$~Core\texttrademark ~i7-2640M CPU {@} 2.80GHz. The aim of the next sections is to study how the uncertainty in some experimental parameters can influence the results. Without presumption of causality, this analysis can highlight parameters that can influence results more deeply than others. The computational cost of solving the full three dimensional problem calls for efficient UQ methodologies that require the minimum number of simulations to make analysis practically feasible.

\subsubsection{One dimensional uncertainties}
Building up on the experience acquired on the two dimensional case and from experimental knowledge, we will focus our attention to the two parameters that are most difficult to match in real experiments, namely the input wave period and height. Due to the lack of information about how accurate experiments can be, we will assume that the input parameters are described by a Gaussian distribution and we will try to evaluate how sensitive the system is to single uncertainties, and, in the next section, to the combination of the two. We will model the wave height and the wave period with Gaussian distributions centered on their nominal values and with $5\%$ standard deviation.

\begin{figure}
  \centering
  \subfloat[]{\label{fig:WhailnShoal3D:UQWaveHeightDistr}\includegraphics[width=1.\textwidth]{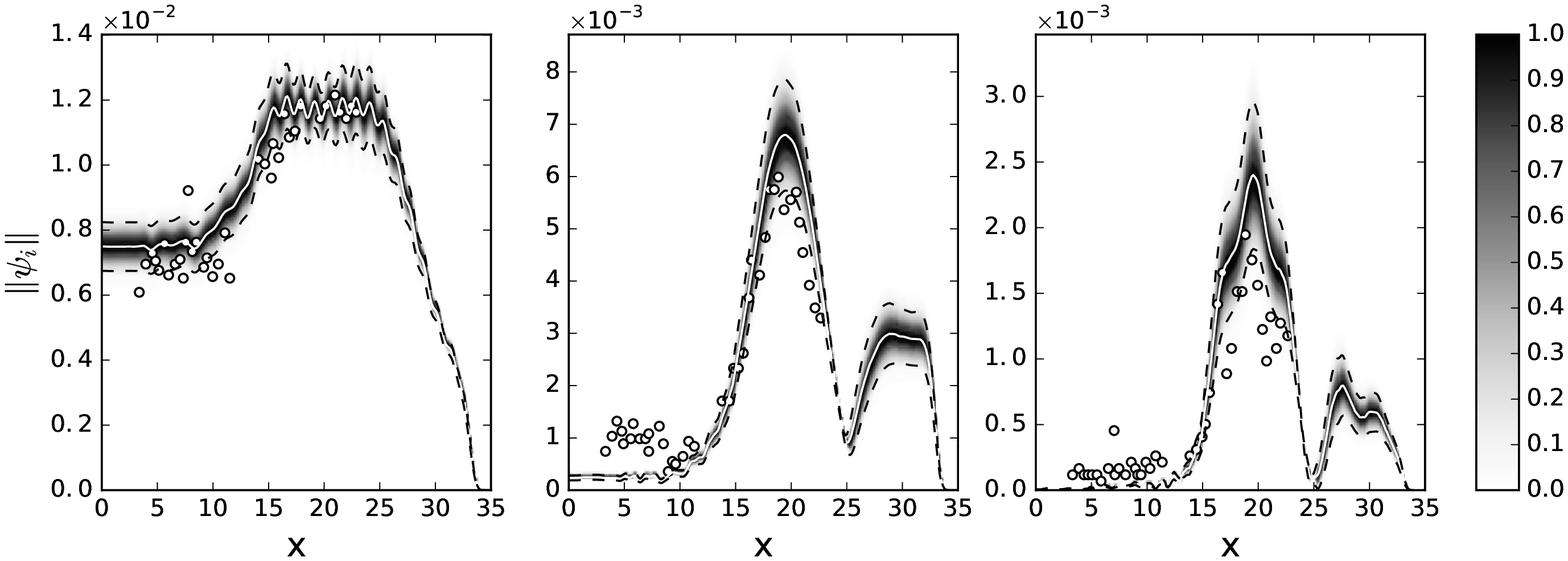}}
  \\
  \subfloat[]{\label{fig:WhailnShoal3D:UQWavePeriodDistr}\includegraphics[width=1.\textwidth]{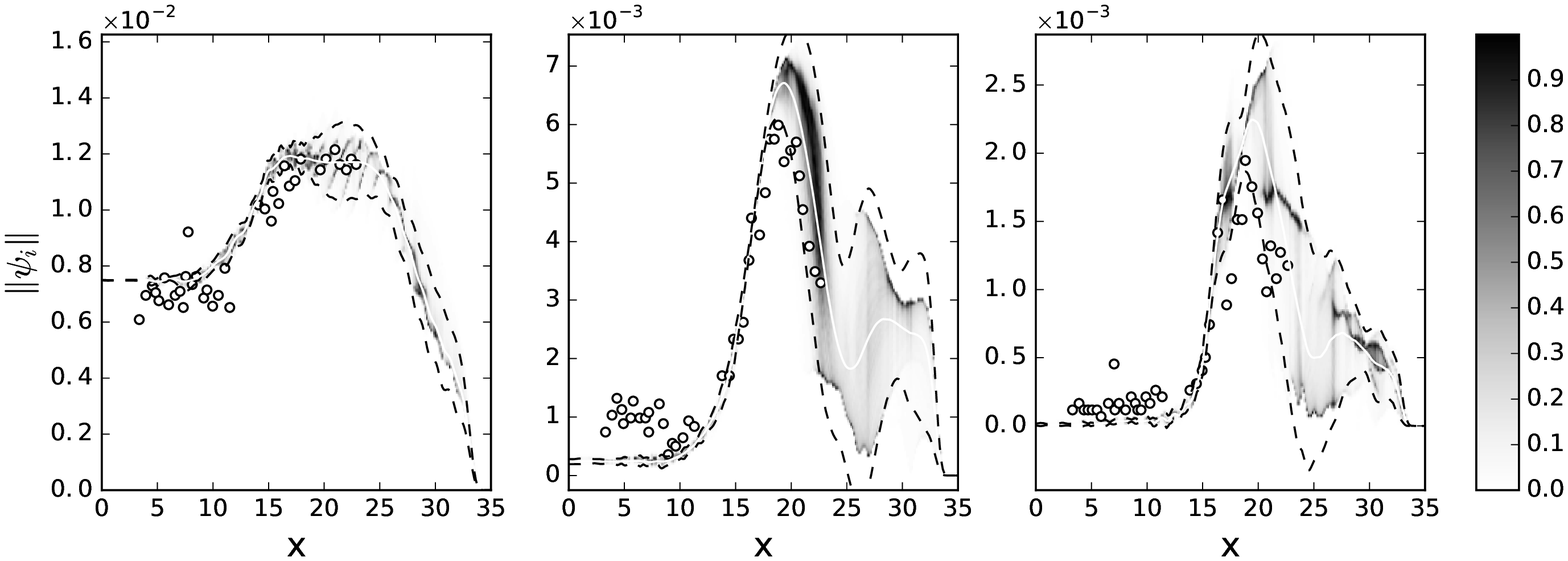}}
  \caption{Reconstructed space-dependent probability distribution function of the three harmonics of the solution of the Whalin test with uncertainty on the wave height -- fig. \protect\subref{fig:WhailnShoal3D:UQWaveHeightDistr} -- and on the wave period -- fig. \protect\subref{fig:WhailnShoal3D:UQWavePeriodDistr}. The white line shows the space-dependent mean, while the dashed lines show the $95\%$ tolerance interval around the mean. The scattered dots show the experimental data results.}
  \label{fig:WhalinShoal3D:UQ1DDistr}
\end{figure}

The SC method for the estimation of the gPC expansion \eqref{eq:gPC:freesurf} is adopted, with the order dictated by the accuracy required. Since the input uncertainties follow a Gaussian distribution, the polynomial basis used for their expansion are the Hermite polynomials. Figure \ref{fig:WhalinShoal3D:UQ1DDistr} shows the mean and the $95\%$ tolerance interval as well as the space-dependent distribution of the harmonics and the fitting with the experimental data for the two parameters considered separately.

% \begin{figure}
%   \centering
%   \subfloat[Wave Height]{\label{fig:WhailnShoal3D:UQWaveHeightCoeff}\includegraphics[width=1.\textwidth]{./Figures/WhalinShoal3D/WaveHeight/UQ_WaveHeight_Normal_000075_SCM_10-Coeff}}
%   \\
%   \subfloat[Wave Period]{\label{fig:WhailnShoal3D:UQWavePeriodCoeff}\includegraphics[width=1.\textwidth]{./Figures/WhalinShoal3D/WavePeriod/UQ_WavePeriod_Normal_01_SCM_20-Coeff}}
%   \caption{Space-dependent decay of the expansion coefficients for the surrogate model  \eqref{eq:UQ:gPCexpansion}.}
%   \label{fig:WhalinShoal3D:UQ1DCoeff}
% \end{figure}

\subsubsection{Two dimensional uncertainty}
% \begin{figure}
%   \centering
%   \includegraphics[width=0.60\textwidth]{./Figures/WhalinShoal3D/WaveHeightPeriod/UQ_WaveHeight_Normal_000075_WavePeriod_Normal_01_SCM_10_20-MeanVarExp-bw}
%   \caption{Uncertainty quantification of the solution of the Whalin test with respect to uncertainties in the wave height and the wave period. The numerical means are plotted as full lines. The shaded areas represent one standard deviation from the mean. The full dots are experimental measurements.}
%   \label{fig:WhalinShoal3D:UQ2DMeanVar}
% \end{figure}
\begin{figure}
  \centering
  \includegraphics[width=1.\textwidth]{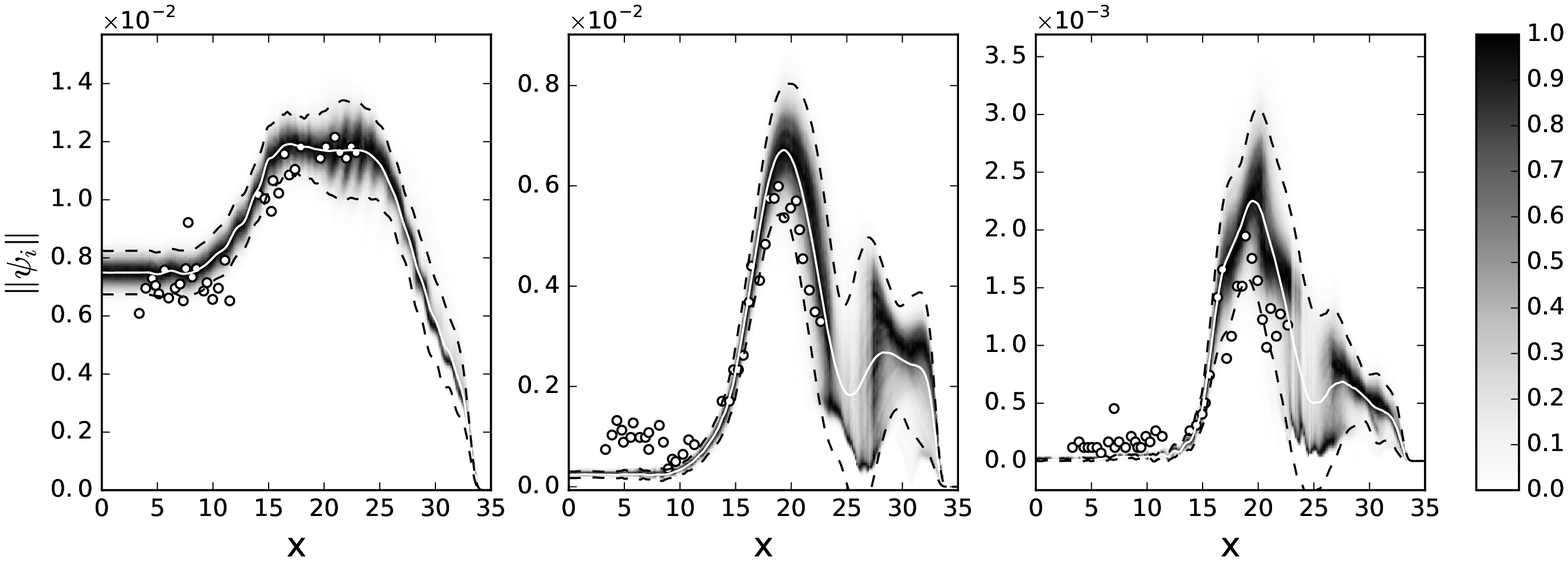}
  \caption{Space-dependent probability distribution function of the Whalin test with two-dimensional uncertainty. The white solid line represents the mean for the three harmonics. The dashed lines show the $95\%$ tolerance interval around the mean. The scattered dots are the experimental measurements.}
  \label{fig:WhalinShoal3D:UQ1DCoeff}
\end{figure}
The same problem setting is now investigated with uncertainty on the wave height and period at the same time. The same distributions used in the one dimensional setting are used here for the uncertainty sources. The SC method of order $5$ is used to compute the space dependent probability distribution of the first three harmonics of the propagated wave. A total of only $36$ deterministic simulations are required to obtain the desired approximation\footnote{The number of quadrature points for a one dimensional Gauss rule of polynomial order $5$ is $6$. The tensorization of this quadrature rule in two dimension leads then to $6^2=36$ evaluation points.}.

Figures \ref{fig:WhalinShoal3D:UQ1DCoeff} shows the space dependent mean and $95\%$ tolerance interval of the first three harmonics, as well as their space dependent probability distribution. Again we can notice that the resulting uncertainty -- measured in variance of the solution -- is not the mere superposition of the variances obtained with one dimensional uncertainties (see fig. \ref{fig:WhalinShoal3D:UQ1DDistr}). The probability distribution of the first three harmonics seem now to include the experimental measurements within some high probability region.

\subsubsection{Uncertain bottom topography}
\begin{figure}
  \centering
  % \subfloat[Bottom Topography]{\label{fig:WhailnShoal3D:DeterministicBottomTopography}\includegraphics[width=0.48\textwidth]{DeterministicBottomTopography-bw}}
  \subfloat[]{\label{fig:WhailnShoal3D:RandomBottomTopography}\includegraphics[width=0.48\textwidth]{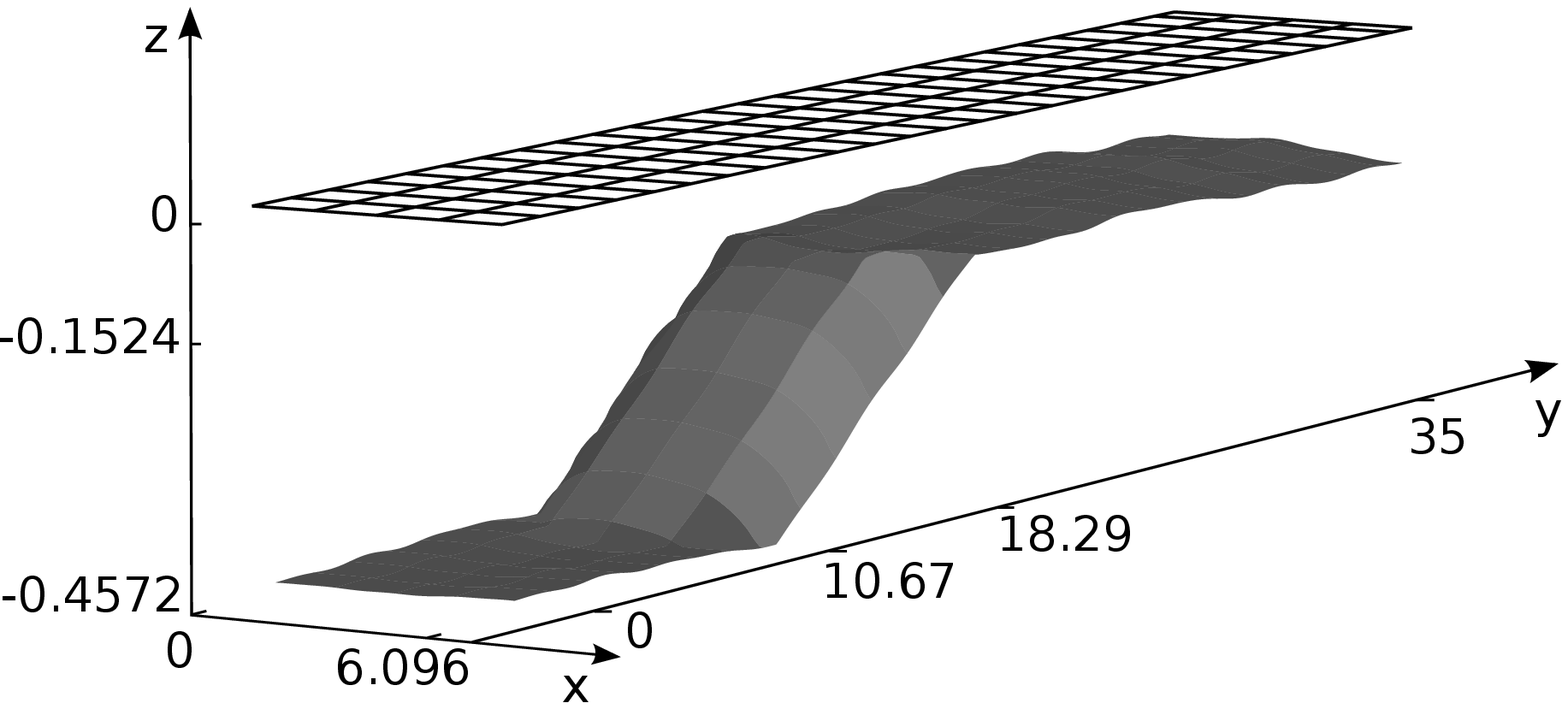}}
  \hspace*{5pt}
  \subfloat[]{\label{fig:WhailnShoal3D:a10-MeanVar}\includegraphics[width=0.48\textwidth]{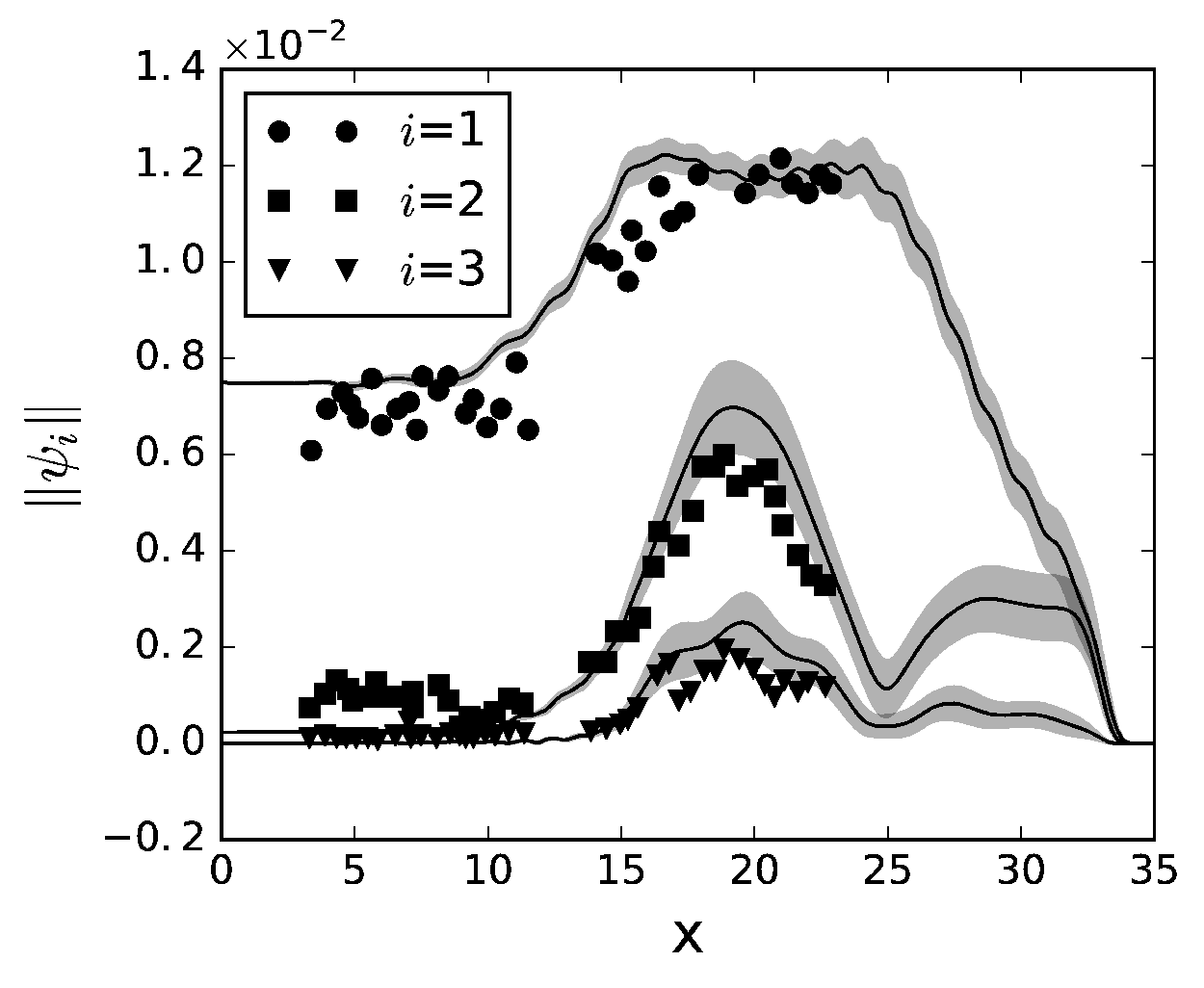}}
  \caption{Experimental setting accounting for uncertainty on the bottom topography and solution of the wave propagation in three dimensions over a semi-circular shoal. Fig. \protect\subref{fig:WhailnShoal3D:RandomBottomTopography}: realization from the Gaussian random field with correlation length $a=10.0$, describing the uncertain bottom topography. Fig. \protect\subref{fig:WhailnShoal3D:a10-MeanVar}: mean and 95\% tolerance interval of the first three harmonics of the numerical solution (full lines), compared with the corresponding experimental measurements at different longitudinal locations in the basin (dots).}
  \label{fig:WhalinShoal3D:Deterministic}
\end{figure}

We model the uncertainty on the bottom topography through the superposition of a Gaussian random field with Ornstein-Uhlenbeck \cite{Uhlenbeck1930} covariance \eqref{eq:KL-exp:exp-cov-kernel} on top of the deterministic bathymetry shown in figure \ref{fig:WhailnShoal3D:DeterministicBottomTopography}. The correlation lengths $a=(30.0, 10.0, 3.0)$ and the total variance of the field $\sigma^2=0.01^2$ are chosen as illustrative examples to characterize the random field. The random fields are set to be isotropic, meaning that the correlation and variability are the same for both of the directions of the two dimensional bottom topography. The random fields are expanded using the KL-expansion \eqref{eq:KL-expansion} capturing $95\%$ of the total variance. This results in truncated KL-expansions with $9$, $43$ and $321$ terms respectively. Figure \ref{fig:WhailnShoal3D:RandomBottomTopography} shows a realization of such topography. Unlike to the previous experiments where the domain was symmetric along its center line, in these experiments the symmetry is lost. Thus, the solution of the system must be computed over the whole domain, resulting in an increased computational cost. The deterministic solution over the whole domain is obtained in approximately 8min on an Intel$^\circledR$~Core\texttrademark ~i7-2640M CPU {@} 2.80GHz.

\begin{figure}
  \centering
  \subfloat[]{\label{fig:WhailnShoal3D:UQField_a30}\includegraphics[width=1.\textwidth]{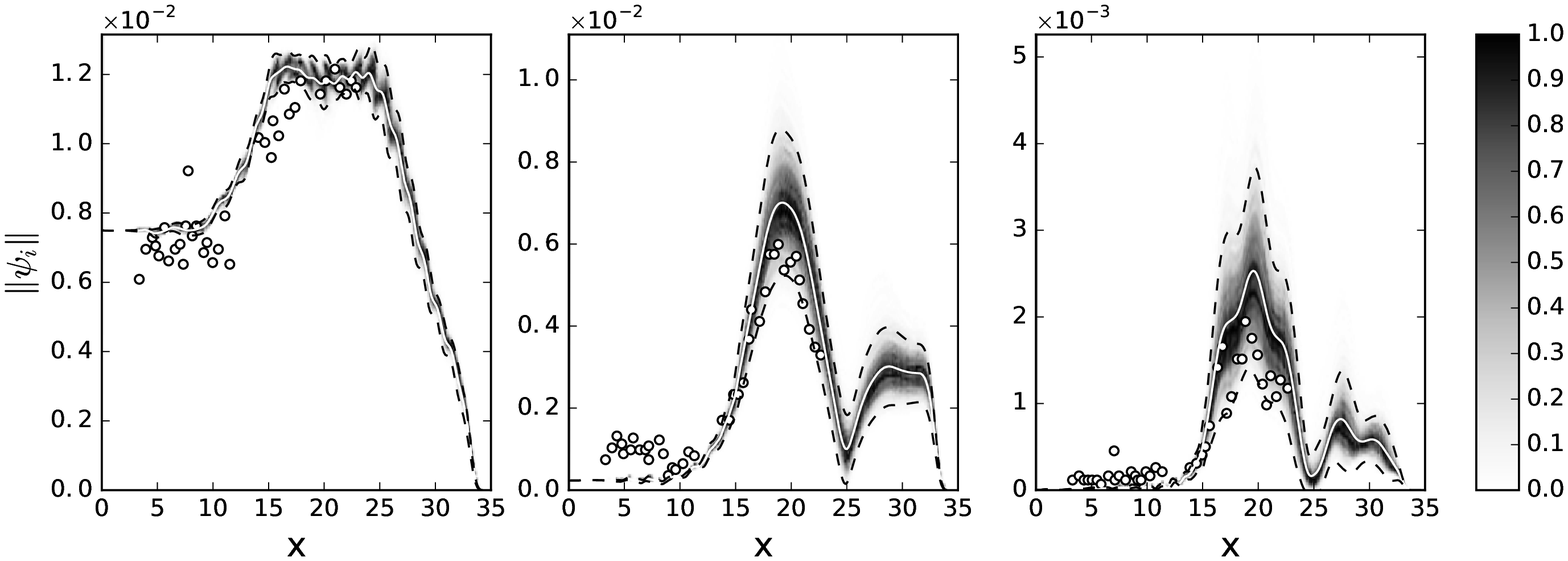}}
  \\
  \subfloat[]{\label{fig:WhailnShoal3D:UQField_a10}\includegraphics[width=1.\textwidth]{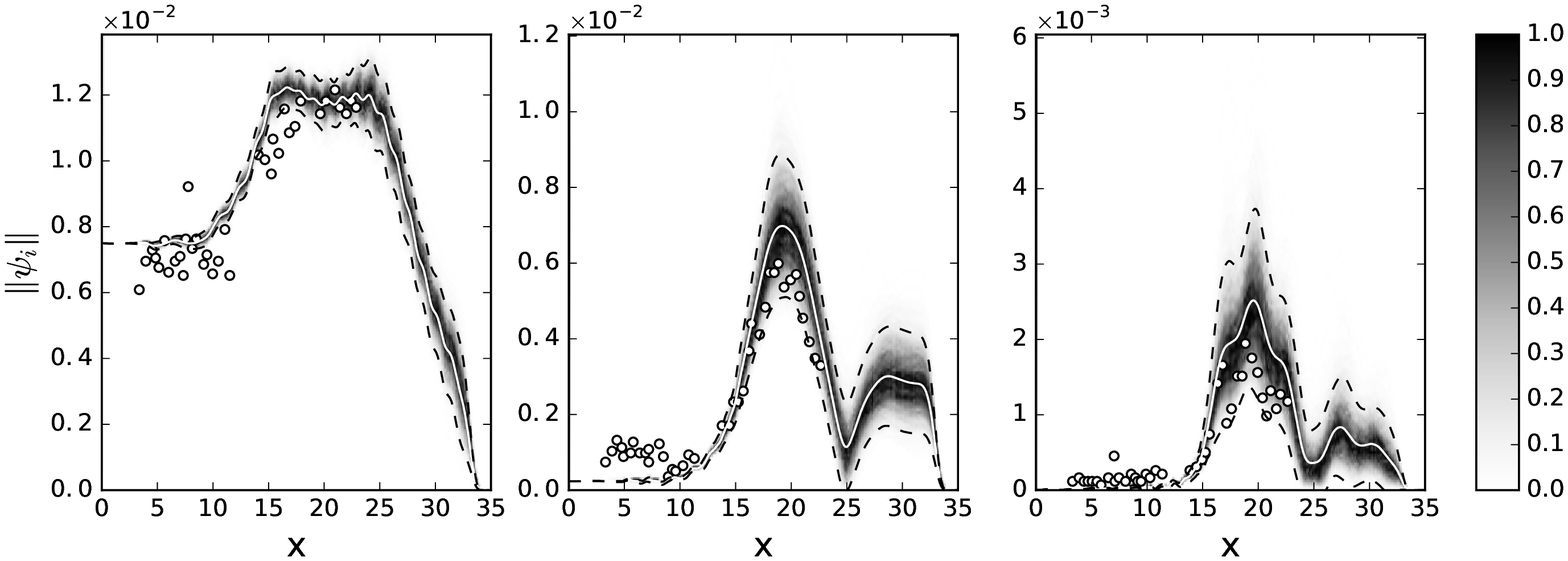}}
  \\
  \subfloat[]{\label{fig:WhailnShoal3D:UQField_a3}\includegraphics[width=1.\textwidth]{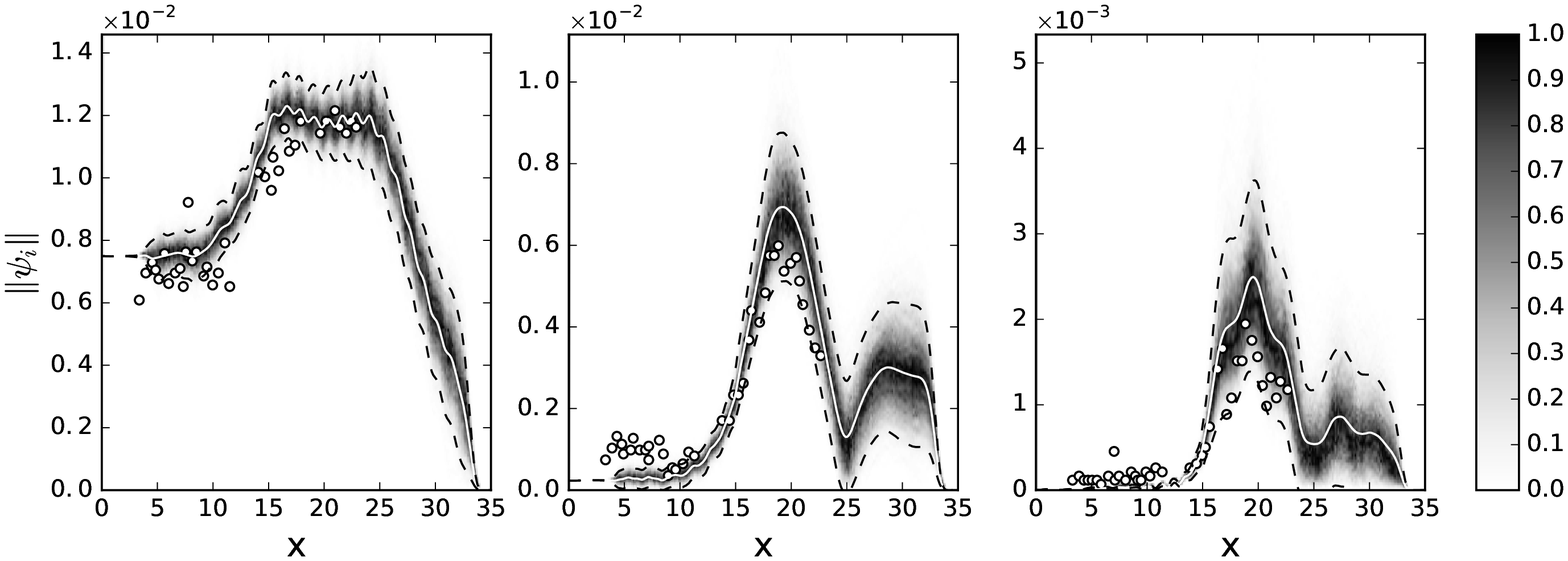}}
  \caption{Figures \protect\subref{fig:WhailnShoal3D:UQField_a30}, \protect\subref{fig:WhailnShoal3D:UQField_a10}, \protect\subref{fig:WhailnShoal3D:UQField_a3} show the approximated space-dependent probability distribution function of the three harmonics of the solution of the Whalin test with uncertain bottom topography described by Gaussian random fields with correlation lengths $a=30.0$, $a=10.0$, $a=3.0$ respectively. The white line shows the space-dependent mean, while the dashed lines show the $95\%$ tolerance interval around the mean. The scattered dots show the experimental data results.}
  \label{fig:WhalinShoal3D:UQbottom}
\end{figure}

The latin hypercube method is used for the three settings, with a number of samples $n=2000$ which is sufficient to estimate the mean up to the second digit of accuracy\footnote{The accuracy estimator is defined similarly to the one used in section \ref{sec:Bartest:Bottom}. For the means and variances $\mu_i(x,y_c),\sigma^2_i(x,y=y_c)$ of the harmonics functions $\{ \hat{f}(x,y=y_c) \}_{i=1}^3$, defined along the center-line of the domain $y=y_c$, the convergence criteria is given by $\max_i\left(\frac{\Vert \sigma_i(x,y) \Vert_\infty / \sqrt{n}}{\Vert \mu_i(x,y) \Vert_\infty}\right)\leq 10^{-2}$.}. 
Figure \ref{fig:WhalinShoal3D:UQbottom} shows the results for the three different correlation lengths of the random field. We can notice that in this case the uncertainty introduced by the field with the shortest correlation length ($a=3.0$) is significantly bigger than the one introduced with longer correlation lengths, in spite of the fact that the total variance of the fields is equivalent. This phenomenon is particularly clear on the first harmonic, where there is a sharp increase in the variance even in the part of the domain with deeper water.

\section{Conclusions}\label{sec:Conclusions}
% \dbnote{I re-wrote the conclusions in light of the new results. I also shortened them, trying to transforming them more to conclusions rather than a summary of the work done. So I addressed the questions:\\a) is it important to account for uncertainties in water waves? \\b) Why do we benefit from non-intrusive PC techniques?\\c) Which are the drawbacks?\\d) What are the conclusions from the experiments?}
The presented numerical experiments of the potential flow model describing water waves have proved to be sensitive to a number of uncertainties. This phenomenon is not surprising, but the analysis showed how relevant its study is for improving the prediction capabilities in coastal and off-shore engineering. The types of uncertainties accounted are those that are likely to appear in experimental settings, such as the input wave characteristics, the water height, and the topography of manufactured basins. \rvnote*{\#6-1}{We made reasonable assumptions regarding} the distribution of such uncertainties, that would otherwise need to be characterized by extensive measurements or by a better description of the believed distributions.

The study of the influence of these uncertainties on problem-dependent quantities of interest benefits greatly from the application of non-intrusive polynomial chaos techniques, such as the stochastic collocation method and the sparse grid method. This is due to three main reasons: i) the convergence properties of these methods allow for super-linear convergence, which is attained on the studied experiments, ii) the calculation of the deterministic solution of the problem is a computationally demanding task, thus restraining the possible number of function evaluation obtainable in reasonable time, iii) the non-intrusive property of the methods allows for the \rvnote*{\#1-1}{reuse} of existing solvers, which have been tuned for High Performance Computing \cite{EngsigKarupEtAl2008,EGNL13}.

On the downside of polynomial chaos methods there is the fact that they are only suitable for problems with a relatively low number of input uncertainties. \rvnote*{\#6-2}{Recalling} that each problem has its own peculiarities (e.g. stochastic dimension, deterministic computational complexity, etc.), we are aware that as the number of random inputs increases, the number of required solutions of the deterministic model increases more than polynomially, making the method not effective. Methods such as sparse grid alleviate this effect, leading to linear growth with respect to the number of random inputs. At the current state of research, pseudo-random sampling techniques are the only one that exhibit a (slow) convergence independent from the number of input uncertainties. 

% It is also important to observe that not all high-dimensional problems are really high-dimensional. One case was shown for the uncertain bottom topography in the submerged bar experiment, where the random field describing the perturbation of the topography has some regularity properties and can be parametrized via the KL-expansion, transforming an ideally infinite dimensional problem to a finite dimensional one. 

The analysis performed on the experimental settings show that the uncertainties on the input wave characteristics and the bottom topography have a relevant effect on the free surface solutions and on the load of these waves on off-shore structures. These effects are amplified when these uncertainties are considered simultaneously, leading to \rvnote*{\#6-3}{nontrivial} transformations of the input probability distribution. The results of such analysis could be considered when explaining some of the discrepancy between numerical solutions and experimental results. In ongoing works we are considering problems with higher stochastic dimensions, resorting to novel techniques for deterministic sampling, such as Adaptive Sparse grid \cite{Conrad2013} and Spectral Tensor Train decomposition \cite{Bigoni}.

The frameworks for random sampling\footnote{\url{https://pypi.python.org/pypi/UQToolbox/}} and for SC method\footnote{\url{https://pypi.python.org/pypi/SpectralToolbox/}}, as well as the results obtained in this work\footnote{\url{http://www2.compute.dtu.dk/~apek/OceanWave3D/}}, are made available on-line and are general enough (non-intrusive) to be applied on both small-scale and large-scale problems with no additional implementation burden.

%The analysis reveals interesting results and in this way adds to traditional deterministic analysis and highlight the efficiency and accuracy attainable with spectral techniques.

%% The Appendices part is started with the command \appendix;
%% appendix sections are then done as normal sections
%% \appendix

%% References
%%
%% Following citation commands can be used in the body text:
%% Usage of \cite is as follows:
%%   \cite{key}          ==>>  [#]
%%   \cite[chap. 2]{key} ==>>  [#, chap. 2]
%%   \citet{key}         ==>>  Author [#]

%% References with bibTeX database:

% \bibliographystyle{spmpsci}
% \bibliographystyle{spbasic}
\bibliographystyle{ieeetr}
\bibliography{biblio}

\end{document}